\documentstyle[psfig]{mn}

\def\etal{{\rm et al.\thinspace}}
\def\eg{{\rm e.g.\ }}

\def\ie{{\rm i.e.\ }}
\def\cf{{\rm cf.\ }}

\def\hii{H{\sc ii}}

\def\halpha{${\rm H}\alpha$}
\def\nH{\hbox{$N_{\rm H}$}}
\def\Mdot{\hbox{$\dot M$}}
\def\Edot{\hbox{$\dot E$}}
\def\h50{\hbox{$h_{50}$\,}}

\def\spose#1{\hbox to 0pt{#1\hss}}
\def\ltsimm{\mathrel{\spose{\lower 3pt\hbox{$\sim$}}
	\raise 2.0pt\hbox{$<$}}}
\def\ltsim{$\mathrel{\spose{\lower 3pt\hbox{$\sim$}}
	\raise 2.0pt\hbox{$<$}}$}
\def\gtsimm{\mathrel{\spose{\lower 3pt\hbox{$\sim$}}
	\raise 2.0pt\hbox{$>$}}}
\def\gtsim{$\mathrel{\spose{\lower 3pt\hbox{$\sim$}}
	\raise 2.0pt\hbox{$>$}}$}

\def\fract#1/#2{\leavevmode\kern.1em                   
   \raise.5ex\hbox{\the\scriptfont0 #1}\kern-.1em
   /\kern-.15em\lower.25ex\hbox{\the\scriptfont0 #2}}
\def\deg{\hbox{$^\circ$}}

\def\Myr{{\rm\thinspace Myr}}
\def\cm{{\rm\thinspace cm}}
\def\erg{{\rm\thinspace erg}}

\def\g{{\rm\thinspace g}}

\def\K{{\rm\thinspace K}}
\def\keV{{\rm\thinspace keV}}
\def\km{{\rm\thinspace km}}

\def\kpc{{\rm\thinspace kpc}}
\def\Lsol{\hbox{$\thinspace {\rm L}_{\odot}$}}
\def\Zsol{\hbox{$\thinspace {\rm Z}_{\odot}$}}

\def\Mpc{{\rm\thinspace Mpc}}
\def\Msol{\hbox{$\thinspace {\rm M}_{\odot}$}}
\def\pc{{\rm\thinspace pc}}
\def\s{{\rm\thinspace s}}

\def\yr{{\rm\thinspace yr}}

\def\ps{\hbox{\s$^{-1}\,$}}

\def\pcc{\hbox{$\cm^{-3}\,$}}
\def\pyr{\hbox{$\yr^{-1}\,$}}

\def\pcm2{\hbox{$\cm^{-2}\,$}}
\def\ergpcm3ps{\hbox{$\erg\cm^{-3}\s^{-1}\,$}}
\def\ergps{\hbox{$\erg\s^{-1}\,$}}

\def\kmps{\hbox{$\km\s^{-1}\,$}}

\def\Lsolppc3{\hbox{$\Lsol\pc^{-3}\,$}}
\def\Msolppc3{\hbox{$\Msol\pc^{-3}\,$}}

\title[Starburst-driven galactic winds: I]
{Starburst-driven galactic winds: I. Energetics and intrinsic X-ray emission}
 
\author[David K.~Strickland and Ian R.~Stevens]
{David K.~Strickland,$^{1}$\thanks{Send offprint requests 
 to D.~K.~Strickland.} \thanks{E-mail: dks@pha.jhu.edu} 
 and Ian R.~Stevens$^{2}$\thanks{E-mail: irs@star.sr.bham.ac.uk} \\  
1. Department of Physics \& Astronomy, The Johns Hopkins University, 
3400 North Charles Street, Baltimore, MD 21218, U.S.A. \\
2. School of Physics \& Astronomy, The University of
Birmingham, Edgbaston, Birmingham, B15 2TT, U.K.} 

\date{Accepted for publication in {\it MNRAS}.}

\pagerange{\pageref{firstpage}--\pageref{lastpage}} 
\pubyear{2000}

\begin{document}

\maketitle
\label{firstpage}

\begin{abstract}

Starburst-driven galactic winds are responsible for the transport
of mass, in particular metal enriched gas, and energy out of galaxies
and into the inter-galactic medium. These outflows
directly affect the chemical evolution of galaxies, and heat and enrich
the inter-galactic and inter-cluster medium.

Currently several basic problems preclude quantitative measurements
of the impact of galactic winds: 
the unknown filling factors of, in particular,
the soft X-ray emitting gas prevents accurate measurements of densities,
masses and energy content; multiphase temperature distributions
of unknown complexity
bias X-ray determined abundances; unknown amounts of energy and mass
may reside in hard to observe $T \sim 10^{5} \K$ and $T \sim 10^{7.5} \K$
phases; and the relative balance of thermal vs. kinetic energy
in galactic winds is not known.

In an effort to address these problems we
have performed an extensive hydrodynamical parameter study of
starburst-driven galactic winds, motivated by
the latest observation data on the best-studied starburst 
galaxy M82. 
We study how the wind dynamics, morphology and X-ray emission 
depend on the host galaxy's ISM distribution,
starburst star formation history and strength, 
and presence and distribution
of mass-loading by dense clouds. 
We also investigate and discuss the influence of finite numerical resolution
on the results of these simulations.

We find that the soft X-ray emission from galactic winds comes 
from low filling factor ($\eta \ltsimm 2$ per cent) gas, which
contains only a small fraction ($\ltsimm 10$ per cent) 
of the mass and energy of the wind, irrespective of whether
the wind models are strongly mass-loaded or not. X-ray observations
of galactic winds do not directly probe the gas that contains
the majority of the energy, mass or metal-enriched gas in the outflow.

X-ray emission comes from a complex phase-continuum of gas, covering a wide
range different temperatures and densities. No distinct phases,
as are commonly assumed when fitting X-ray spectra, are
seen in our models. Estimates of the properties of
the hot gas in starburst galaxies based on fitting simple spectral
models to existing X-ray spectra
should be treated with extreme suspicion. 

The majority of the thermal and kinetic
energy of these winds is in a volume filling hot, $T \sim 10^{7} \K$, component
which is extremely difficult to probe observationally due to its
low density and hence low emissivity. Most of the total
energy is in the kinetic energy of this hot gas, a factor
which must be taken into account when attempting to constrain
wind energetics observationally.
We also find that galactic winds are efficient at transporting
large amounts of 
energy out of the host galaxy, in contrast to their inefficiency
at transporting mass out of star-forming galaxies.

\end{abstract}


\section{Introduction}
\label{sec:introduction}

Starbursts, episodes of intense star formation lasting $\ltsimm 10^{8}$
years, are now one of the cornerstones of the modern view of galaxy formation
and evolution. Starbursts touch on almost all aspects of extra-galactic
astronomy, from the processes of primeval galaxy formation at high 
redshift to being a significant mode of star formation even in the present
epoch, and covering systems of all sizes from 
dwarf galaxies to the dust-enshrouded starbursts in ultraluminous 
merging galaxies (see Heckman 1997 for a recent review).

An inescapable consequence of a starburst is the driving of a powerful
galactic wind (total energy $10^{54} \le E \le 10^{58} \erg$, velocity 
$v \gtsimm 10^{3} \kmps$) from the host galaxy into the 
inter-galactic medium (IGM) due to the return
of energy and metal-enriched gas into the inter-stellar medium (ISM)
 from the large numbers of 
massive stars formed during the burst (Chevalier \& Clegg 1985, hereafter CC; 
McCarthy, Heckman \& van Breugel 1987).

The best developed theoretical model for starburst-driven galactic winds,
elaborated over the years by various workers (see the
review by Heckman, Lehnert \& Armus 1993. See also
Tomisaka \& Bregman 1993; Suchkov \etal
1994), is of outflows of supernova-ejecta and swept-up ISM driven by the
the mechanical energy of multiple type II supernovae and stellar
winds from massive stars. This paradigm is very successful at
explaining almost all of the observed properties of galactic winds,
and can reproduce quantitatively what is known of the kinematics 
and energetics of observed local starburst-driven outflows.

Galactic winds are unambiguously detected in many local edge-on starburst
galaxies (Lehnert \& Heckman 1996), and their presence can even be inferred in
starbursts at high redshift (\eg Pettini \etal 1999). 
Filamentary optical emission line gas, soft thermal X-ray emission 
and non-thermal radio emission, all extended preferentially along 
the minor axis of the galaxy and emanating in a loosely collimated flow from
a nuclear starburst, are all classic signatures of a galactic wind.
In the closest and brightest
edge-on starburst galaxies the outflow can be seen in all phases of the ISM,
from cold molecular gas to hot X-ray emitting plasma (Dahlem 1997).

Galactic winds 
are of cosmological importance in several ways:
\begin{enumerate}
\item The transport of metal-enriched gas out of galaxies by such winds
affects the chemical evolution of galaxies and the IGM. This effect
may be extremely important in understanding the chemical evolution of
dwarf galaxies where metal ejection efficiencies are expected
to be higher (Dekel \& Silk 1986; Bradamante, Matteucci \& D'Ercole 1998).

\item Galactic winds may also be responsible for reheating the IGM, evidence
of which is seen in the entropy profiles of gas in the inter-cluster
medium (ICM) of groups and clusters (Ponman, Cannon \& Navarro 1999).
A substantial fraction of the metals now in the ICM were probably
transported out of the source galaxies by early galactic winds 
(\eg Loewenstein \& Mushotzky 1996).

\item Galactic winds are an extreme mode of the 
``feedback'' between star formation
and the ISM. This feedback is a necessary, indeed vital, ingredient
of the recipes for galaxy formation employed in todays cosmological
N-body and semi-analytical models of galaxy formation. An aspect
of feedback where galactic winds will have an important effect,
and where the existing prescriptions for feedback need updating
with less ad hoc models, is in the escape of hot gas from haloes 
of galaxies, in particular low mass galaxies. This directly affects
the faint end of the galaxy luminosity function in  semi-analytical
galaxy formation models (\eg Kauffman, Guiderdoni \& White 1994;
Cole \etal 1994; Somerville \& Primack 1999), 
as recently discussed by Martin (1999). 
\end{enumerate}

Assessing the importance of starburst-driven winds 
quantitatively requires
going beyond what is currently known of their properties.
It is necessary to make more
quantitative measurements of parameters such as mass loss rates, 
energy content and chemical abundances, and how these relate to
the properties of the underlying starburst and host galaxy.

This in turn requires a deeper understanding the basic physics 
of such outflows,
and the mechanisms underlying the multi-wavelength
emission we see. In particular, the origin of the soft X-ray emission
seen in galactic winds is currently uncertain, with several
different models currently being advanced
(as we shall discuss below). This uncertainty in what we are actually
observing makes estimating the
total mass and energy content of these winds difficult.

Recent theoretical and hydrodynamical models (De Young \& Heckman 1994;
Mac Low \& Ferrara 1999; D'Ercole \& Brighenti 1999) suggest that
starburst-driven winds are in general not efficient at ejecting 
significant amounts of the host galaxy's ISM, in the single burst
scenarios that have been explored until now.
More complex star formation histories have yet to be explored numerically,
but qualitative arguments suggest mass loss rates will be even lower
than in single burst scenarios (Strickland \& Stevens 1999).
This is beginning to overturn the popular concept of catastrophic
mass loss in dwarf galaxies due to galactic winds 
advanced by Dekel \& Silk (1986) and Vader (1986). 

Despite the sophistication of these and other
recent models, it has not been shown that these simulations
reproduce the observed kinematics, energetics and emission properties of
any real starburst-driven outflow. A very wide range of model
parameters can produce a bipolar outflow, and only a relatively small
number of models have been run which explore only a limited parameter space.
More rigorous tests and comparisons of the {\em observable} properties
of the different theoretical models 
against the available observational data are now
required to judge the relative successes and failures of the
current theory.

Observational attempts to directly measure the mass and energy content
of galactic winds, using optical or X-ray observations
(\cf Martin 1999; Read, Ponman \& Strickland 1997; 
Strickland, Ponman \& Stevens 1997) have been made.
Unfortunately these may only be accurate to an order of magnitude, given
that the volume filling factors of the cool $T \sim 10^{4} \K$ and hot
$T \sim 10^{6.5} \K$ gas phases that are probed by these observations
are unknown. 
This uncertainty in filling factor affects all observational studies of
the hot gas in starburst galaxies, such as Wang \etal (1995), Dahlem
\etal (1996), Della Ceca, Griffiths \& Heckman (1997) to name only
a few.

The soft X-ray emission from galactic winds and
starburst galaxies is well fit by thermal plasma
models of one or more components with temperatures in the range
$kT \sim 0.1$ to $1.0 \keV$ (\cf Read \etal 1997; Ptak \etal 1997; 
Dahlem, Weaver \& Heckman 1998 among many others). 

A variety of models have been put forward to explain 
the soft X-ray emission from galactic winds. Currently
the origin and physical state of the emitting gas is not clear,
either observationally or theoretically.
There is little disagreement
that the diffuse soft X-ray emission comes from some form of hot gas.
The main uncertainties
lie in the filling factor and thermal distribution of this gas.
These in turn affect the degree to which soft X-ray observations
provide a good probe of the important properties of galactic winds that
we need to measure --- the mass and energy content and the chemical
composition.

Different models for the origin of the soft X-ray emission
from galactic winds
range from shock-heated clouds (of low volume filling factor)
embedded in a more tenuous wind (\eg CC),
through conductive interfaces between hot and cold gas
(\eg D'Ercole \& Brighenti 1999), to emission from a volume-filling
hot gas where the wind density has been increased by the hydrodynamical
disruption of clouds overrun by the wind (\eg Suchkov \etal
1996).

In M82 the X-ray emission occupies a similar area in projection to both
the emission line filaments (Watson, Stanger \& Griffiths 1984, 
Shopbell \& Bland-Hawthorn 1998)
and to the radio emission (Seaquist \& Odegard 1991). 
Although  existing X-ray observations
of M82 and other galactic winds
do not have the spatial resolution necessary to constrain the exact
relationship between the emission line gas and the hot gas,
the general similarity in the two spatial distributions have
prompted models where the soft X-ray emission comes from
shock-heated clouds (\cf Watson \etal 1984;
CC). In this hypothesis both the
optical line emission and the soft X-ray emission
come from clouds shocked by a fast tenuous, and presumably hotter,
wind that the clouds are embedded in.  The wind drives
fast shocks into less tenuous clouds causing soft thermal X-ray 
emission, and slower shocks into denser clouds causing optical emission.
The clouds
occupy very little of the total volume, but dominate the
total emission. 
The distribution of clouds within the wind
 hence determines both the 
observed distribution of optical and X-ray emission.
The temperature of the X-ray-emitting gas is determined by the
speed of the shock waves driven into them, which is then
determined by the density of the clouds and the density and
velocity of the wind running into them.
Two dimensional hydrodynamical models of galactic winds (\eg
Tomisaka \& Ikeuchi 1988; 
Tomisaka \& Bregman (1993); Suchkov \etal 1994, here after
TI, TB and S94 respectively) strongly favor
interpretations of the X-ray emission coming from shock-heated
ISM overrun by the wind.

In this model we do not see the ``wind'' itself,
as it is too tenuous to emit efficiently enough to be detected.
If this model is correct, then X-ray observations do not directly 
probe the heavy element-enriched wind fluid that drives the outflow
and contains most of the total energy.

D'Ercole \& Brighenti (1999) suggest that S94's conclusion, that
the majority of soft X-ray emission in their hydrodynamical
simulations arise from shocked disk and halo gas, was incorrect.
They point out that the numerically unresolved interfaces between
cold and hot gas have the correct temperature and density to 
produce large amounts of soft X-ray emission. Such regions are
almost inevitable in hydrodynamical simulations, and would be 
very difficult to distinguish from
regions of cold disk gas shock-heated by the surrounding hot
wind material.

In reality thermal conduction can lead to physically broadened
interface regions, which could be a significant source of soft thermal
X-ray emission in galactic winds. Such conductive interfaces
are believed to dominate the X-ray emission in wind-blown
bubbles (Weaver \etal 1977), which are very similar to the
superbubbles young starbursts blow.

Fabbiano (1988) and Bregman, Schulman \& Tomisaka 
(1995) explicitly interpret the
X-ray emission from M82 in terms of it being from an adiabatically
expanding hot wind, in contrast to shock-heated
clouds model above. The temperature and density of the gas in such
a model of a volume-filling X-ray emitting wind is determined by the
energy and mass injection rates within the starburst, and also
by the outflow geometry which controls the degree of adiabatic
expansion and cooling the wind experiences. If this model
is correct then
soft X-ray observations provide a good probe of the hot gas driving the
outflow, and hence of the metal abundance, mass and energy content
of starburst-driven winds.

CC had explicitly rejected the wind itself being the
source of the X-ray emission seen in M82 by Watson \etal (1984).
The problem is that, for reasonable estimates of the wind's mass and energy
injection rate based on M82's supernova (SN) rate, the outflow
has a very low density. 
As the X-ray emissivity is proportional to the square of
the density, the resulting X-ray luminosity is  extremely low. 
For example, for a SN rate of $0.1 \pyr$ with a resultant wind mass injection
rate of $\Mdot_{\rm w} \sim 1 \Msol \pyr$, and a starburst region
of radius $R_{\star} = 150 \pc$, the resulting
total $0.1$ -- $2.4 \keV$ 
X-ray luminosity from within a radius $R = 10 \times R_{\star}$ is only
$3.2 \times 10^{37} \ergps$, about 60 per cent of which comes from
within the starburst region itself. This is a factor $10^{-5}$ times
the starburst's wind energy injection rate,
and considerably less than the
observed  $0.1$ -- $2.4 \keV$ luminosity of M82's wind of $L_{\rm X} \sim
2 \times 10^{40} \ergps$ (Strickland \etal 1997).

It is possible to rescue the concept of a volume filling wind
fluid being responsible for the observed X-ray luminosity if the wind
has been strongly mass loaded (\ie additional mass has been efficiently mixed
into the wind) --- a theoretical model presented by Suchkov \etal (1996,
here after S96)
and subsequently explored further by Hartquist, Dyson
\& Williams (1997). 
Increasing the mass injection rate into the wind by a factor $N$ increases the
wind density by $N^{1.5}$ and emissivity by a factor $N^{3}$,
as not only is there more mass in the wind but its outflow velocity
is lower.

Bregman \etal (1985) used {\it ROSAT} HRI observations of M82 to argue that
the observed X-ray surface brightness distribution was consistent
with a well-collimated adiabatically expanding hot gas. However,
analysis of the spectral properties of a set of regions along
M82's wind using {\it ROSAT} PSPC data by Strickland \etal (1997)
shows that the entropy of the soft X-ray emitting gas
 increases with distance from the
plane of the galaxy, which is inconsistent with an adiabatic outflow
model. 

A conservative assessment would be that X-ray observations do not
strongly constrain the origin of the soft X-ray emission in galactic winds
beyond that it is from a hot thermal plasma. The existing observations
are broadly consistent with any of the models advanced above:
shocked clouds, thermal conduction or mass-loading.

Even assessing theoretically 
the relative importance of processes such as mass-loading
as compared to shock-heating or thermal conduction has not been
possible up until now. S96 argued that M82's wind must be mass-loaded
to produce the required soft X-ray luminosity and temperatures.
However, their mass loaded wind  simulations 
did not include the interaction of the wind with the ambient
ISM, which S94 had showed was capable of providing the observed
X-ray luminosity.

Uncertainties in the filling factor are not the only problems
affecting the interpretation of soft X-ray data on galactic winds.
Understanding the temperature distribution of the X-ray emitting
gas is also an important, if relatively unexplored, theoretical aspect
of galactic winds.
Deriving plasma properties from X-ray spectra requires 
fitting a spectral model that is a good approximation to the
true emission process. Failure to do so can lead to models
that fit the data well but give meaningless results.

A good example of this is X-ray derived metal abundances, where many
{\it ROSAT} and {\it ASCA} studies of starburst galaxies report
extremely low metal abundances, between $0.05$ to $0.3$ times
Solar (\cf Ptak 1997, Ptak \etal 1997, Tsuru \etal 1997, Read \etal 1997)
for the soft thermal plasma components.
We believe this to be primarily an artefact of using overly simplistic
spectral models to fit the limited-quality data available from these
missions. Consistent with the idea that the X-ray emission comes
from a complex range of temperatures, Dahlem \etal (1998) have 
shown that, when using multiple hot plasma components to represent the
soft thermal emission from galactic winds, most of the galaxies
in their sample could be fit using Solar element abundances.
Strickland \& Stevens (1998), using simulated {\it ROSAT} PSPC observations
of wind-blown bubbles (physically very similar to superbubbles
and the early stages of galactic winds), show that under-modeling
the X-ray spectra leads to severely underestimating the metal
abundance. 

Failure to correctly fit one parameter such as the metal abundance
can also severely bias other plasma properties. For example, the
derived emission measure (proportional to the density
squared integrated over the volume) is approximately inversely
proportional to the fitted metal abundance
for gas in the temperature range $T = 10^{5}$ to $10^{7} \K$. 
Underestimating the
metal abundance then leads to overestimates of the gas density,
and hence gas mass and energy content.

 In an effort to address many of the issues described above,
we have performed the most detailed hydrodynamical simulations
of galactic winds to date. We investigate a larger volume of model parameter 
space than any previous numerical study, inspired by the latest
observational studies of the archetypal starburst galaxy M82. We
study the variation in the properties of these galactic winds due to
the starburst star formation history and intensity, the host galaxy
ISM distribution and the presence and distribution of mass-loading
of the wind by dense clouds. 

Our aim in this series of papers is to go beyond the previous
hydrodynamical simulations of galactic winds, and to:
\begin{enumerate}
\item Explore 
      in a more systematic manner the available model
      parameter space, focusing on M82 (the best-studied starburst
      with a galactic wind), using 2-D hydrodynamical
      simulations run at high numerical resolution.
\item Investigate several important aspects of galactic winds that
      are currently uncertain or have previously not been devoted
      much attention:
  \begin{itemize}
    \item  The origin and filling factor of the
      soft X-ray emitting gas, along with the temperature distribution
      of this gas.
    \item The wind energy budget and energy transport efficiency, and
      the degree to which the energy content can be probed by soft
      X-ray observations.
    \item Wind collimation. Producing the observed properties of
      well collimated winds with narrow bases seems to be a
      problem for the simulations of TI, TB and S94, as has been
      pointed out by Tenorio-Tagle \& Mu\~{n}oz-Tu\~{n}\'{o}n (1997, 1998).
  \end{itemize}
\item Assess the effect of finite numerical resolution and other numerical
      artefacts on the results of these simulations.
\item Directly compare the observable properties of these models (primarily
      concentrating on their soft X-ray emission) to the available 
      observational data for M82 and other local starburst galaxies.
\end{enumerate}

In paper II we discuss the {\em observable} X-ray properties of the winds 
in these simulations. Focusing on artificial {\it ROSAT} PSPC 
X-ray images and spectra (the PSPC has been the best X-ray instrument
for the study of superbubbles and galactic winds due to it's
superior sensitivity to diffuse soft thermal emission over other 
instruments such as the {\it ROSAT} HRI or the imaging spectrometers 
upon {\it ASCA}) we consider (a) the success of these models
at reproducing the observed X-ray properties of M82, and
(b) the extent to which such X-ray observations and
standard analysis techniques can 
allow us to derive the true properties of the hot plasma in
these outflows.

In common with the previous simulations of TI, TB, S94 and S96 we shall
base our models largely on the nearby ($D = 3.63 \Mpc$, Freedman \etal 1994)
starburst galaxy M82. M82 is the best studied starburst galaxy, and after
NGC 253 and NGC 1569, it is the closest starburst with a galactic wind. 
As the observational constraints on M82's starburst and galactic wind
are the best available for any starburst galaxy, we choose to concentrate
on models aiming to reproduce M82's galactic wind.

\section{Numerical modeling}
\label{sec:model_param}

The primary variables that will affect the growth and evolution of
a superbubble and the eventual galactic wind in a starburst galaxy
are the ISM density distribution,
the strength and star-formation history of the starburst, and the
presence, if any, of additional interchange processes between the
hot and cold phases of the ISM, such as hydrodynamical or conductive
mass loading.  

As a detailed exploration of such a multidimensional parameter space
for all possible galactic winds
would be prohibitively expensive computationally,
we choose to focus on simulations of M82's galactic wind, 
given that is the best studied starburst
galaxy. Our aim to to attempt to roughly bracket the range of
possible ISM distributions, starburst histories 
and mass-loading occurring in M82, in a series of ten simulations
based on recent observational studies
of this fascinating galaxy. Two further simulations explore the 
degree to which finite numerical resolution affects our computational
results.

The majority of this section explores our choice of model parameters,
preceded by information on the hydrodynamical code and analysis
methods we use.

\subsection{Hydrodynamic code}

Our simulations of starburst-driven galactic winds have been performed
using {\sc Virginia Hydrodynamics-1} (VH-1), a high resolution
multidimensional astrophysical hydrodynamics code developed by
John Blondin and co-workers (see Blondin 1994; Blondin \etal 1990; 
Stevens, Blondin \& Pollock 1992).
VH-1 is based on the piecewise parabolic method (PPM) of Collela and Woodward 
(1984), a third-order accurate extension of Godunov's (1959) method.

VH-1 is a PPM with Lagrangian remap (PPMLR) code, in that as it
sweeps through the multi-dimensional computational grid of
fluid variables it remaps
the fixed Eulerian grid onto a Lagrangian grid, solves the Riemann problem
at the cell interfaces, and then remaps the updated fluid variables
back onto the original Eulerian grid. Moving to a Lagrangian frame 
simplifies solving the Riemann problem, and results in a net
increase in performance over PPM codes using only an Eulerian grid.

For the purposes of these simulations VH-1 is run in 2-D, assuming cylindrical
coordinates ($r$, $z$, $\phi$) and azimuthal symmetry around the $z$-axis.
Radiative cooling, mass and energy injection from the starburst region, 
and mass deposition due to the hydrodynamical ablation of dense unresolved
clouds are incorporated using standard operator splitting techniques. 
As in the previous 
simulations of TI, TB and S94 only one quadrant of the flow is calculated,
symmetry in the $z$ and $r$-axes is assumed and determines the boundary 
conditions that operate along these edges of the computational grid. Material
is allowed to flow in or out along the remaining two grid boundaries.

Although dependent on the exact model parameters, we have
generally us a computational grid
formed of a rectangular grid of 
400 uniform zone along the $r$-axis,
covering a physical region $\sim 6 \kpc$ long, by 800 equal-sized
zones along the $z$-axis covering $\sim 12 \kpc$. Simulations
are run until the outermost shock of the galactic wind flows off
the computational grid, typically at $t \sim 10$ -- $15 \Myr$
after the start of the starburst.

\begin{figure}
\centerline{
 \psfig{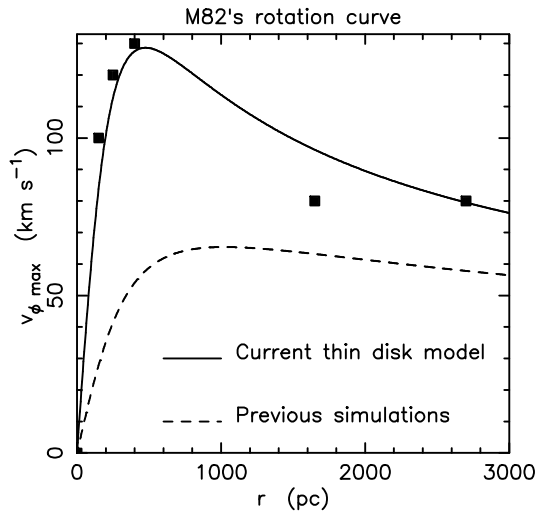}
	}
\caption[M82's rotation curve compared to those assumed in the simulations]
 {M82's rotation curve (filled squares, from G\"otz \etal 1990) 
 compared to the rotation curves 
  $v(r) =   (r \partial \Phi_{\rm tot} / \partial r )^{1/2}$,
resulting from the mass distributions assumed in
 our thin disk models and the simulations of Tomisaka \& Ikeuchi (1988),
 Tomisaka \& Bregman (1993) and Suchkov \etal (1994).}
\label{fig:rotcurve}
\end{figure}

\subsection{Data analysis}
\label{sec:hydro_analysis}
Our results are based on the analysis of data files of the
fluid variables $\rho$, $P$ and ${\vec v}$ that VH-1 produces at
$0.5 \Myr$ intervals.

We make a distinction between the intrinsic properties of the wind
(\eg the ``true'' gas temperatures, densities, masses, 
energy content and luminosities, which are described in this paper)
and the observable X-ray properties (covered in the forthcoming
companion paper). The observed properties
need not necessarily directly reflect the true, intrinsic, wind properties
due to the complications of projection, instrumental limitations
and the systematic and statistical errors inherent in real observations.

We analyse only the material within the wind, \ie swept-up disk 
and halo ISM along with enriched starburst
ejecta.  The undisturbed ISM is ignored. Quantitatively the wind is
distinguished from the undisturbed ISM by the shock that marks its outer
boundary. The 2-D fluid variables are converted into a 3-D dataset covering
both poles of the wind by making use of the
assumed symmetries around the $z$-axis and in the plane of the galaxy.

\subsection{Calculation of intrinsic wind properties}
\label{sec:analysic_intrinsic}
The value of the fluid variables are assumed to be constant within each
computational cell for the purposes of all the data analysis performed
on these simulations. We do not use the parabolic interpolation used 
by VH-1's hydrodynamic scheme 
to reconstruct the variation of the fluid variables within each cell
from the cell averaged values in our data analysis. Numerically
summing the appropriate quantities, over those cells within the 3-D
dataset that are within the region defined as being the galactic wind,
gives the total integrated wind properties such as mass, volume, 
energy content and X-ray luminosity. X-ray properties such
as emissivities or instrument-specific 
count rates are obtained assuming a mean mass
per particle of $10^{-24} \g$ and collisional ionisation equilibrium,
using either the MEKAL hot plasma code (Mewe, Kaastra \& Liedahl 
1995) or an updated
version of the  Raymond-Smith plasma code (Raymond \& Smith 1977), 
along with published instrument
effective areas and the Morrison \& McCammmon (1983) absorption
coefficients.

In addition to studying the total integrated values of wind volume, 
mass, emission measure, and similar properties, we can investigate
how these values are distributed as a function of gas temperature,
density or velocity within the wind.

\subsubsection{Starburst disk and halo properties}

Many of the ``classic'' starburst galaxies with clear
galactic winds such as M82, NGC 253, NGC 3628 and NGC 3079 are almost
edge-on, allowing the low surface brightness optical and X-ray
emission from  the wind to be seen above the sky background. Had these
galaxies been face on (\eg the starburst galaxy M83)
it would have been much more difficult to distinguish
emission from the wind from the X-ray and optically 
bright disks of these galaxies.

Previous simulations such as TB and S94 have only considered the total
properties of galactic winds integrated over the entire volume of the
outflow. We feel it is important to distinguish between the properties
of the gas associated with the galactic wind within the disk of
the host galaxy, and the gas clearly above the disk (in what would be
observationally identified as the wind in an edge-on starburst).
A more ambitious study might divide the wind up into several different
sections, as done in Strickland \etal (1997), but for now we will only
consider two separate regions: disk or halo.

The same techniques of analysis used on the entire wind described above
are applied to gas within the plane
or disk of the galaxy (defined as being the region with $|z| \le 1.5 \kpc$)
and gas above the plane in the halo ($|z| > 1.5 \kpc$). This allows us
to study the spatial variation of the wind properties in a simple but
meaningful manner.

\subsection{ISM distribution and gravitational potential}
\label{sec:ism_distribution}

We adopt the philosophy of TB and S94 in setting up an initial ISM
distribution of a cool disk and hot halo, in rotating hydrostatic equilibrium
under the applied gravitational field. Although the ISM  in
starburst galaxies is not static or in equilibrium, we choose a 
static solution to prevent the dynamics of an arbitrary 
non-static ISM distribution from affecting the wind dynamics.

The ISM modeled in these 2-D numerical simulations and in TI, TB \& S94
is essentially a single-phase volume filling gas.
Although in reality the ISM of late-type galaxies is clearly
multi-phase we make the assumption that the most important phase,
with respect to the ISM's interaction with the starburst-driven wind,
is that ISM component
that occupies the majority of the volume. In the disk this might be cool
$\sim 10^{4} \K$ gas and coronal gas in the halo. 
Mass-loaded simulations such as S96 treat the interaction of the wind
with dense, low volume filling factor clouds. Although the majority of the 
mass in the ISM is in molecular clouds 
these occupy only a small fraction of the
total volume, even within a starburst region such as the centre of M82.
Lugten \etal (1986) find that the molecular gas within the nuclear
region of M82 is comprised of small clouds, thin filaments or sheets
with a total volume filling factor $\eta \sim 10^{-3}$.

To investigate the effect of the ISM on the collimation and confinement of the
wind, and how this alters the observable X-ray properties of the galactic
wind, we study two different ISM distributions. We adopt TB's ISM distribution
as representative of a thick disk with dense cool gas high 
above the plane of the galaxy, which provides substantial wind collimation. 
The gravitational potentials used by TB and S94 
do not reproduce M82's observed rotation curve, as can be seen in 
Fig.~\ref{fig:rotcurve}. The maximal rotational velocity in TB and S94's models
is a factor two lower than that observed in M82. Consequently, 
with less enclosed mass,
the gravitational force felt by the wind in S94 and TB's simulations is too
low. For our thin disk model we remedy this by adding the potential due to
a Miyamoto \& Nagai (1975) disk to the stellar spheroid used in TB.
The resulting disk using this combined potential 
is significantly thinner than that of TB.

The large scale ISM distribution in M82 is very poorly known, 
and the majority of investigations into the ISM in M82 have 
focussed on the wind or the conditions deep within the starburst 
region. The current ISM distribution in M82 may not be representative of
the initial ISM distribution prevailing at the time of the birth 
of the starburst. We do not claim these ISM distributions used in our 
simulations correspond to the true ISM in a starburst galaxy such as M82, 
but they capture essential features which we wish to investigate, namely
the effect of collimation on the wind dynamics and the resulting X-ray 
emission.

The initial static ISM distribution is computed as a solution to the
steady state momentum equation
\begin{equation}
 ({\vec v} \cdot \nabla){\vec v} = 
 - \frac{1}{\rho} \nabla P - \nabla \Phi_{\rm tot},
\label{equ:ss_momentum}
\end{equation}
where ${\vec v}$, $\rho$, $P$ and $\Phi_{\rm tot}$ are the gas velocity,
density, pressure and total gravitational potential respectively.

We assume the ISM is supported predominantly by rotation in the plane of the
galaxy, so the rotational velocity $v_{\phi}$
(\ie azimuthal component of velocity) in the plane of the galaxy  is given by
\begin{equation}
  v_{\phi}(r) = e_{\rm rot} 
        \left( r \frac{\partial \Phi_{\rm tot}}{\partial r} \right)^{1/2}.
\end{equation}
If the disk were entirely supported by rotation then $e_{\rm rot} = 1$. In
these simulations a small fraction of the gravitational force in the plane
is supported by the turbulent and thermal pressure of the ISM. For
the thick disk we choose $e_{\rm rot} = 0.9$, and for the thin disk 
$e_{\rm rot} = 0.95$.

To reduce the rotational velocity of the gas
 above the plane (as seen in NGC 891
by Swaters, Sancisi \& van der Hulst [1997] for example), 
and have a non-rotating galactic halo we assume a simple 
model where the rotational support (and hence $v_{\phi}$) drops off
exponentially with increasing height $z$ above the plane of the
galaxy
\begin{equation}
e = e_{\rm rot} \exp (-z / z_{\rm rot}),
\label{equ:erot}
\end{equation}
where the scale height for this reduction in rotational velocity
$z_{\rm rot} = 5 \kpc$ as used in TB.

In 2-D numerical simulations such as these which assume cylindrical symmetry
around the $z$-axis (\ie azimuthally symmetric) only the $r$ and $z$ components
of the fluid variables $\rho$, $P$ and ${\vec v}$ are calculated. Rotational
motion is simulated by solving a modified form of 
Eqn.~\ref{equ:ss_momentum}:
\begin{equation}
 - \frac{1}{\rho} \nabla P - \nabla \Phi_{\rm eff} = 0,
\end{equation}
where $\Phi_{\rm eff}$ is the effective potential, the sum of the true
gravitational potential and the centrifugal potential arising from
the incorporation of the rotational motions. Similarly the 
force required to hold the disk in equilibrium against the pressure gradient
in these 2-D simulations is not just the true gravitational
force ${\bf g} = - \nabla \Phi_{\rm tot}$ but the effective gravitational force
 ${\bf g_{\rm eff}} = - \nabla \Phi_{\rm eff}$.

The gravitational potential used in our thick disk model is identical to that
used by TB, a single spherically symmetric component 
identified with the central stellar spheroid.
Note that there are typographical errors in Eqns.~2, 3, 4 \& 7 of TB, so
the equations presented here differ from those presented in TB.

We use a King model to describe the stellar spheroid,
\begin{equation}
\rho_{\rm ss}(\omega) = \frac{\rho_{0}}{[1 + (\omega/\omega_{0})^{2}]^{3/2}},
\label{equ:king}
\end{equation}
where $\rho_{\rm ss}(\omega)$ is the stellar density as a function of the
radial distance from the centre $\omega$, $\rho_{0}$ the central 
density and $\omega_{0}$ the core radius. 
The gravitational potential $\Phi_{\rm ss}$ due to this stellar 
spheroid is then
 \begin{equation}
  \Phi_{\rm ss}(\omega) = - \frac{G M_{\rm ss}}{\omega_{0}}
  \left[ \frac{ \ln \{ ( \omega / \omega_{0} ) +
  \sqrt{1 + (\omega / \omega_{0})^{2}} \} }{\omega / \omega_{0}} \right],
 \end{equation}
where $\omega = \sqrt{r^{2} + z^{2}}$ is the distance from the nucleus, and
we define $M_{\rm ss} = 4 \pi \rho_{\rm ss} \omega_{0}^{3}$.
Following TB, we choose 
$M_{\rm ss} = 1.2 \times 10^{9} \Msol$ and $\omega_{0} = 350 \pc$.

For the thin disk model we retain a King model to represent the central
stellar spheroid, and use a Miyamoto \& Nagai (1975) 
disk potential to represent
the disk of the galaxy,
\begin{equation}
\Phi_{\rm disk}(\omega) = - \frac{G M_{\rm disk}}
        {\sqrt{r^{2} + (a + \sqrt{z^{2} + b^{2}} )^{2}}}, 
\end{equation}
where $a$ and $b$ are the radial and vertical scale sizes of the disk.
The total potential in the thin disk model is
$\Phi_{\rm tot} = \Phi_{\rm ss} + \Phi_{\rm disk}$.
To approximately reproduce M82's rotation curve (Fig.~\ref{fig:rotcurve}) 
we use 
$M_{\rm ss} = 2 \times 10^{8} \Msol$, $\omega_{0} = 350 \pc$,
$M_{\rm disk} = 2 \times 10^{9} \Msol$, $a = 222 \pc$ and $b = 75 \pc$.

We do not incorporate an additional massive dark matter halo component
into the current set of simulations, as M82's rotation curve is
well described by the chosen stellar spheroid plus Miyamoto \& Nagai
disk model. We are primarily interested in the behaviour of the winds
over the initial $\sim 20 \Myr$ of the starburst, a period during which the
gravitational effects of any dark matter halo will be negligible. 
Dark matter haloes may
shape the long term behaviour of material in weak winds 
(\eg the \mbox{2-D} simulations of winds in dwarf galaxies by 
Mac Low \& Ferrara 1999), and the fate of 
slowly moving gas dragged out of the disk in starburst-driven winds
like M82. These simulations are not designed to investigate the long term 
fate of this gas, but the dynamics and properties of the gas in
the observed galactic winds which have dynamical ages of $\sim 10^{7} \yr$.

In common with TB, we incorporate a disk ISM of central density 
$n_{\rm disk, 0} = 20 \pcc$ and a tenuous halo of central density 
$n_{\rm halo, 0} = 2 \times 10^{-3} \pcc$ in both models. 
The initial ISM density and pressure is then given by
\begin{equation}
 \begin{array}{ll}
\rho_{\rm disk}(r,z) & = \rho_{\rm disk, 0} \, \times \\
                       & \exp  \left[ - \frac{\Phi_{\rm tot}(r,z) 
        - e^{2} \Phi_{\rm tot}(r,0) - (1-e^{2})\Phi_{\rm tot}(0,0)}
        {c_{\rm s, disk}^{2}} \right],
 \end{array}
\label{equ:rho_disk}
\end{equation}
\begin{equation}
 \begin{array}{ll}
\rho_{\rm halo}(r,z) & = \rho_{\rm halo, 0} \, \times \\
 	& \exp \left[ - \frac{\Phi_{\rm tot}(r,z) 
        - e^{2} \Phi_{\rm tot}(r,0) - (1-e^{2})\Phi_{\rm tot}(0,0)}
        {c_{\rm s, halo}^{2}} \right],
\label{equ:rho_halo}
 \end{array}
\end{equation}
\begin{equation}
\rho(r,z) = \rho_{\rm disk}(r,z) + \rho_{\rm halo}(r,z), 
\end{equation}
and
\begin{equation}
P(r,z) = \rho_{\rm disk}(r,z) c_{\rm s, disk}^{2} 
        + \rho_{\rm halo}(r,z) c_{\rm s, halo}^{2},
\end{equation}
where $e$  quantifies the fraction of rotational support of the ISM as
given in Eqn.~\ref{equ:erot}, and $c_{\rm s, disk}$ and $c_{\rm s, halo}$
are the sound speeds in the disk and halo respectively. 
The complexity of Eqns.~\ref{equ:rho_disk} 
\& \ref{equ:rho_halo}
is due to incorporating rotational support through an effective potential. 
Without the centrifugal potential, these equations would be of the form 
$\rho = \rho_{0} \exp [ - \{ \Phi_{\rm tot}(r,z) - 
\Phi_{\rm tot}(0,0)\} / c_{\rm s}^{2}]$.

Radio recombination line tracing gas in \hii~regions within the starburst
region by Seaquist \etal (1996) imply the density of thermal electrons
with $T_{\rm e} = 10^{4} \K$ is between $10$ -- $100 \pcc$. Bear
in mind that observational evidence from a wide range of sources suggests
the ISM within the starburst region is extremely non-uniform, from
dense molecular clouds to very hot coronal gas. The disk ISM in
these models is chosen to represent one phase of this multiphase ISM,
the phase which presents the starburst-driven wind with the greatest
physical opposition, \ie the phase which has the greatest volume
filling factor. Dense clouds which are enveloped and overrun by the
wind can be represented within these simulations, 
although not resolved by the computational grid, by mass-loading as
discussed in Section~\ref{sec:massload}.

In place of our ignorance of the properties of the haloes galactic 
winds expand into we envision a hot tenuous halo enveloping M82.
As star formation rates in M82 appear to have been elevated since
since the nearest encounter with its neighbor M81 
approximately $200 \Myr$ ago
(Cottrell 1997) chimneys above OB associations, 
a galactic fountain or even previous galactic winds all could have
created such a hot halo.

As the complex 3-dimensional, multiphase, turbulent structure of the ISM 
can not be treated in these simulations, we follow the lead of
TI, TB and S94 in increasing the isothermal sound speed of the 
disk gas to simulate the turbulent pressure support seen in real disks 
(\eg Norman \& Ferrara 1996).
Without this increased pressure the resulting disks
are extremely thin. Following TB we set $c_{\rm s, disk} = 30 \kmps$
and $c_{\rm s, halo} = 300 \kmps$ in both thick and thin disk models.
This corresponds roughly to a disk temperature of 
$T_{\rm disk} = 6.5 \times 10^{4} \K$ and halo of 
$T_{\rm halo} = 6.5 \times 10^{6} \K$. 

This imposes a minimum allowed temperature of $T_{\rm disk}$ on the gas
in these simulations, so we can not directly treat the cooler
$T \sim 10^{4} \K$ gas that is responsible for the optical emission lines
in galactic winds. Currently we only make qualitative
comparisons between the dynamics of the coolest gas in our simulations
and the cool gas seen in M82's and other starburst's winds.
This minimum temperature does not affect the dynamics and emission properties
of X-ray emitting gas we are primarily interested in.

\subsection{Starburst history and mass and energy injection rates}
\label{sec:starburst_history}
The actual star formation history in M82 is only crudely known. 
The high optical extinction towards the nucleus and edge-on
inclination make it difficult to investigate
the properties of stars or star clusters in the starburst nucleus, except using
IR observations. Even in relatively nearby and unobscured 
dwarf starburst galaxies such as NGC 1569 and NGC 5253 the history of
all but the most recent star formation is a subject of debate (see
Calzetti \etal [1997] with reference to NGC 5253, 
or Gonz\'{a}lez Delgado \etal [1997] for NGC 1569). 

As we are  interested in the effect of the star formation history on the
wind dynamics and observable X-ray properties, we shall explore a few simple
star formation histories. These are motivated by the near-IR spectroscopic
study of individual stellar clusters in M82's starburst nucleus 
by Satyapal \etal (1997), dynamical arguments concerning the total
mass of stars formed in the starburst by McLeod \etal (1993), and radio
estimates of the current SN rate by Muxlow \etal (1994) and Allen
\& Kronberg (1998). We briefly review what these observations can tell
us about the total mass of stars formed in M82's current starburst, and
its star formation history.

\subsubsection{IR observations of individual super star clusters}
Satyapal \etal (1997) use  de-reddened Br$\gamma$ and CO band imaging to
study 12 unresolved stellar clusters within $r = 270 \pc$ of the nucleus.
In comparison with instantaneous starburst models 
(appropriate for individual clusters), they used the 
CO indices and Br$\gamma$ equivalent widths of the individual clusters
to estimate the cluster ages, and the ionising photon fluxes from the
extinction corrected Br$\gamma$ line flux.

We used this information, along with the 
instantaneous starburst evolutionary synthesis models
of Leitherer \& Heckman (1995; henceforth LH95), to estimate the
initial clusters masses assuming a Salpeter (1955) IMF extending between 
$1$ -- $100 \Msol$.

Note that the ages of individual clusters 
differ between the two age estimators, with ages based on the CO index 
generally being a few Myr older than those based on the 
Br$\gamma$ equivalent widths.  Ages inferred from the CO index 
were in the range $5$ -- $10 \Myr$, while those from  $W$(Br$\gamma$)
lay in the range $4$ -- $8\Myr$. The relative ages of the different clusters
are also inconsistent between the two methods. 

For a given ionising photon flux the initial cluster mass is a
sensitive function of the assumed age, due to the rapid evolution
of the extremely luminous stars at the high end of the initial mass function.
Hence the uncertainty in cluster ages leads to a large uncertainty
in the mass of stars formed in the starburst.

The $W$(Br$\gamma$)-derived ages and ionising photon fluxes yield a 
total initial mass of stars formed in
the starburst of $M_{\rm SB} \sim 8 \times 10^{6} \Msol$. 
From the LH95 models the peak SN rate associated with these stars
is $\sim 8 \times 10^{-3}~{\rm SN} \pyr$. 

Using the CO index instead to derive the cluster ages yields 
$M_{\rm SB} \sim 1.3 \times 10^{8} \Msol$, with a peak SN rate of
 $\sim 0.13~{\rm SN} \pyr$. 
If the IMF extends down to $M_{\rm low} = 0.1 \Msol$ the total
mass  $M_{\rm SB} \sim 3 \times 10^{8} \Msol$, in agreement with the
value of  $M_{\rm SB} \sim 2.5 \times 10^{8} \Msol$ quoted in Satyapal
\etal (1997). This is $\sim 40$ per cent 
of the total dynamical mass within the
starburst region. 

\subsubsection{Dynamical limits on the total mass of stars formed}
McLeod \etal (1993) present a simple dynamical argument that places an
upper limit of the total mass of stars formed in M82's starburst
of $\ltsimm 2.5 \times 10^{8} \Msol$. The total dynamical mass within $500 \pc$
is $M_{\rm tot} \approx 7 \times 10^{8} \Msol$, of which approximately
$10^{8} \Msol$ is gas. Only a fraction of the remaining mass can be due
to stars formed in the current starburst activity, as there must have been
a pre-existing stellar population. McLeod's argument is based on the
following scenario for the triggering of M82's starburst:
The starburst was probably triggered by a close interaction of M82 with
M81 approximately $200 \Myr$ ago (Cottrell 1977), which lead to gas
losing angular momentum and falling into the nucleus over a time-scale of
$\sim 100 \Myr$. Eventually self gravity within the gas will trigger 
the strong burst of star formation. 
Self-gravitation in the ISM will have become
important before the gas mass equals the mass of pre-existing stars within the
nucleus, so it is unlikely that the total mass of gas and stars formed in
the starburst exceeds half the current dynamical mass within the starburst
region. 

This upper limit of $2.5 \times 10^{8} \Msol$ is very similar to Satyapal
\etal's (1997) estimate of the starburst mass assuming conservatively 
a Salpeter IMF extending between $0.1$ -- $100 \Msol$, where
very low mass stars consume most of the mass.
If we assume a top heavy IMF (\ie biased against
low mass stars) we can get a very powerful starburst that consumes only
a small amount of gas, but would violate observational constraints
on the starburst luminosity.

\subsubsection{The current SN rate}
 Radio observations of the 50 or so SNR's
within the central kiloparsec of M82 can be used to estimate the 
current (\ie within the last several thousand years) SN rate. 
As a simple rule of thumb, the SN rate between 
$t \sim 3$ -- $40 \Myr$ for an instantaneous burst forming $10^{6} \Msol$
of stars is $\sim 10^{-3} \pyr$ (using the LH95 models and
assuming a Salpeter IMF between $M_{\rm low} = 1$ and 
$M_{\rm up} = 100 \Msol$).

Muxlow \etal (1994) estimate
a SN rate of $0.05 \pyr$ assuming the SNRs are still freely expanding
at $v = 5000 \kmps$. If the SNRs are not in free expansion the SN rate
should be reduced. This SN rate corresponds to a total starburst mass of
$\sim 5 \times 10^{7} \Msol$.

Based on the oldest and largest SNR in M82, Allen \& Kronberg (1998)
estimate the SN rate to be $\gtsimm 0.016 \pyr$.
This is a lower limit
as other large (and hence faint) SNRs may be missing from current SNR surveys.

Note that the current 
SN rate is not particularly sensitive to the star formation
history, as the SN rate for an instantaneous burst of stars
is approximately constant between $\sim 3$ to $\sim 40 \Myr$.  

\begin{figure}
 \centerline{
   \psfig{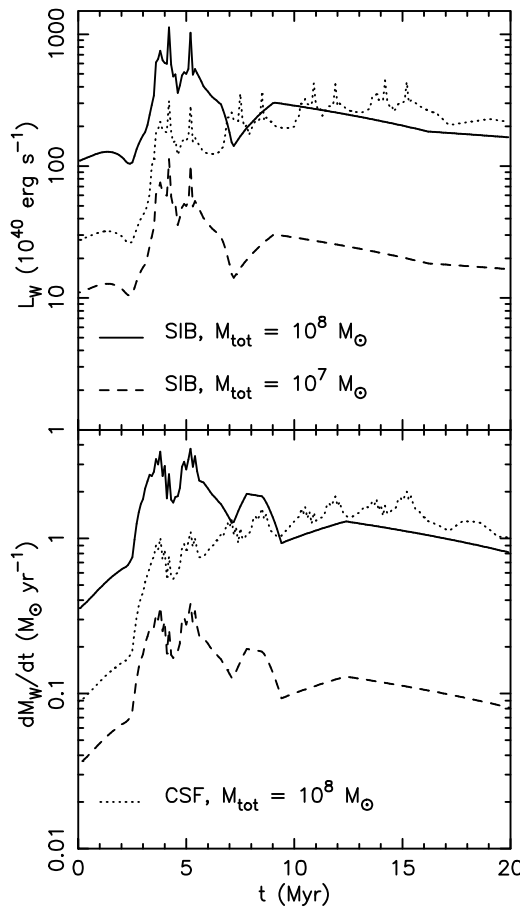}
	}
\caption[Starburst energy and mass injection rates]
 {Starburst energy injection (top panel) and mass injection 
 (bottom panel) rates adopted in these simulations
 (see Section~\ref{sec:starburst_history}).}
\label{fig:mass_and_energy}
\end{figure}

\subsubsection{Starburst models used in these simulations}

The observational evidence considered above suggests that although the
exact star formation history and mass in M82 may not be accurately known,
it is possible to bracket the true solution. 

We assume Solar metal abundances
and a Salpeter IMF between $1$ -- $100 \Msol$ in all these mass
estimates below. This is purely for convenience as the LH95 models 
assume this IMF. All the other mass estimates have been 
converted to this IMF. 

The total mass of the starburst
clusters inferred from Satyapal \etal's (1997) Br$\gamma$ equivalent
widths, $M_{\rm tot} \sim 8 \times 10^{6} \Msol$,
and Allen \& Kronberg's (1998) lower limit on the SN rate,
$M_{\rm tot} \gtsimm 1.6 \times 10^{7} \Msol$, suggest a lower limit on the
starburst mass of $M_{\rm tot} \sim 10^{7} \Msol$.

The dynamical arguments of McLeod \etal (1993) suggest an upper limit on
the total mass recently converted into stars in M82 centre of 
$M_{\rm tot} \ltsimm 10^{8} \Msol$, consistent with the mass inferred by
Satyapal of $M_{\rm tot} \sim 10^{8} \Msol$ based on the CO index 
derived ages. The recent SN rate inferred by Muxlow \etal (1994) implies 
a starburst mass of $M_{\rm tot} \sim 5 \times 10^{7} \Msol$.

The ages inferred from Satyapal \etal's (1997) CO indices and $W$(Br$\gamma$),
although marginally inconsistent with each other, suggest a spread in ages
for the bright clusters of between $4$ -- $10 \Myr$ old.

We choose three simple SF histories to explore the effects of M82's possible
SF history on the dynamics and X-ray emission from the galactic wind (see
Fig.~\ref{fig:mass_and_energy}):
\begin{enumerate}
\item A single instantaneous starburst (SIB) of total mass $M_{\rm tot} = 
  10^{8} \Msol$. This model can
  be considered as the most powerful starburst consistent with the 
  observational constraints.
\item A weaker single instantaneous starburst (SIB) 
  of total mass $M_{\rm tot} = 10^{7} \Msol$, consistent with the least
  powerful starburst suggested observationally.
\item A starburst of total mass $M_{\rm tot} = 10^{8} \Msol$, but with
  a slightly more complex SF history (CSF) 
  spread over a period of $10 \Myr$ rather than a SIB. With the same total mass
  as the power SIB, this model allows us to investigate the effect of
  a more gradual mass and energy injection history.
\end{enumerate}

\subsubsection{Energy thermalisation efficiency}

We assume that 100 per cent
of the mechanical power from stellar winds and SNe is available
to power the wind, \ie the immediate radiative energy losses within 
the starburst region are negligible. The value of this
``thermalisation efficiency'' is by no means well understood
either observationally or theoretically, and will depend on the
specific environment (for example, whether the SN goes off
in or near a molecular cloud or in a pre-existing low density cavity).

Thornton \etal (1998) perform a parameter study to assess the radiative
losses of supernova remnants expanding into 
uniform media of different densities. For an SNR of age $10^{5} \yr$
(a sufficient time for any SNR in a nuclear starburst region to
have interacted or merged with neighboring SNRs) radiative losses
range from negligible (ISM number densities $\sim 10^{-2} \pcc$)
to $\ge 90$ per cent (number densities $\sim 10 \pcc$) of 
the initial supernova energy.

Although the majority of the 
mass of the ISM in a starburst region is in dense molecular clouds,
these occupy only a small fraction of the
total volume ($\sim 10^{-3}$, see Lugten \etal 1986)
within a starburst region such as the centre of M82.
Hence on average
young supernova remnants in M82 will interact 
with tenuous gas, with reasonably low radiative losses.

Observations of the properties of local starburst driven winds
already strongly argue for high thermalisation efficiencies.
From a theoretical basis,
Chevalier \& Clegg (1985) explicitly state that the requirement
to drive a galactic wind is a high thermalisation efficiency.
Two simple arguments based on M82's wind will suffice. 
For the purposes of the following arguments alone, we shall assume
a very simple model for the starburst 
of 0.1 SNe $\pyr$ for $10^{7} \yr$. This gives a total mechanical
energy injection of $E = 10^{57} \erg$ and an average mechanical power of
$\Edot = 3 \times 10^{42} \ergps$ (assuming 100 per cent thermalisation).
\begin{enumerate}
\item The thermal energy of the $T \sim 5 \times 10^{6} \K$ gas is 
  $\sim 3\times 10^{56} \eta^{1/2} \erg$ (Strickland \etal 1997, where
 $\eta$ is the volume filling factor of the X-ray-emitting gas).
  Hence to first order the thermal energy content of the wind is
  $\sim 30$ per cent
  of the total energy released by SNe and stellar winds.
  If we instead assume the X-ray emission comes from shocked-clouds 
  of low filling factor (\cf Chevalier \& Clegg 1985), 
  then the total energy of the wind may be even greater, as
  remainder of the volume
  must be occupied by tenuous but energetic gas.  
  This estimate does not include the kinetic energy of the wind, which we 
  shall show later to larger than the thermal energy content 
  of the wind.
\item The soft X-ray luminosity of the hot gas in M82 is 
  $L_{\rm X} = 2 \times 10^{40} \ergps$ (Strickland \etal 1997; Dahlem \etal
  1998) in the {\it ROSAT} band. If the thermalisation
  efficiencies were as low as the $\sim 3$ per cent value argued by 
  Bradamante \etal (1998)
  then galactic winds must extremely efficient at radiating
  X-rays. This is inconsistent with the estimated X-ray emitting gas
  cooling times (\cf Read \etal 1997) and it would be extremely
  difficult to drive a wind at all given such strong radiative losses.
\end{enumerate}

For galactic winds the immediate thermalisation efficiency 
within the starburst region must then be 
reasonably high, \ie between 10 -- 100 per cent.
We therefore follow the lead of previous simulations
of galactic winds and assume 100 per cent
of the mechanical energy from SNe and stellar winds can be used to
drive the wind.

\subsubsection{The starburst region}

At each computational time step we inject the appropriate amount of mass
and energy uniformly within the computational cells corresponding to
the starburst region.

For the thin disk model the chosen starburst region is a
cylinder $150 \pc$ in radius, extending to a height of $30 \pc$ 
above and below the plane of the galaxy.
This corresponds to $10 \times 2$ computational cells in the quadrant of the
flow actually modeled (the other three quadrants are by symmetry identical
to the one calculated in the simulations).

For the thick disk model we use a spherical starburst region of radius
$150 \pc$. As the scale height of the ISM above the starburst region 
is greater than $150 \pc$ this 
does not constitute a significant difference between
the thick and thin disk models. We did perform an additional simulation
with the thin disk model with a spherical starburst region, and confirmed
that the dynamics and properties were almost identical to those in
the default cylindrical starburst region models.

As discussed in Strickland \& Stevens (1999)
the picture of energy and mass being injected uniformly into a
 starburst region, and driving a single wind and bubble into the ISM,
is overly simplistic. Individual super star clusters (Meurer \etal 1995; 
O'Connell \etal 1995) blow strong winds 
into the surrounding starburst region, which
interact with the complex ISM structure of molecular clouds, SNRs and
winds from other massive stars and clusters. The inferred ages of
the star clusters in NGC 5253 (Gorjian 1996; Calzetti \etal 1997)
and M82 (Satyapal \etal 1997) suggests that the formation of massive
star clusters propagates across or outwards through the starburst region.
Sadly, simulating this much detail requires computational
resources far in excess of what is currently available.

\subsection{Metallicity and radiative cooling} 
\label{sec:cooling}

For the purposes of calculating the radiative cooling of the gas in the
simulations, as well as calculating the X-ray emission, we assume Solar
metallicity and collisional ionisation equilibrium.

The metallicity of the cool ambient ISM and the hot gas in M82 is uncertain.
Optical or infrared observations suggest a metal abundance similar to
Solar, while X-ray observations of the hot gas give an abundance
of less than one third of the Solar value.

Measurements of infrared fine-structure lines (Puxley \etal 1989) show
the abundances of argon and neon in M82's ISM are $Z = 1.0\pm{0.5} \Zsol$.
O'Connell \& Mangano (1978) find optical emission line ratios typical
of \hii~regions with metallicity similar to or slightly higher than Solar.     

In principle X-ray observations could directly 
measure the metal abundance in the hot gas responsible for the observed
soft thermal X-ray emission.  X-ray determined metal 
abundances of the hot gas in starburst galaxies (including M82)
using {\it ROSAT}
 and {\it ASCA} generally give low abundances
$Z \ltsimm 0.3 \Zsol$ (\cf Ptak 1997), with the 
Iron abundance depressed relative to the $\alpha-$process elemental 
abundances (the {\it ROSAT} PSPC is only sensitive to the 
Iron abundance through the strong Fe-L complex at $E \sim 0.8 \keV$).
This may just be an artefact of using overly simplistic spectral models
to fit X-ray spectra arising from multiphase hot gas (see 
Dahlem \etal [1998] for observational evidence for this argument,
or Strickland \& Stevens [1998] for supporting theoretical modeling).

We use a parametrized form of the total emissivities for gas in the 
temperature range $10^{4.5}$ -- $10^{8.5} \K$ from a recent version of the
Raymond \& Smith (1977) hot plasma code to implement radiative cooling
in VH-1. The temperature is updated each computational time step using
a fully implicit scheme as described by Strickland \& Blondin (1995).
Gas is prevented from cooling below 
$T = T_{\rm disk} = 6.5 \times 10^{4} \K$ to prevent the
artificially hot disk from cooling and collapsing.
 
We restrict the cooling rate at unresolved interfaces between hot diffuse
gas and cold dense gas, as the finite width of sharp features on the
computational grid can lead
to anomalously high cooling rates. 
At any unresolved density gradients we use the minimum volume cooling rate
in the immediate vicinity, a similar scheme to that used by 
Stone \& Norman (1993).

\begin{figure*}
 \centerline{
   \psfig{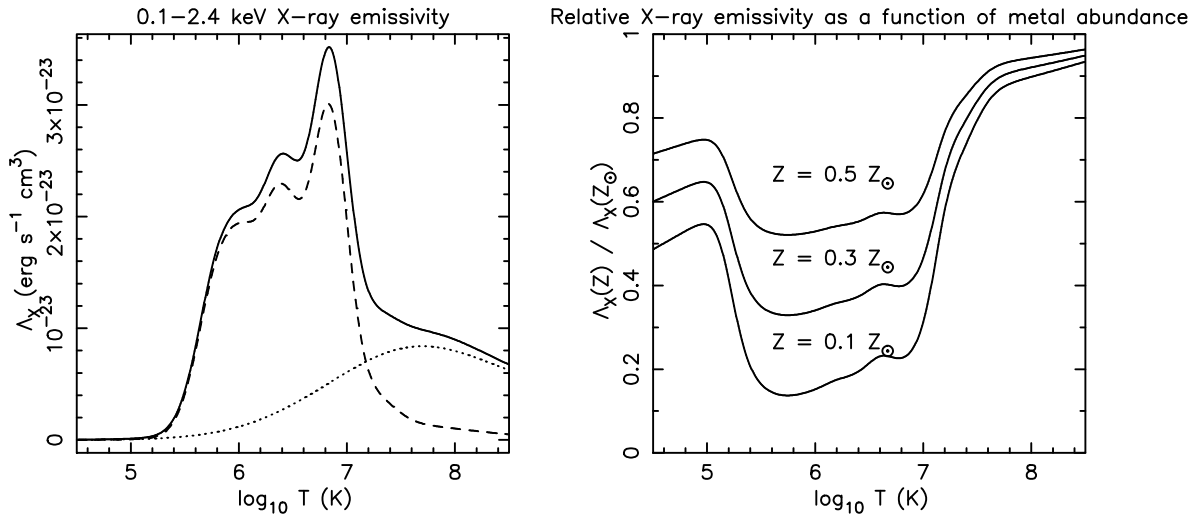}
	}
\caption[X-ray emissivity as a function of temperature and metal abundance]
 {X-ray emissivity $\Lambda_{\rm X}$ in the $0.1$ -- $2.4 \keV$ band 
 as a function of temperature and metal abundance, assuming Solar abundance
 ratios and collisional ionisation equilibrium. (a) The total X-ray emissivity
 from the Raymond-Smith hot plasma code (Raymond \& Smith 1977) for
 Solar metallicity (solid line), decomposed into the contribution from H and He
 only (dotted line) and metal ions (dashed line).
 (b) The soft X-ray emissivity for a hot plasma of metal abundance 
 $Z = 0.1, 0.3$ and $0.5 \Zsol$, relative to the emissivity of a Solar 
 metallicity hot plasma. For temperatures between
 $3 \times 10^{5} \K$ and $10^{7} \K$ 
 line emission from metals dominates the emission.
 X-ray emission from low metallicity hot plasmas is significantly depressed
 in this temperature range due to the lack of line emission.}
\label{fig:lambdax}
\end{figure*}

The total cooling rate and X-ray emissivity of a hot gas 
in the temperature range $3 \times 10^{5}$ -- $10^{7} \K $
is a strong function of its metal abundance, as shown in
Fig.~\ref{fig:lambdax}. Hence different explanations
for the origin of the soft X-ray
emission in galactic winds also imply that we expect different
metal abundances for this gas and hence differing emissivities
for given density and temperature.
For example S94 argue that the majority of the 
X-ray emission from starburst driven winds 
is due to shocked disk material, which would have significantly lower
metal abundance than the SN-enriched starburst ejecta.
The metallicity of the gas in the wind will strongly 
affect its cooling rate and
X-ray luminosity. This is unlikely to affect the dynamics of the 
X-ray emitting gas as it is an inefficient radiator of its thermal
energy. The main effect of the assumed metallicity on the X-ray
properties will be on the absolute normalisation of the X-ray luminosity,
detector count rates and X-ray surface brightness from our simulations.

In the absence of more conclusive observational estimates of the 
metal abundance of the ISM in M82, we shall assume all the gas in these
simulations is of Solar abundance. In paper II we shall
investigate any biases in X-ray determined metallicities
of the hot gas in galactic winds, by
producing and analysing artificial X-ray observations 
from the multiphase gas distributions in these simulations.

\subsection{Mass-loading}
\label{sec:massload}

A starburst-driven wind, in its expansion through the ISM as a superbubble 
and post-blowout as a galactic wind, will overrun and envelop clumps
and clouds that are denser than the ambient ISM. Once inside the bubble or
wind, conductive, hydrodynamical or even photo-evaporative
processes in the shocked or free wind regions will evaporate, ablate and
shock-heat these clouds. This will add cool material into the hot X-ray
emitting regions, and hence potentially altering the energetics and 
observational properties of the wind. This addition of mass into the flow
is termed ``mass-loading'' (Hartquist \etal 1986).
Unlike the interaction of the wind with the ambient, inter-cloud, medium,
which adds mass to the outside of the superbubble/wind, mass-loading
from clouds adds material gradually into the hot interior.

Conductive mass-loading is the evaporation of clouds by the 
thermal conduction of hot electrons from the hot plasma penetrating 
and heating the clouds (\cf Cowie \etal 1981). 
Hydrodynamical mass-loading is the ablation and physical
destruction of clouds by hydrodynamical processes as the hot plasma flows
past dense clouds, either sub-or-supersonically (Hartquist \etal 1986). 
Our models do not incorporate thermal conduction, so we concentrate
on a hydrodynamically mass-loaded model. Although the physics of the two
processes are quite different, our simple model captures the essential feature
of mass-loading which is the addition of additional cold material into the
hot interior of the flow. 
Our aim is to study the effects of a simple but physically motivated
model of the interaction of the wind with dense clouds.

A dense cloud embedded in a subsonic flow will experience pressure
differences along its surface that lead to it expanding perpendicularly to the
direction of the surrounding flow (the head of the cloud experiences the ram
pressure plus the thermal pressure of the tenuous flow, whereas the sides of
the cloud only experience the thermal pressure of the surrounding 
flow). Rayleigh-Taylor
(RT) and Kelvin-Helmholtz (KH) instabilities will remove material from the 
perimeter of the expanding cloud. The ablation rate of the cloud (\ie its
mass-loading rate) is proportional to the expansion speed of the cloud
divided by the size of the mixing region between the cloud and the wind.

Dense clouds embedded in supersonic flows are crushed by shocks driven into
them by the wind, before being disrupted by pressure differences in a similar
manner to clouds in subsonic flows.

Physical arguments based on the picture given above (Hartquist \etal 1986; see
Arthur \& Henney 1996 for a numerical treatment of mass-loaded SNRs) suggest 
that the mass-loading rate \hbox{$\dot q$} 
of a flow
that ablates small, denser clouds depends on the Mach number $\cal M$ of the
upstream flow for subsonic flows, but not for supersonic flows which are
ablated at a maximum mass-loading rate $Q$, i.e.
\begin{equation}
  {\dot q}  = \left\{ \begin{array}{ll}
        Q \times {\cal M}^{4/3} & \mbox{if  ${\cal M} < 1.0$}  \\
        Q             &  \mbox{if  ${\cal M} \ge 1.0$}.
        \end{array} \right.
\label{equ:qdot}
\end{equation}

In practice the maximum mass-loading rate $Q$ 
depends on both the properties of the cloud and the flow density and
velocity.
The maximum mass-loading rate per unit volume over the cloud is
\begin{equation}
  Q = a 
        \left( \frac{\rho_{\rm w}  v_{\rm w}  k T_{\rm cl}  \rho_{\rm cl}^{2}}
        { \mu m_{\rm H}  R_{\rm cl}^{3}} \right)^{1/3},
\label{equ:mass_load}
\end{equation}
where $a$ is a constant of order unity, $\rho_{\rm w}$ and $v_{\rm w}$
the density and velocity of the flow the cloud is embedded in, and
$T_{\rm cl}$, $\rho_{\rm cl}$, $R_{\rm cl}$ the cloud temperature,
density and radius.

Despite the simplicity of this analytical treatment, numerical simulations of
clouds ablated by tenuous flows (Klein, McKee \& Collela 1994) 
support this model of mass-loading. For the purposes of these simulations
we shall consider two different models of mass-loading of a starburst-driven
galactic wind, based loosely on the central and distributed mass-loading model
used by S96.

\subsubsection{Central mass-loading}
S96's steady state mass-loaded wind models 
(with no ISM apart from the clouds) suggested that all mass-loading in M82 was
confined to the starburst region itself. Given the observed molecular ring
at a radius of $\sim 250 \pc$ from the centre, and that large masses of 
molecular material must have existed within the starburst region to form the
young stars, it is not unreasonable to expect the majority of cloud
material to exist within the starburst region.

As in S96's model we simulate this central mass-loading by ignoring the
detailed cloud mass-loading rates given above, and increasing the mass 
deposition rate in the starburst by a factor 5. This results in a typical
mass injection rate of $\sim 5 \Msol \pyr$, similar to the
models S96 considered most successful.

\subsubsection{Distributed mass-loading}
The alternative to a central reservoir of cloud material are clouds  
distributed throughout the disk of the galaxy. We assume all clouds have the
same density, 
size and temperature, irrespective of
their position within the galaxy. As these clouds are destroyed by the
wind these properties do not alter, but only the total mass in clouds is 
reduced. The only cloud property we allow to vary with spatial position
over the disk is the
local cloud volume filling factor, which then 
controls the local mass in clouds. This cloud filling factor is assumed to
remain constant with time, the total mass in clouds reducing with time
as the wind overruns them and destroys them.

The maximum mass-loading rate $Q$ is a relatively weak function of
the wind and cloud properties (as can be seen in Eqn.~\ref{equ:mass_load}). 
Rather than
explicitly calculate $Q$ as a function of the local flow variables at
every computational cell and time step, we fix this maximum mass-loading rate
over the entire grid.
This allows us to explicitly control the minimum cloud destruction time-scale
\begin{equation}
\tau_{\rm cl} = \rho_{\rm cl} / Q,
\end{equation}
where $\rho_{\rm cl}$ is the density within a cloud.

We choose $Q$ and $n_{\rm cl}$ to give a minimum cloud destruction time 
scale that is scientifically interesting. If the
cloud destruction time-scale  $\tau_{\rm cl} \ll \tau_{\rm dyn}$  (where
$\tau_{\rm dyn} \sim 10 \Myr$ is the dynamical age 
scale of the galactic wind),
then clouds will be destroyed almost instantaneously by the outer 
shock of the wind. All the cloud mass will be added to the outermost
part of the wind, and the mass-loading will be almost identical to
the evolution of a wind in a slightly denser medium. 
If $\tau_{\rm cl} \gg \tau_{\rm dyn}$, then almost no mass-loading 
will occur. Hence the case of $\tau_{\rm cl} \sim \tau_{\rm dyn}$ is 
the most interesting as far as the effects of mass-loading on the properties
of galactic winds is concerned. We therefore set $\tau_{\rm cl} = 10 \Myr$.

We assume the clouds are distributed identically to the 
high filling factor ambient ISM, in rotating hydrostatic equilibrium.
All clouds are assumed to have number density $n_{\rm cl} = 10^{3} \pcc$ 
and temperature $T_{\rm cl} = 10^{3} \K$, and apart from the assumed rotational
motion, are at rest with respect to the starburst region.
The total mass in clouds is almost identical 
in the thick and thin disk models,
with a central cloud filling factor of $\eta_{\rm cl} = 4 \times 10^{-2}$
in the thin disk models and $\eta_{\rm cl} = 1.3 \times 10^{-2}$ 
in the thick disk models. The original mass of cloud material 
within the central $500 \pc$ is $\sim 3 \times 10^{7} \Msol$, and within the
volume occupied by the entire galactic wind at an age $t \sim 15 \Myr$
the original cloud mass is $\sim 3 \times 10^{8} \Msol$. As the minimum
cloud destruction time is $\tau_{\rm cl} = 10 \Myr$, not all
of this cloud mass will have been added into the flow. 

At each computational step we calculate the local Mach number ${\cal M}$
and calculate the cloud mass-loading rate ${\dot q}$ from 
Eqn.~\ref{equ:qdot}. Note that this is the mass-loading rate per unit volume
of cloud material, so the total mass-loading rate at any position is this
 ${\dot q}$ multiplied by the local cloud filling factor.
We computationally track the mass remaining in clouds to ensure
mass-loading ceases in regions where all the clouds have been destroyed.

\begin{table*}
\caption[Models used starburst-driven wind parameter study]
 {Model parameters for the twelve starburst-driven galactic wind simulations
 studied. The thick and thin disk ISM and gravitational potential parameters
 are given in Table.~\ref{tab:model_ism} }
 \label{tab:model_param}
 \begin{center}
 \begin{tabular}{lcccccccc}
 \hline
 Model & ISM & Starburst  & SFH & $R_{\star}$ & $H_{\star}$ 
        & Mass-loading & Grid size & Cell size \\
       & model          & mass ($\Msol$)  &     & ($\pc$)     & ($\pc$)
        &               & (cells, r$\times$z)                 & ($\pc$) \\
 \hline
 tbn\_1 & thick & $10^{8}$ & SIB & 150 & - & none 
        & $400 \times 800 $ & $14.6 \times 14.6$ \\
 tbn1a  & thick & $10^{8}$ & SIB & 150 & - & none 
        & $400 \times 800 $ & $7.3 \times 7.3$ \\
 tbn1b  & thick & $10^{8}$ & SIB & 150 & 60 & none 
        & $640 \times 1280 $ & $4.9 \times 4.9$ \\
 tbn\_2 & thick & $10^{7}$ & SIB & 150 & - & none 
        & $400 \times 800 $ & $14.6 \times 14.6$ \\
 tbn\_6 & thick & $10^{8}$ & SIB & 150 & - & central 
        & $400 \times 800 $ & $14.6 \times 14.6$ \\
 tbn\_7 & thick & $10^{8}$ & CSF & 150 & - & none 
        & $400 \times 800 $ & $14.6 \times 14.6$ \\
 tbn\_9 & thick & $10^{8}$ & SIB & 150 & - & distributed 
        & $400 \times 800 $ & $14.6 \times 14.6$ \\
 mnd\_3 & thin & $10^{8}$ & SIB & 150 & 60 & none 
        & $480 \times 800 $ & $14.6 \times 14.6$ \\
 mnd\_4 & thin & $10^{7}$ & SIB & 150 & 60 & none 
        & $480 \times 800 $ & $14.6 \times 14.6$ \\
 mnd\_5 & thin & $10^{8}$ & SIB & 150 & 60 & central 
        & $480 \times 800 $ & $14.6 \times 14.6$ \\
 mnd\_6 & thin & $10^{8}$ & CSF & 150 & 60 & none 
        & $480 \times 800 $ & $14.6 \times 14.6$ \\
 mnd\_8 & thin & $10^{8}$ & SIB & 150 & 60 & distributed 
        & $480 \times 800 $ & $14.6 \times 14.6$ \\
 \hline
 \end{tabular}
 \end{center}
\end{table*}

\subsection{Model parameter study}
\label{sec:my_models}

We have chosen the following set of models to investigate how the dynamics
and observational properties of starburst-driven galactic winds depend
on the host galaxy's ISM distribution, the starburst strength and history,
and the presence and distribution of mass-loading by dense clouds.
Although only comprising 12 simulations, 
we believe this to be the most detailed and systematic theoretical
study of galactic winds to date. The model parameters for these simulations
are summarised in Tables.~\ref{tab:model_param} \& \ref{tab:model_ism}. 

\subsubsection{Thick disk models}

The thick disk models have the same ISM distributions as TB's simulations,
although run on a higher resolution computational grid.
The resulting thick collimating disk allows us to investigate
the effect of strong wind collimation by dense gas high above the plane
of the galaxy on the wind dynamics, morphology and X-ray emission. 

\begin{enumerate}

\item {\bf Model tbn\_1} has a powerful starburst forming $10^{8} \Msol$
of stars 
(assuming a Salpeter IMF between $1$ -- $100 \Msol$) instantaneously within
a spherical starburst region of radius $150 \pc$. 
To investigate the interaction of the wind
with the ambient ISM alone no mass-loading is
included in this simulation.

\item {\bf Model tbn1a} has identical model parameters to model tbn\_1, but
is run on a higher resolution grid of twice the resolution to the 
other models (each cell is $7.3 \pc \times 7.3 \pc$), although only covering a
smaller physical region. This allows us to investigate the effects of limited
numerical resolution of the wind properties.

\item {\bf Model tbn1b} has triple the resolution of model tbn\_1, with
cells $4.9 \pc \times 4.9 \pc$ in size. As with model tbn1a the
aim to investigate the influence of numerical resolution.
Unlike the other thick disk models the starburst region in this model
is a more realistic cylindrical region also used in the thin disk
models described below.

\item {\bf Model tbn\_2} is almost identical to model tbn\_1 with a single 
instantaneous starburst (SIB), except the starburst is only one tenth 
as powerful at that in model tbn\_1. This starburst represents a lower
limit on the power of the starburst in M82. In comparison with model
tbn\_1 this simulation allows us to investigate how wind properties and 
dynamics scale with starburst power. 

\item {\bf Model tbn\_6} has the same starburst as in model tbn\_1, but
mass-loading of the wind by dense clouds occurs within the starburst region
(central mass-loading). This is modeled by increasing the mass injection
rate from the starburst by a factor of 5.

\item {\bf Model tbn\_7} has a more complex SF (CSF) 
history than the instantaneous
starbursts used in the other models. The total mass of stars formed in the
starburst is $10^{8} \Msol$, as in model tbn\_1, but the star formation is
spread over a period of $10 \Myr$. This results in a more gradual deposition
of mass and energy by the starburst, allowing us to investigate the effects
of the history of mass and energy injection on the wind dynamics.

\item {\bf Model tbn\_9} has a SIB as in model tbn\_1, but also incorporates
mass-loading distributed throughout the disk 
as discussed in Section~\ref{sec:massload}. In combination with model tbn\_6
these mass-loaded simulations allow us to investigate both the
 effect of mass-loading on starburst-driven winds 
in combination with the wind's interaction with the
ambient high filling factor ISM (unlike S96's mass-loaded simulations, where
the wind did not interact with the ambient ISM), and how the
distribution of the cloud material affects the wind dynamics.

\end{enumerate}

\subsubsection{Thin disk models}

The thin disk models include a more realistic gravitational potential than
the one used in TB \& S94's simulations (and the thick disk models). This
new gravitational potential approximately reproduces M82's observed rotation
curve (Fig.~\ref{fig:rotcurve}). The deeper potential results in a much thinner
disk, with less collimation of the wind and lower gas density above the plane
of the galaxy. We use a larger computational grid 
of $480 \times 800$ cells covering a physical region 
$7.0 \kpc \times 11.6 \kpc$ to allow for the greater radial expansion
of the wind in this less collimating ISM distribution. 

\begin{enumerate}
\item {\bf Model mnd\_3} differs only from model tbn\_1 in its thin disk ISM 
distribution and more realistic gravitational potential, and its cylindrical
starburst region of radius $150 \pc$ and height $60 \pc$ (\ie extends to 
$z = \pm{30} \pc$). The starburst is a SIB of total mass $10^{8} \Msol$.

\item {\bf Model mnd\_4} is a weaker SIB of mass $10^{7} \Msol$, but otherwise
is identical to model mnd\_3.

\item {\bf Model mnd\_5} is a centrally mass-loaded wind with otherwise 
identical model parameters to model mnd\_3. As in the centrally mass-loaded
thick disk model tbn\_6 the starburst mass deposition rate has been increased
by factor 5. 

\item {\bf Model mnd\_6} explores the same complex SF history as model tbn\_7
but in a thin disk ISM.

\item {\bf Model mnd\_7} is a thin disk model with a SIB of mass
$10^{8} \Msol$ that incorporates
 distributed mass-loading. The total mass in clouds is
very similar to model tbn\_9, although the clouds are distributed within a
thin disk.

\end{enumerate}

\begin{table}
\caption[Thick and thin disk ISM parameters]
 {Thick and thin disk ISM parameters}
 \label{tab:model_ism}
 \begin{center}
 \begin{tabular}{lcc}
 \hline
 Parameter & Thick disk models & Thin disk models \\
 \hline
 $M_{\rm ss}$ ($\Msol$) & $1.2 \times 10^{9}$ & $2 \times 10^{8}$ \\
 $\omega_{0}$ ($\pc$)   & 350                 & 350               \\
 $M_{\rm disk}$ ($\Msol$) & -                 & $2 \times 10^{9}$ \\
 $a$ ($\pc$)              & -                 & 222 \\
 $b$ ($\pc$)              & -                 & 75 \\
 $n_{\rm disk, 0}$ ($\pcc$)   & $20$            & $20$ \\
 $c_{\rm s, disk}$ ($\kmps$)   & $30$            & $30$ \\
 $T_{\rm disk, 0}$ ($\K$)   & $6.5 \times 10^{4}$ & $6.5 \times 10^{4}$ \\
 $n_{\rm halo, 0}$ ($\pcc$)   & $2 \times 10^{-3}$ & $2 \times 10^{-3}$ \\
 $c_{\rm s, halo}$ ($\kmps$)   & $300$            & $300$ \\
 $T_{\rm halo, 0}$ ($\K$)   & $6.5 \times 10^{6}$ & $6.5 \times 10^{6}$ \\
 $e_{\rm rot}$          & 0.90                & 0.95 \\
 $Z$ ($\Zsol$)          & 1.0                 & 1.0 \\
 \hline
 \end{tabular}
 \end{center}
\end{table}

\section{Results}
\label{sec:hydro_results}

\begin{table*}
\begin{minipage}{175mm}
\caption[Galactic wind properties in all the models at $t = 7.5 \Myr$]
 {Galactic wind properties in all the models
 at $t = 7.5 \Myr$. Note that these
 values account for both lobes of the bipolar wind, except the quoted
 sizes, which are measured from the centre of the galaxy. The input
 parameters for the different models can be found in 
 Tables~\ref{tab:model_param} and \ref{tab:model_ism}.}
\label{tab:main_results}
 \begin{tabular}{lccccccccccccc}
  \hline
Property & Units & tbn\_1 & tbn1a & tbn1b & tbn\_2 & tbn\_6 & tbn\_7 & tbn\_9 
	 & mnd\_3 & mnd\_4 & mnd\_5 & mnd\_6 & mnd\_7 \\ 
  \hline
$L_{\rm W}^{a}$ & $10^{40} \ergps$
	& 206.0 & 206.0 & 206.0 & 20.6 & 206.0 & 210.0 & 206.0
	& 206.0 & 20.6 & 206.0 & 210.0 & 206.0 \\
$L_{\rm X,soft}^{b}$  & $10^{40} \ergps$
	& 18.78 & 12.08 & 6.92
	& 0.84 & 57.49 & 14.45 & 36.86 
	& 0.63 & 0.07 & 22.38 & 0.88 & 3.67 \\
$L_{\rm X,hard}^{c}$ & $10^{38} \ergps$
	& 8.22 & 8.12 & 6.14
	& 0.19 & 51.76 & 3.89 & 13.96 
	& 5.92 & 1.77 & 40.18 & 2.11 & 7.09 \\
$F_{\rm X,int}^{d}$ & count $\ps$
	& 30.68 & 19.05 & 10.76 
	& 1.37 & 79.1 & 24.00 & 50.12
	& 0.83 & 0.08 & 21.36 & 1.27 & 5.61 \\
$F_{\rm X,abs}^{e}$ & count $\ps$
	& 1.65 & 1.40 & 1.00 
	& 0.05 & 11.85 & 1.02 & 7.16
	& 0.13 & 0.02 & 9.29 & 0.07 & 0.74 \\
$E_{\rm inj}^{f}$ & $10^{56} \erg$
	& 7.10 & 7.10 & 7.10 & 0.71 & 7.10 & 2.53 & 7.10
	& 7.10 & 0.71 & 7.10 & 2.53 & 7.10 \\
$E_{\rm th}^{g}$ & $10^{56} \erg$
	& 1.55 & 1.35 & 1.15 
	& 0.15 & 0.75 & 0.65 & 1.53
	& 2.47 & 0.44 & 0.81 & 1.15 & 2.57 \\
$E_{\rm th, z>1.5}^{g}$ & $10^{56} \erg$
	& 1.06 & 0.73 & 0.80 
	& 0.02 & 0.17 & 0.27 & 0.53
	& 2.31 & 0.40 & 0.61 & 1.03 & 2.28 \\
$E_{\rm KE}^{h}$ & $10^{56} \erg$
	& 2.73 & 2.51 & 3.21 
	& 0.19 & 3.13 & 1.08 & 2.74
	& 4.86 & 0.35 & 5.75 & 1.78 & 4.70 \\
$E_{\rm KE, z>1.5}^{h}$ & $10^{56} \erg$
	& 2.04 & 1.71 & 2.58 
	& 0.04 & 1.51 & 0.51 & 1.63
	& 4.27 & 0.30 & 4.28 & 1.22 & 4.05 \\
$M_{\rm inj}^{f}$ & $10^{7} \Msol$
	& 1.20 & 1.20 & 1.20 & 0.12 & 1.20 & 0.43 & 1.20 
	& 1.20 & 0.12 & 1.20 & 0.43 & 1.20 \\
$M_{\rm gas}^{i}$ & $10^{7} \Msol$
	& 56.5 & 54.8 & 49.4 
	& 19.5 & 56.8 & 30.9 & 61.1
	& 18.9 & 8.0 & 21.1 & 11.7 & 24.9 \\
$M_{\rm gas, z>1.5}^{i}$ & $10^{7} \Msol$
	& 9.6 & 6.7 & 7.4 
	& 0.4 & 6.1 & 2.5 & 7.5
	& 4.0 & 0.9 & 5.8 & 2.0 & 5.4 \\
$\eta_{\rm warm}^{j}$ & -
	& 0.10 & 0.12 & 0.11 
	& 0.20 & 0.38 & 0.13 & 0.17
	& 0.21 & 0.12 & 0.23 & 0.20 & 0.08 \\
$f_{\rm M, warm}^{k}$ & -
	& 0.06 & 0.04 & 0.04 
	& 0.013 & 0.05 & 0.03 & 0.04
	& 0.04  & 0.03 & 0.13 & 0.05 & 0.12 \\
$f_{\rm TH, warm}^{l}$ & -
	& 0.11 & 0.10 & 0.11 
	& 0.06 & 0.17 & 0.07 & 0.09
	& 0.02 & 0.04 & 0.17 & 0.03 & 0.07 \\
$f_{\rm KE, warm}^{m}$ & -
	& 0.21 & 0.22 & 0.19 
	& 0.11 & 0.31 & 0.15 & 0.14
	& 0.33 & 0.16 & 0.30 & 0.27 & 0.05 \\
$\log n_{\rm e, warm}^{n}$ & $\pcc$
	& -0.77 & -0.80 & -0.98 
	& -1.06 & -0.52 & -0.62 & -0.63
	& -2.15 & -2.30 & -1.54 & -1.80 & -1.52 \\
$v_{\rm warm}^{o}$ & $\kmps$
	& 621 & 668 & 802 
	& 433 & 745 & 646 & 607
	& 1936 & 1143 & 732 & 1579 & 289 \\
$\eta_{\rm hot}^{p}$ & -
	& 0.81 & 0.75 & 0.83 
	& 0.55 & 0.32 & 0.72 & 0.77
	& 0.71 & 0.82 & 0.55 & 0.80 & 0.87 \\
$f_{\rm M, hot}^{q}$ & -
	& 0.03 & 0.03 & 0.04 
	& 0.005 & 0.03 & 0.02 & 0.07
	& 0.18  & 0.11 & 0.08 & 0.15 & 0.18 \\
$f_{\rm TH, hot}^{r}$ & -
	& 0.57 & 0.45 & 0.66 
	& 0.23 & 0.61 & 0.40 & 0.78
	& 0.83 & 0.88 & 0.76 & 0.89 & 0.81 \\
$f_{\rm KE, hot}^{s}$ & -
	& 0.49 & 0.41 & 0.59 
	& 0.16 & 0.15 & 0.46 & 0.50
	& 0.56 & 0.74 & 0.53 & 0.66 & 0.89 \\
$\log n_{\rm e, hot}^{t}$ & $\pcc$
	& -2.04 & -1.91 & -2.02 
	& -2.15 & -0.79 & -1.82 & -1.31
	& -2.65 & -2.76 & -1.50 & -2.74 & -2.40 \\
$v_{\rm hot}^{u}$ & $\kmps$
	& 804 & 742 & 836 
	& 660 & 591 & 889 & 735
	& 772 & 644 & 331 & 693 & 815 \\
$z_{\rm max}^{v}$ & $\pc$
	& 7088 & 5826 & 6217 
	& 3471 & 4025 & 5469 & 6329
	& 8444 & 5338 & 6635 & 6796 & 7831 \\
$r_{\rm max}^{w}$ & $\pc$
	& 2888 & 2392 & 2460 
	& 1327 & 1546 & 1983 & 2115
	& 5513 & 3675 & 4404 & 4696 & 5119 \\
$r_{\rm base}^{x}$ & $\pc$
	& 1167 & 1096 & 1102 
	& 695 & 1259 & 841  & 1220
	& 1220 & 763 & 1395 & 904 & 1045 \\
  \hline
 \end{tabular}
$^{a}$ Starburst mechanical energy injection rate averaged over the
	period between $t = 6.5$ and $7.5 \Myr$. \\
$^{b}$ Intrinsic soft X-ray luminosity in 
	the {\it ROSAT} $0.1$ -- $2.4 \keV$ band.\\
$^{c}$ Intrinsic hard X-ray luminosity in the 
	$2.4$ -- $15.0 \keV$ energy band.\\
$^{d}$ {\it ROSAT} PSPC count rate assuming no absorption 
	and distance $D = 3.63 \Mpc$ to M82.\\
$^{e}$ {\it ROSAT} PSPC count rate assuming a uniform hydrogen column of 
	$\nH = 4 \times 10^{20} \pcm2$ (the Galactic column density
        towards M82, Stark \etal 1992) 
        and distance $D = 3.63 \Mpc$ to M82.\\
$^{f}$ Total energy and mass injected from SNe and stellar winds in
        the starburst up to this time.\\
$^{g}$ Total thermal energy within the entire wind ($E_{\rm TH}$), 
       and within those parts of the wind lying above $z = 1.5 \kpc$ 
       ($E_{\rm TH, z>1.5}$).\\
$^{h}$ Total kinetic energy within the entire wind ($E_{\rm KE}$), 
	and within those parts of the wind lying above $z = 1.5 \kpc$
       ($E_{\rm KE, z>1.5}$).\\
$^{i}$ Total gas mass within the entire wind and within the wind lying
 above $z = 1.5 \kpc$\\
$^{j}$ Volume filling factor of the warm ($5.5 \le \log T (\K) < 6.5$) gas.\\
$^{k}$ to $^{m}$ Fraction of total gas mass ($f_{\rm M}$), 
	thermal energy ($f_{\rm TH}$) and kinetic energy ($f_{\rm KE}$)
 	in warm gas.\\
$^{n}$ Root mean square electron density of warm gas within the wind.\\
$^{o}$ Volume-averaged velocity of warm gas.\\
$^{p}$ to $^{u}$ As $i$ -- $o$ but for hot ($6.5 \le \log T (\K) < 7.5$) gas.\\
$^{v}$ Maximum extent of the wind (the position of the outermost shock) 
	along the minor axis measured 
	from the nucleus of the galaxy. \\
$^{w}$ Maximum radial extent of the wind, measured from the minor axis.\\
$^{x}$ Maximum radial extent of the wind in the plane of the galaxy 
        (\ie at $z = 0$). \\
\end{minipage}
\end{table*}

We shall concentrate on three main topics in this present paper:
(a) wind growth and outflow geometry, in particular the
issues of wind collimation and confinement; (b) the origin
and physical properties of the soft X-ray emitting gas in
these winds, in particular the filling factor of the X-ray
dominant gas, and (c) the previously 
unexplored aspects of wind energetics and energy transport
efficiencies.

The observable X-ray properties of these models, \ie simulated X-ray imaging
and spectroscopy, will be in the second paper of this series. 
Also deferred to Paper II is the discussion of which model parameters
seem best to describe M82's observed properties.

Table~\ref{tab:main_results} provides a general compilation
of physically interesting wind properties in all twelve of the models at a
fairly typical epoch of wind growth, $7.5 \Myr$ after the start
of the starburst.

For descriptive convenience we shall define gas temperatures
in the following terms. In general, ``cool'' gas has temperatures
in the range $4.5 \le \log T (\K) < 5.5$, ``warm'' gas lies in the
range $5.5 \le \log T (\K) < 6.5$, ``hot'' gas has 
$6.5 \le \log T (\K) < 7.5$ and ``very hot'' gas has temperatures
$7.5 \le \log T (\K) < 8.5$.

\subsection{Wind growth and outflow geometry}

In this section we shall concentrate on the intrinsic
morphology of the wind, in particular opening angles and the radius of the
wind in the plane of the galaxy, as well as the qualitative wind structure
in comparison to standard wind-blown bubbles. 
We shall consider the information X-ray surface brightness morphology
provides separately in Paper II. 

Morphological information on the wind geometry in M82 is primarily based
upon optical and X-ray observations. Optical emission line studies 
such as Heckman \etal (1990), G\"{o}tz \etal (1990) \& McKeith \etal (1995)
constrain the cool-gas outflow geometry strongly within $z \ltsimm 2 \kpc$ 
of the plane of the galaxy, using spectroscopy and imaging.
Narrow-band optical imaging can trace the wind out to $z \sim 6 \kpc$.
X-ray observations by the {\it ROSAT} PSPC and 
HRI also trace the warm and hot phases of the
wind out to $z \sim 6 \kpc$ from the plane, 
but suffer from poor resolution
and point source confusion near the plane of the galaxy.

\begin{figure*}
 \centerline{
   \psfig{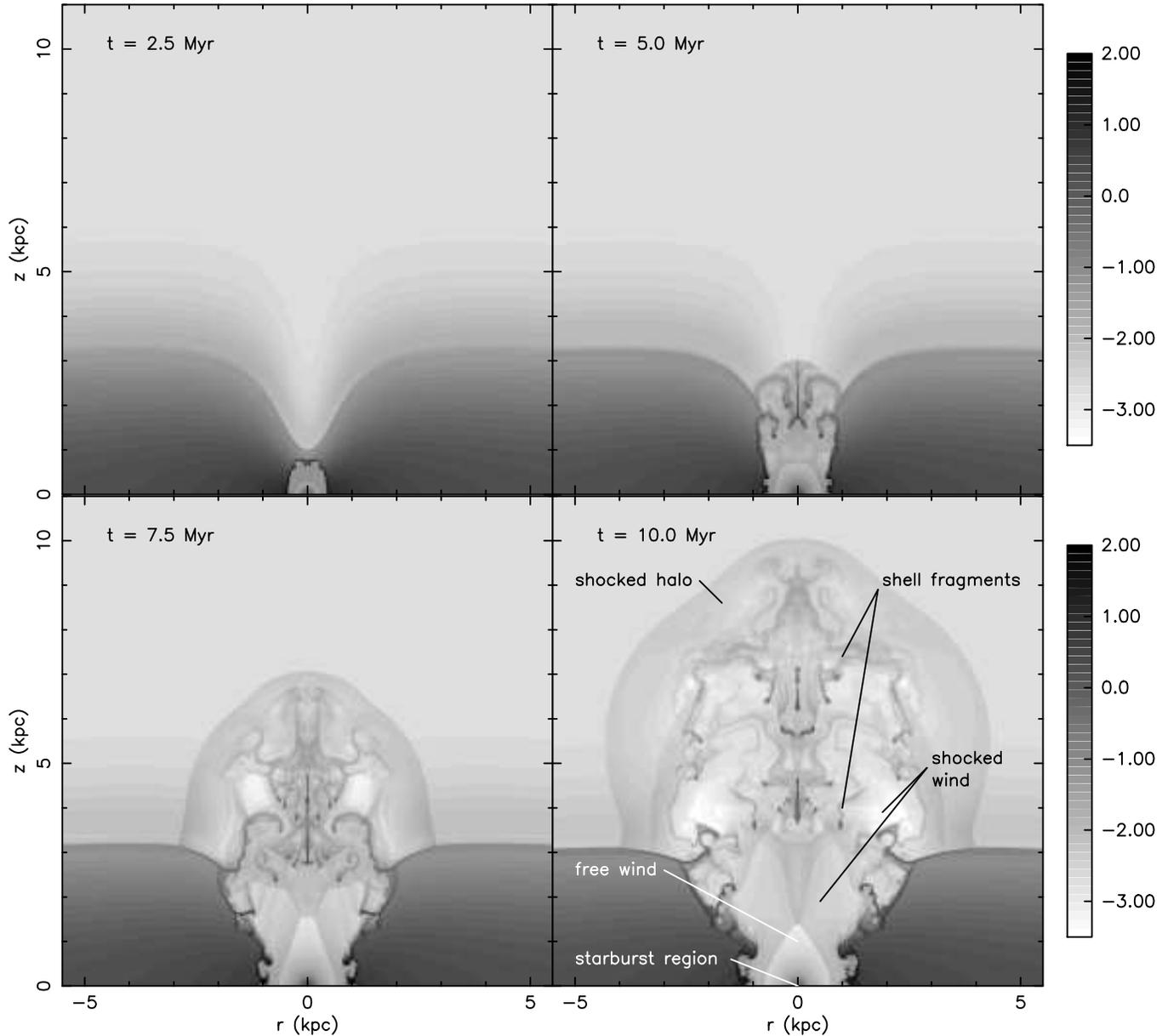}
	}
\caption[Wind number density distributions for model tbn\_1 at four epochs]
 {The logarithm of the wind number density in the thick disk 
 model tbn\_1 at four epochs. The intensity scale 
 extends between $-3.5 \le \log n (\pcc) \le 2.0$, 
 and number densities above or below 
 this range are shown as black or white respectively. The final panel labels
 some important features of the wind: 
 the starburst region in the nucleus of the galaxy, the freely expanding
 wind, shocked wind, shocked disk and halo gas, and superbubble shell 
 fragments. Note that these
 numerical simulations assume cylindrical symmetry around the $z$-axis,
 so the shell fragments are actually annuli. 
 Note the very small region of 
 freely expanding wind, the initially cylindrical geometry of the wind
 opening out with time into a truncated cone, and the increasing size of
 the wind in the plane of the galaxy.}
\label{fig:tbn_1_lnum}
\end{figure*}

\begin{figure*}
 \centerline{
   \psfig{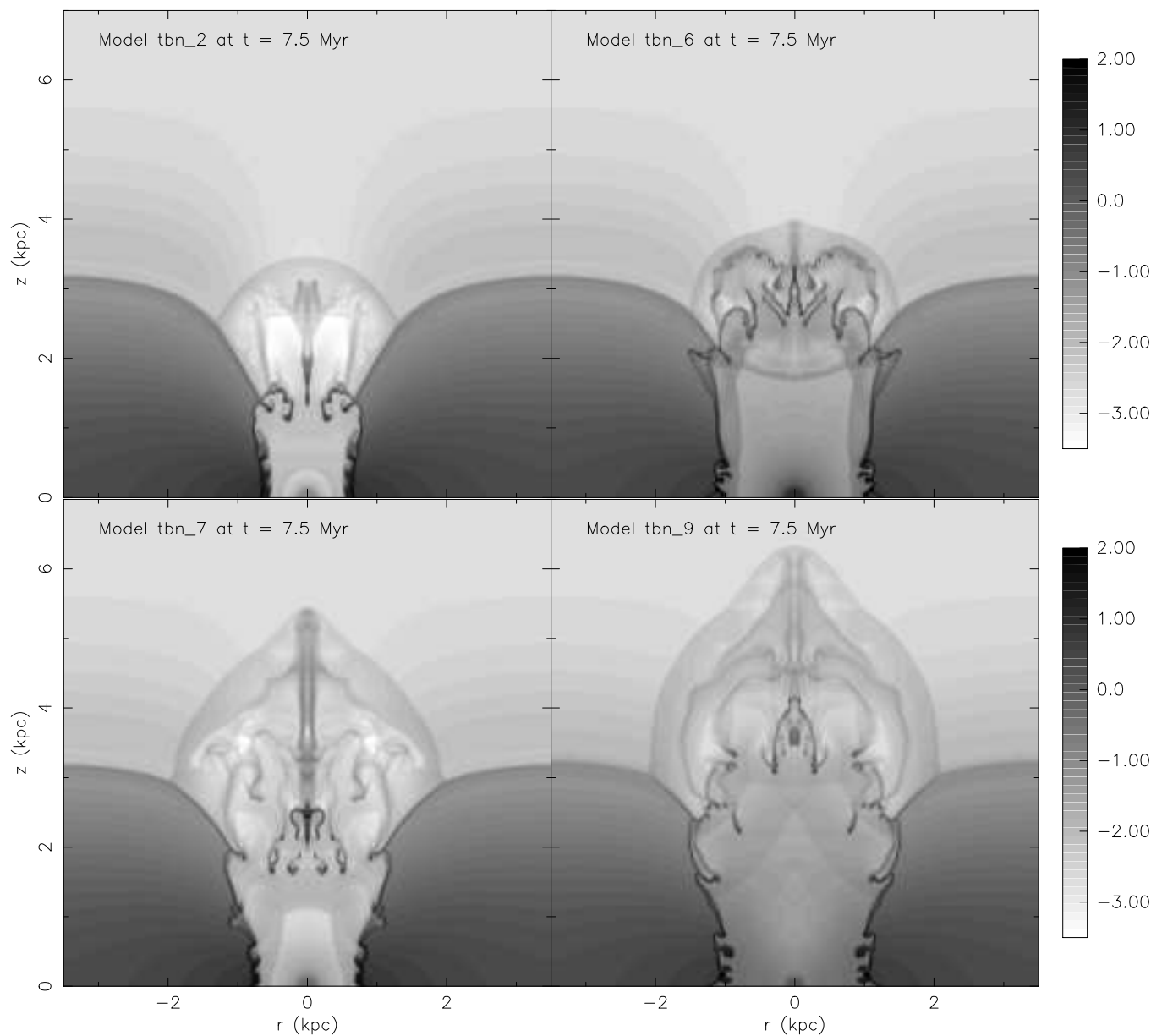}
	}
\caption[Wind number density distributions for model tbn\_2 at six epochs]
 {The logarithm of the wind number density in the thick disk 
 models tbn\_2, tbn\_6, tbn\_7 \& tbn\_9 at $t = 7.5 \Myr$.
 The intensity scale extends between $-3.5 \le \log n (\pcc) \le 2.0$, 
 and number densities above or below 
 this range are shown as black or white respectively. The weaker starburst
 tbn\_2 (a factor 10 less powerful than model tbn\_1) leads to a smaller but 
 otherwise structurally similar wind to model tbn\_1.
 In model tbn\_6 the cooler and denser wind (due to the 
 additional mass injection within the starburst region) 
 significantly alters the structure of the 
 regions of shocked wind material within the galactic wind, and reduces
 its expansion rate. The more gradual energy injection rate of model tbn\_7
 compared to model tbn\_1 also reduces 
 the wind's growth rate, but
 otherwise this model is very similar to model tbn\_1.
 The interior structure of the wind in model tbn\_9 
 is much more homogeneous than
 the other thick disk models due the enhanced wind density due to distributed 
 mass-loading,
 and it is difficult to distinguish separate free wind and shocked 
 wind regions.}
\label{fig:tbn_2_lnum}
\end{figure*}

\subsubsection{Wind density structure}

\begin{figure*}
 \centerline{
   \psfig{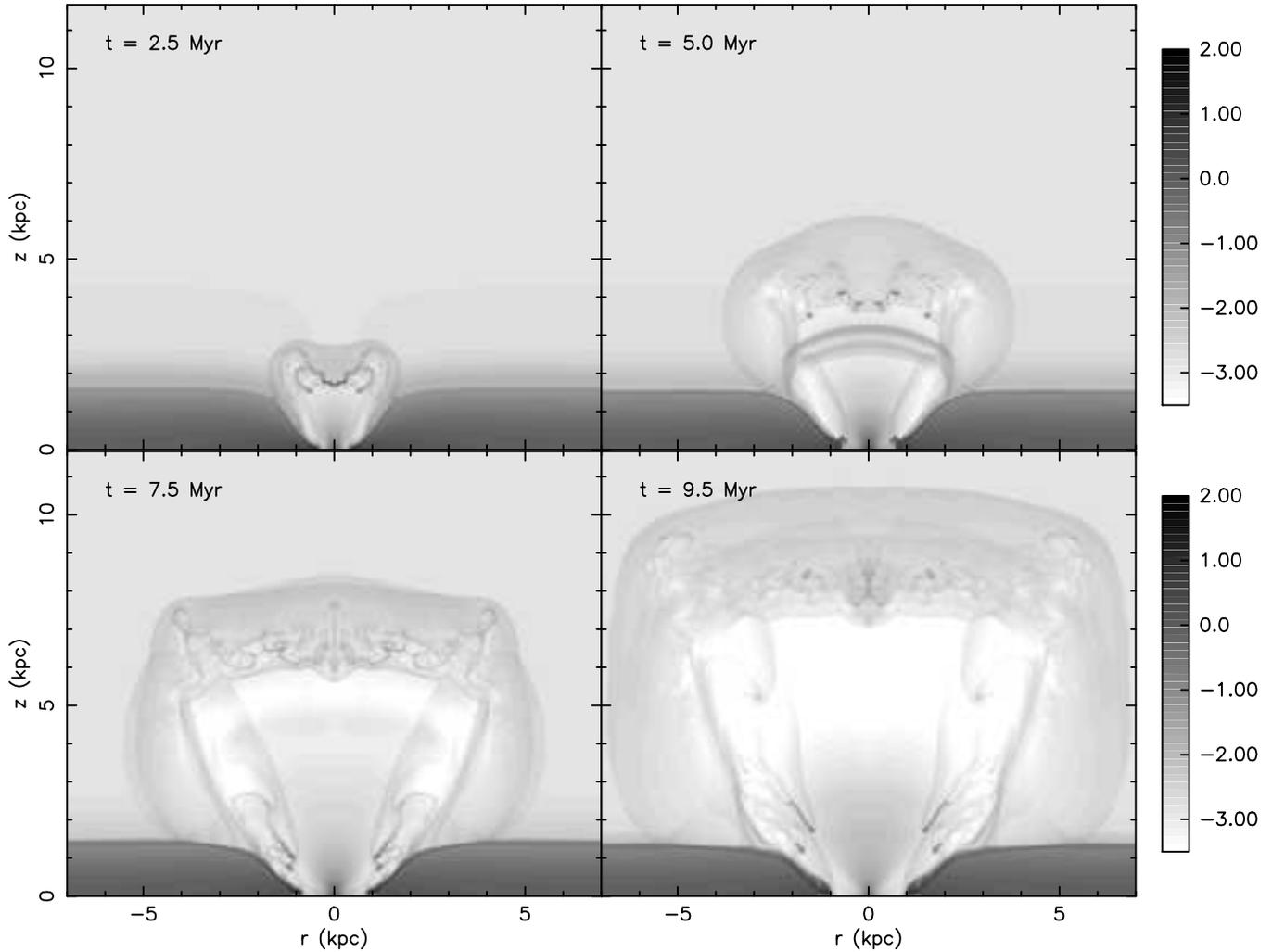}
	}
\caption[Wind number density distributions for model mnd\_3 at five epochs]
 {The logarithm of the wind number density in the thin disk 
 model mnd\_3 at four epochs. The intensity scale 
 extends between $-3.5 \le \log n (\pcc) \le 2$, 
 and number densities above or below 
 this range are shown as black or white respectively. 
 In comparison with the thick disk model tbn\_1 note 
 the very large volume occupied by the freely expanding wind and the
 large opening angle. Due to the lack of disk material above the starburst
 the importance of shocked disk and shell fragments is reduced in these
 thin disk models.}
\label{fig:mnd_3_lnum}
\end{figure*}

\begin{figure*}
 \centerline{
   \psfig{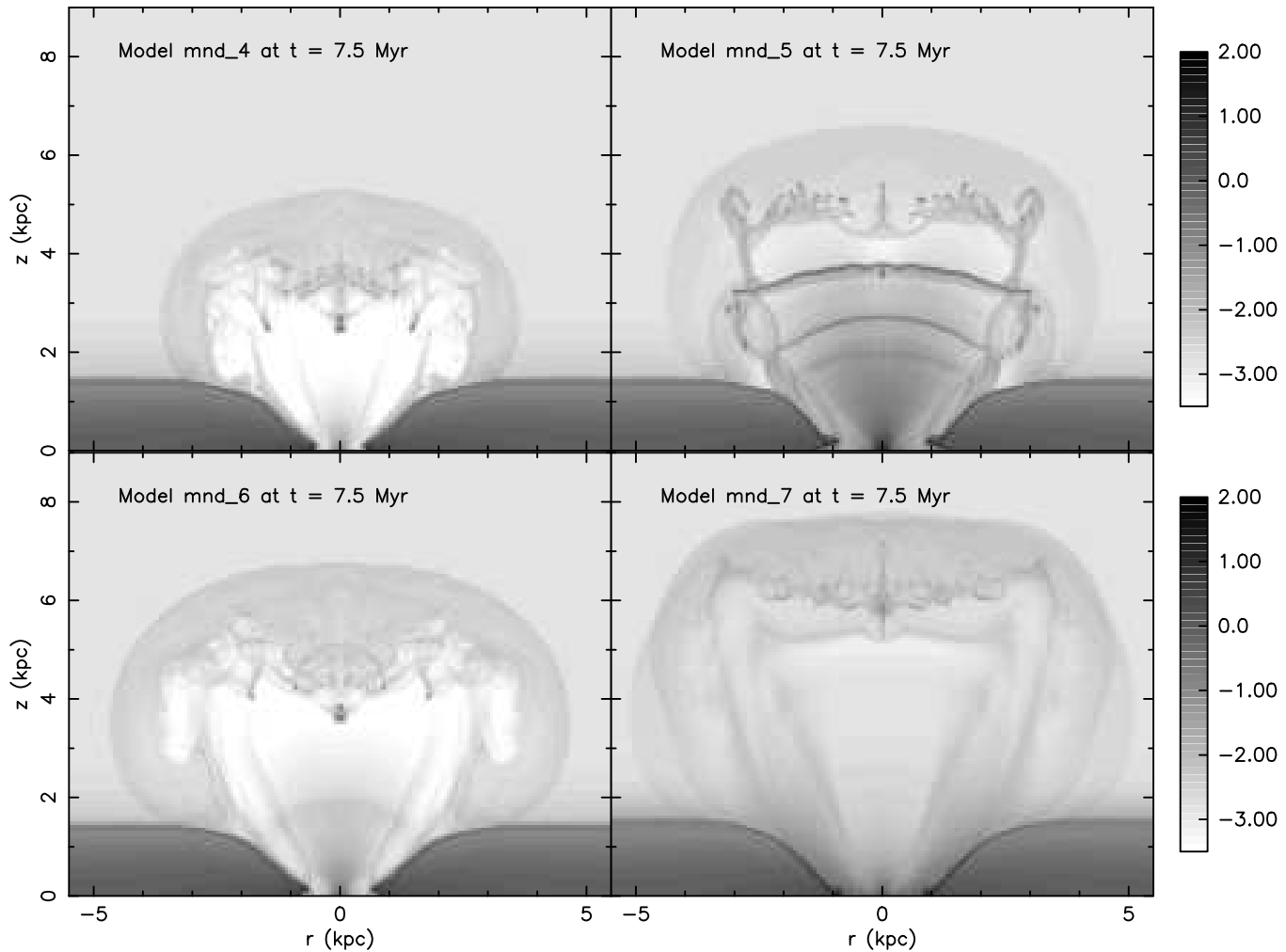}
	}
\caption[Wind number density distributions for model mnd\_4 at five epochs]
 {The logarithm of the wind number density in the thin disk 
 models mnd\_4, mnd\_5, mnd\_6 \& mnd\_7 at $t = 7.5 \Myr$. The 
 intensity scale extends between $-3.5 \le \log n (\pcc )\le 2$, 
 and number densities above or below 
 this range are shown as black or white respectively. The weak starburst
 in model mnd\_4
 leads to a smaller free wind region, and the majority of the outflow
 volume is occupied by shocked halo material. 
 In model mnd\_5, as in model tbn\_6, the cooler and denser mass-loaded
 wind alters the structure of the 
 regions of shocked wind material within the galactic wind.
 The sharp features seen propagating out in the free
 wind are due to periods of increased mass injection in the instantaneous
 starburst model used (see Fig.~\ref{fig:mass_and_energy}). The simple
 model for central mass-loading used assumes the total mass-loading rate in
 the starburst region is just a multiple of the starburst mass injection
 rate. A more realistic model of starburst history and mass-loading would
 lead to a smoother mass injection rate.
 The more gradual energy injection rate of model mnd\_6 compared to the
 instantaneous burst model mnd\_3 reduces the wind's growth rate slightly, but
 otherwise this model is very similar to model mnd\_3.
 In model mnd\_7 the density of the
 free wind is increased due to distributed mass-loading.
 As the clouds are assumed to be distributed in the same way
 as the ambient ISM the majority of the mass-loading occurs not near the
 $z$-axis but near the outer edge of the wind.
 }
\label{fig:mnd_4_lnum}
\end{figure*}

As an effective visual method of illustrating galactic wind evolution
and growth, and some of the differences between the various models, 
we reproduce grey-scale images
of $\log$ number density in the $r$-$z$ plane in 
Figs.~\ref{fig:tbn_1_lnum} -- \ref{fig:mnd_4_lnum}. These show the
wind at $2.5 \Myr$ intervals up to $t = 10 \Myr$ in models
tbn\_1 and mnd\_3, and at $t = 7.5 \Myr$ in all the other
models excepting models tbn1a \& tbn1b.

We shall briefly describe the evolution and structure 
of the wind in model tbn\_1 (see Fig.~\ref{fig:tbn_1_lnum}), 
a single instantaneous starburst occurring in the thick disk ISM, and
then discuss the differences in 
the evolution of the wind in the other models to this
model. 
Differences between model tbn\_1 and the higher resolution models
tbn1a and tbn1b are discussed in Section~\ref{sec:results_numres}
along with a general discussion of the effects of finite numerical resolution.

At $t = 2.5 \Myr$ the starburst-driven superbubble has yet to blow out of
the thick disk, although it is elongated along the minor axis. Although
difficult to see when shown to scale alongside the later stages of the
wind, the superbubble has a standard structure of starburst region, free
wind, shocked wind and a denser cooler shell of swept-up and shocked disk
material.

By $t = 5 \Myr$ the superbubble has blown out, the dense superbubble
shell fragmenting under Rayleigh-Taylor (RT) instabilities. The superbubble
shell was RT-stable
as long as it was decelerating, but the negative density gradient along the 
$z$-axis and the sudden influx of SN energy at $t \sim 3 \Myr$
rapidly accelerate the shell along the minor axis 
after $t \sim 3 \Myr$. The internal structure of the wind 
is more complex than the superbubble described above.
A new shell of shocked halo matter forms, but given its
 low density and high temperature it never cools to form a dense shell
as in the superbubble phase. The re-expanding shocked wind can be seen
to be ablating the shell fragments. Note also the structure of the
reverse shock terminating the free wind region, which is no
longer spherical. The oblique nature of this shock away from the minor
axis acts to focus material out of the plane of the galaxy, as 
first described by TI. The combination of the disk density gradient
and this shock-focusing make the wind cylindrical at this stage. 

Note that the complex structure of the wind 
after blowout and shell fragmentation means that it is not meaningful
to model the emission from interior of a galactic wind 
using the standard Weaver \etal (1977) similarity solutions.
Nevertheless, semi-analytical models based on the thin-shell
approximation (\eg Mac Low \& McCray 1988; Silich \& Tenorio-Tagle 1998)
can be used to explore the location of the outer shock
of the wind with good accuracy, provided the radiative losses
from the interior of the wind are not significant.
Tracking the location of the outer shock is perhaps only 
important in assessing if a superbubble of a set mechanical 
energy injection rate can blow out of a given ISM distribution.
Calculations of observable properties and the long term fate of the
matter in any outflow do require the use of multidimensional hydrodynamical
simulations.

The wind geometry within the disk is similar to a truncated cone
at $t = 7.5 \Myr$. In the halo the outer shock propagating in the halo
becomes more spherical as the anisotropy of the ISM reduces with
increasing distance along the $z$-axis. The structure of the shocked wind
region is becoming even more complex, as it interacts with the superbubble
shell fragments. The shell fragments are steadily being carried out of the
disk, although being slowly spread out over a larger range of $z$ with time.
The shocked wind also interacts with the disk,
removing disk gas and carrying it slowly out of the disk.

At $t = 10 \Myr$ the wind has increased in size, 
but remains qualitatively very similar
in structure to the wind at $t = 7.5 \Myr$. Shell fragments are spread between
$3.5 \le z \le 7 \kpc$, a large range given their common
origin in the superbubble shell.
Long tails can be seen extending from the shell fragments, their curling
shapes tracing the complex flow pattern within the shocked wind.
Regions of expansion followed by new internal shocks can also be seen
within the shocked wind, a marked difference from the structure
of conventional wind blown bubbles.

With mass and energy injection rates from the starburst reduced by
a factor 10 from model tbn\_1, the wind in model tbn\_2 evolves 
at a slower pace than model tbn\_1 (see Fig.~\ref{fig:tbn_2_lnum}). 
The eventual
wind structure is very similar to that of model tbn\_1,
although disruption of the disk appears reduced and the wind
remains more cylindrical within the disk. From standard self-similar
wind blown bubble theory (Weaver \etal 1977) the size of a pressure-driven
bubble is only a weak function of the energy injection rate, 
$R \propto L_{\rm w}^{1/5}$, so we might expect model tbn\_2 to be
$\sim 60$\% the size of model tbn\_1 at any given epoch.  
In practice model tbn\_2 is less than 60\% the size of model tbn\_1
at the same epoch, which may be due the increased relative importance of
radiative cooling in depressurising tbn\_2's wind, given that
the cooling time-scales are the same in both simulations.

The wind in model tbn\_6 (Fig.~\ref{fig:tbn_2_lnum})
grows at slower rate than 
model tbn\_1 despite having exactly the same energy injection history.
In this case the reduced growth must be due to the central mass-loading used.
The mass injection rate within the starburst has been increased by
a factor 5 from that due to stellar winds and SNe alone, representing
the entrainment of dense gas remaining from the star formation.
The central mass-loading increases the density of the free wind
by a factor $5^{3/2} \approx 11$, due to the additional mass and
the reduced outflow rate
(as the energy per particle is less), and reduces
the temperature of the gas by a factor $5$ from $\sim 1.5 \times 10^{8} \K$
to $\sim 3 \times 10^{7} \K$. Despite the  enhanced density of the material
ejected from the starburst, the wind retains recognisable
features of shell fragments and shocked halo, although their dynamics
and morphology have clearly been altered by the denser, slower wind
fluid in this model.

Model tbn\_7 (Fig.~\ref{fig:tbn_2_lnum})
was chosen to investigate the effect of a more
gradual energy and mass injection history than the instantaneous 
starburst used in the other models (see Fig.~\ref{fig:mass_and_energy}). 
Apart from the resulting slightly slower growth, the structure
of the wind in this model is very similar to that in model tbn\_1.

The distributed mass-loading in model tbn\_9 significantly
increases the density of what would be the free and/or shocked wind
regions. Note the apparent lack of an obvious shock separating
the free wind and shocked wind regions
in this model (Fig.~\ref{fig:tbn_2_lnum}). This is not unexpected,
as distributed mass-loading has the interesting property of increasing
the Mach number of subsonic flows while reducing the Mach number in
supersonic flows, to produce a flow with a Mach number of order unity
(Hartquist \etal 1986). As in model tbn\_6 the wind's growth is slightly
slower than the non-mass-loaded model tbn\_1.

The wind structure in model mnd\_3 (Fig.~\ref{fig:mnd_3_lnum}) 
is significantly different
from model tbn\_1 due to the thin, less-collimating disk used
in this and the following models. The lack of substantial amounts of dense
gas above the starburst region allows very rapid blowout of the wind,
easily shattering a much less massive superbubble shell. The net result is
a wind with a larger opening angle (due to the lack of any significant
collimation by the disk), a very large free wind region, and an indistinct
region of shocked wind and shocked disk material
surrounded by the standard shocked halo
region. The few superbubble fragments rather rapidly loose definition
and are mixed in with the shocked wind by $t \sim 7.5 \Myr$. Due to the
early blowout into the halo and lack of collimation the wind is much
more spherical than the thick disk models.

The morphology of the 
weak starburst model mnd\_4 (Fig.~\ref{fig:mnd_4_lnum})
is generally similar to that of model mnd\_3, although with two
interesting exceptions. The much less energetic wind is not able to
punch out into the halo as effectively as the wind in model mnd\_3.
A weak, almost spherical, shock wave does propagate out into the halo,
leading to a wind with much of the volume being shocked halo material.
Also note the free wind is confined to a narrow region along the
$z$-axis by shocked wind material flowing up out of the disk. This
flow of shocked wind is also present in the other thin disk models,
and is responsible for their peculiar, almost box like, morphology.

The structure and growth of the central mass-loaded thin disk 
model mnd\_5, the complex star formation
history model mnd\_6, the distributed
mass-loading model mnd\_7 (see Fig.~\ref{fig:mnd_4_lnum}) are related to
the basic thin disk model mnd\_3 in the same way as
the equivalent thick disk models are related to model tbn\_1. 

The main differences
between the thin disk models and the thick disk models are the
lack of collimation in the thin models, the resulting large regions
of freely expanding wind, and less complex shocked wind and disk regions
with shell fragments that are disrupted and mixed into the flow at an
earlier stage.

\subsubsection{Wind growth}

\begin{figure*}
 \centerline{
   \psfig{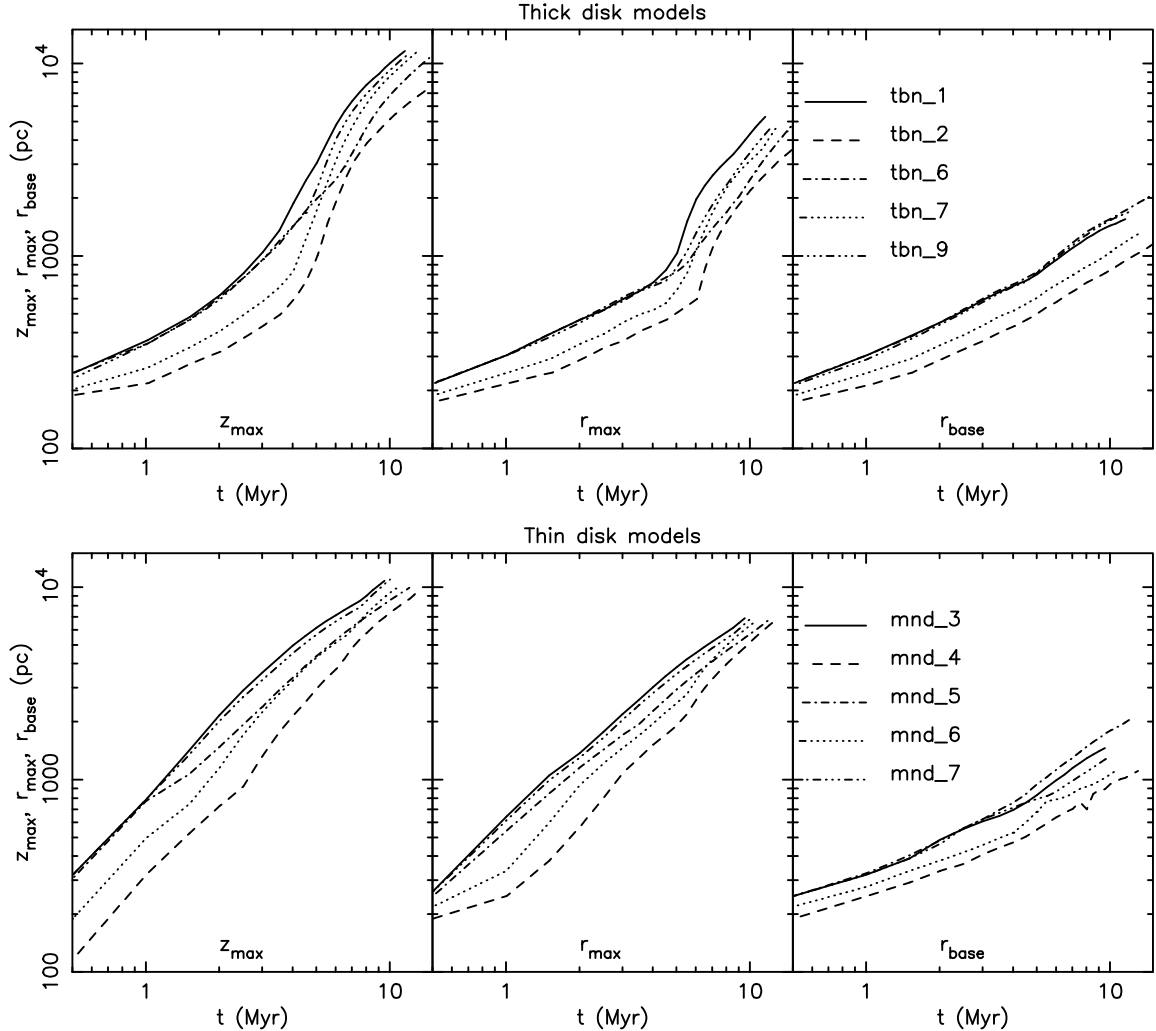}
	}
\caption[Vertical and radial growth of all the wind models]
 {Vertical and radial growth of all the wind models as a function of time.
 The initial evolution of the maximum vertical size $z_{\rm max}$,
 maximum radial size $r_{\rm max}$ and size of the wind in the plane of the
 galaxy $r_{\rm base}$ for the thick disk models, 
 obeys the $R \propto t^{0.6}$ law expected of confined pressure-driven 
 bubbles. Post blow out (which occurs almost instantaneously in the thin
 disk models) $z_{\rm max}$ and $r_{\rm max}$ grow more rapidly,
 as the wind accelerates down the density gradient into the halo.}
 \label{fig:wind_growth}
\end{figure*}

The vertical and radial growth of the galactic winds as a function of time 
in these models is shown in Fig.~\ref{fig:wind_growth}. The
maximum vertical extent of the wind (almost invariably size of
wind on the $z$-axis) $z_{\rm max}$, maximum radial extent $r_{\rm max}$
and radius of the base of the wind $r_{\rm base}$ 
in all the models can be seen to be well behaved power laws
(or broken power laws) in time.

The vertical and radial growth rates in the thick disk models clearly
differ from those in the thin disk models. The thick disk models
all show initially slow vertical ($z_{\rm max}$) and radial growth
($r_{\rm max}$), followed by rapid acceleration. The wind begins to accelerate
along the $z$-axis after $t \sim 3 \Myr$, and later along
the $r$-axis, $t \sim 5 \Myr$. The initial evolution is similar to that
predicted by the Weaver \etal (1977) model of a pressure driven bubble
in a constant density medium, \ie $R \propto t^{0.6}$. The later phase
of acceleration
is caused by both the superbubble ``running down'' the ISM density gradient
leading into the halo, and the dramatic increase in mechanical
energy injection rate
at $t \sim 3 \Myr$ due to the first SNe. At later times the wind
should return to the $R \propto t^{0.6}$ expansion law once in the almost
constant density halo. The initial stages of this final deceleration may be
what is seen in Fig.~\ref{fig:wind_growth} 
at late times in models tbn\_2 and tbn\_7.

In contrast the wind in the thin disk models shows no evidence for 
periods of increased or decreased acceleration over the simulation. 
Measured over the period from $t = 0.5 \Myr$
until the end of the simulation the wind is constantly accelerating
both vertically 
($1.10 \le \partial \log z_{\rm max} / \partial \log t \le 1.37$) 
and radially ($1.04 \le \partial \log r_{\rm max} / \partial \log t \le 1.25$)
in all the thin disk models. The thin disk allows the wind to blow out
very early in its evolution, and it gradually runs down the density
gradient into the halo.

In all the models (both thin and thick disk) 
the growth of the wind in the plane of the galaxy 
($r_{\rm base}$) is well approximated by the expansion law of
a constant power bubble expanding into a uniform medium. The 
base of the wind expands at a rate 
$\partial \log r_{\rm base} / \partial \log t \sim 0.6$, measured
over the period $t = 0.5 \Myr$ until the end of the simulation.
The range in slope around the expected value of 0.6 is small,
from 0.53 (model mnd\_6) to 0.67 (model tbn\_9), and does not
deviate in any clear
 systematic manner due to mass-loading, the ISM distribution
or the star formation history. 

There is no evidence for a change in 
$\partial \log r_{\rm base} / \partial \log t$ once the wind has
broken out, as might be expected if blowout leads to a depressurisation
of the wind. The evolution of the
hot gas in the plane of the galaxy does not appear
to be affected by the blowout over the period covered by our simulations.
 
\subsubsection{Wind collimation and opening angles}
\label{sec:results_collimation}
 
The density distributions
 shown in Figs.~\ref{fig:tbn_1_lnum} -- \ref{fig:mnd_4_lnum}
clearly demonstrate that the main factor controlling the wind morphology
is the disk ISM distribution. The thick disk ISM models produce
initially cylindrical winds that evolve into collimated truncated conical winds
with low opening angles. The thin disk models invariably produce
conical winds with large opening angles.

\begin{figure}
 \centerline{
   \psfig{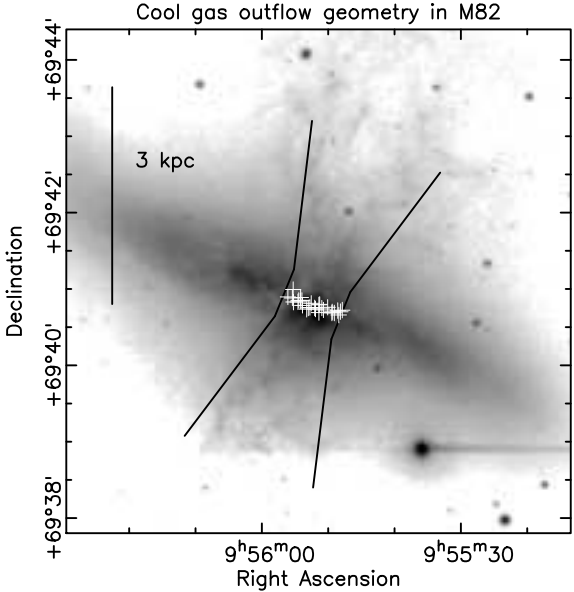}
	}
\caption[Inferred line of sight outflow geometry 
 compared to \halpha~morphology]
 {The wind outflow geometry in M82 based on the work of 
 G\"{o}tz \etal (1991) and McKeith \etal (1995),
 overlaid on a \halpha~image. The young supernova remnants (Muxlow 
 \etal 1994) marking the starburst region are shown as a 
 series of white crosses.}
\label{fig:outflow_geom}
\end{figure}

These outflow geometries should be compared to that inferred from optical
observations of M82 (see Fig.~\ref{fig:outflow_geom}), of an initially
conical wind ($r \sim 420 \pc$ for $z \le 330 \pc$) flaring out above
$z = 330 \pc$ into a cone of opening angle\footnote{Note that we
quote the full opening angle, and not the half opening angle, which
is also commonly used in the literature.}
 $\theta \sim 30\deg$ (based on
G\"{o}tz \etal 1990; McKeith \etal 1995, scaled to our assumed distance
of $3.63 \Mpc$ to M82).

Measuring the wind opening angles in our models as a function of time,
it is clear that the thin disk models fail to produce
such a well collimated wind. The opening angle is 
$\theta_{\rm thin} \sim 90\deg$ in model mnd\_3 at all epochs,
which is typical of all the thin disk models.

In comparison the thick disk models (deliberately chosen to provide a
collimating ISM distribution) are much more successful, although
even in these models the opening angles can become too large to be
a good match for M82 at late times. 
The opening angles in the thick disk models 
do show some variation between the different models, but typically
the wind is initially spherical in its superbubble phase, becoming
much more cylindrical (\ie $\theta \sim 0\deg$) as it begins to blow out.
Post-blowout the opening angles are typically 
$\theta_{\rm thick} \sim 40\deg$, but this masks a general increase
from $\theta \sim 0\deg$ at $t \sim 5 \Myr$ to $\theta \sim 60\deg$
at $t \sim 10$ -- $15 \Myr$. More gentle energy injection histories,
as in models tbn\_2 and tbn\_7, do give {\em slightly} 
lower opening angles at a given
epoch than in model tbn\_1, confirming S94's conclusion that weak
initial winds can reduce later disruption of the disk by the more energetic
phases of the starburst. Note that this is a {\em weak} effect
in the simulations we consider,
and the main factor affecting wind geometry is the initial ISM distribution,
not the starburst energy injection history.

It is clear that a thick, collimating ISM distribution seems necessary to
reproduce the observed narrow, low opening angle, wind in M82. Even our thick
disk model does not provide sufficient collimation.

\subsubsection{Confinement}
\label{sec:results_confinement}
TT's criticism
that the size of the base of the wind is too large in TB and S94's
simulations remains true in these simulations. 
The radius of the wind in the plane of the galaxy grows
larger than that observed in M82, and shows no signs of slowing down
as can be seen Fig.~\ref{fig:wind_growth}.

G\"{o}tz \etal's (1990) observations limit the radius of the wind
to $r \sim 420 \pc$ at $z \sim 110 \pc$ above the plane of the galaxy.
This radius is very similar to the extent of the SN remnants that measure
the current SN rate (see Fig.~\ref{fig:outflow_geom}).

In contrast $r_{\rm base} \sim 1500 \pc$ in model mnd\_3 at $t = 9.5 \Myr$,
a factor ten larger than the assumed starburst region. All the simulations
have base radii in the range $r_{\rm base} \sim 1$ -- $2 \kpc$ 
at $t \sim 10 \Myr$. A dramatic
reduction in starburst power by a factor ten (model tbn\_1 to model tbn\_2)
only reduces $r_{\rm base}$ from $\sim 1400 \pc$ to $\sim 850 \pc$ at 
$t = 10 \Myr$. 

By $t = 5 \Myr$ the base of the wind is too large
in all the models, so to explain this problem away with the
ISM distributions we have used would require M82's starburst 
to be very young, \ie $t < 5 \Myr$. This is
difficult to justify observationally, and as we shall show in Paper II, the 
observed X-ray extent of the wind requires a slightly older wind 
($t \gtsimm 7.5 \Myr$).

TT constructed steady state models of bipolar outflows from starburst
galaxies, where the ram pressure of infalling dense molecular gas confines the
radius of the base of the wind to a fixed position. Although this
solves the confinement problem, their model requires unphysically
large masses of gas to be falling in along the plane of the galaxy.
For example, in the model published in Tenorio-Tagle \& Mu\~noz-Tu\~n\'on
(1998), the mass of the ISM within central kiloparsec is $M_{\rm gas}
\sim 5 \times 10^{9} \Msol$, where observations 
of M82 limit the total mass of the ISM to be
$\ltsimm 10^{8} \Msol$ within the same radius.

Dense molecular gas, even if not falling into the nuclear region, 
may nonetheless be important for confining the base of the wind.
CO observations (Nakai \etal 1987) reveal a molecular ``ring''
extending in radius between $r \sim 100$ -- $400 \pc$, at the outer
edge of which spurs of molecular material emerge perpendicular
to the plane and extend $\sim 500 \pc$ from the disk. Much of the
gas mass within the central kiloparsec is probably within this molecular
gas. It may be that this molecular ring can provide the wind with
enough resistance to slow its expansion in the plane of the galaxy.
Although this molecular gas does not have a volume filling
factor of order unity as envisioned by TT (see Lugten \etal [1986]), 
its areal filling factor may
be high, and it undoubtedly could strongly mass-load the wind in the
plane of the galaxy. Further simulations explicitly including a molecular
ring are currently in progress to explore this possibility.

Somewhat more speculatively, magnetic fields might
collimate the wind (\eg de Gouveia Dal Pino \& Medina Tanco 1999).
We do not believe pursuing this option is currently
necessary, given that the alternatives have not yet been explored.
Nevertheless, we explore this idea further when discussing
magnetic fields in Section~\ref{sec:limitations}.

\begin{figure*}
 \centerline{
   \psfig{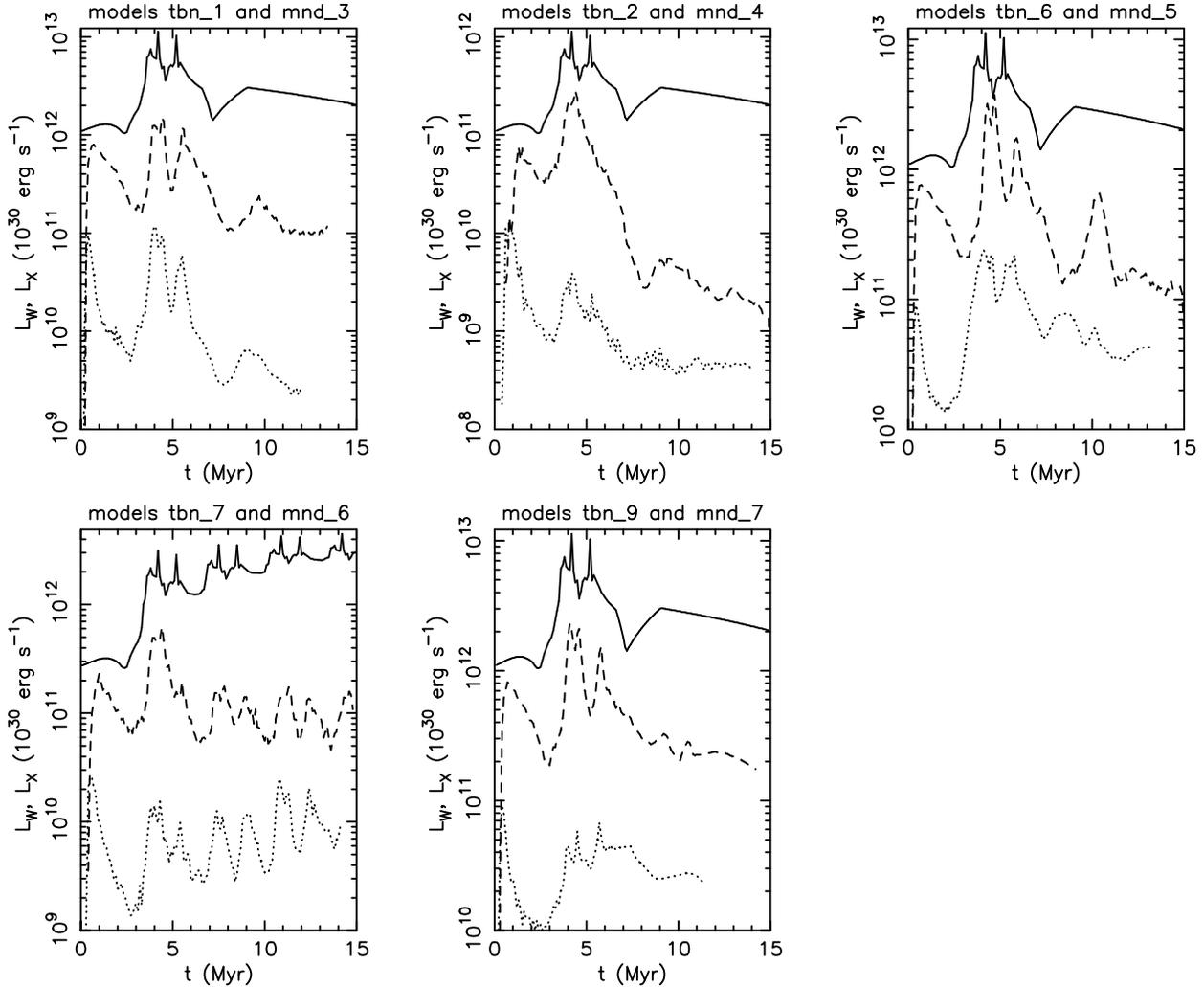}
	}
\caption[Soft X-ray luminosities for the simulated galactic winds compared
 to the starburst energy injection rate]
 {Soft X-ray luminosities in the {\it ROSAT} $0.1$ -- $2.4 \keV$ band 
 for the simulated galactic winds compared to the starburst 
 energy injection rate $L_{\rm W}$. The five panels show $L_{\rm W}$ 
 (solid line) and $L_{\rm X}$ for each of the starburst models. Equivalent
 models differing only in having thick (dashed line) or thin disk 
 (dotted line) ISM distribution have been grouped together. Note the overall
 similarity in form between $L_{\rm X}$ and the $L_{\rm W}$ --- periods of 
 increased starburst energy injection are closely followed by periods of
 increased soft X-ray emission. The initial spike of high X-ray luminosity
 in all the models for $t \ltsimm 2 \Myr$ is a numerical artefact 
 due to poor numerical 
 resolution when the superbubble is very young.}
 \label{fig:lx_lw}
\end{figure*}

\subsection{X-ray emission from galactic winds}
\label{sec:results_xorigin}

\subsubsection{Efficiency of soft X-ray emission}
\label{sec:results_softlx}

\begin{figure}
 \centerline{
   \psfig{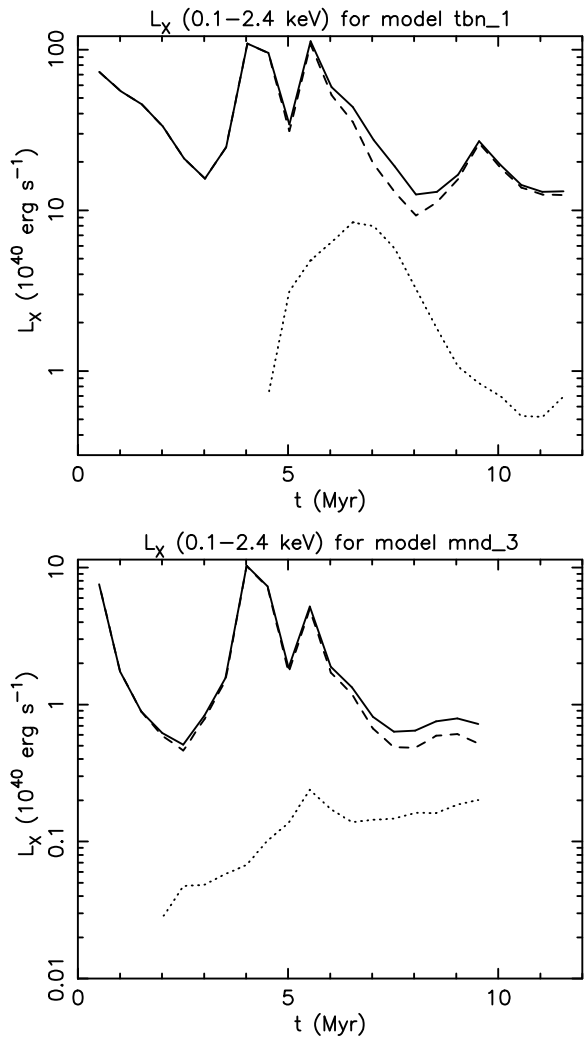}
	}
\caption[Soft X-ray luminosity as a function of time and position in models
 tbn\_1 and mnd\_3]
 {Soft X-ray luminosity as a function of time and position in models
 tbn\_1 and mnd\_3. The total {\it ROSAT} $0.1$ -- $2.4 \keV$ band X-ray
 luminosities are shown as a function of time (solid line), 
 along with $L_{\rm X}$ decomposed into emission from within the disk 
 ($|z| \le 1.5 \kpc$, dashed line) and halo ($|z| > 1.5 \kpc$, dotted
 line). Half or more of the total soft X-ray emission is produced within the
 plane of the galaxy in all the models we have run.}
\label{fig:lx_diskhalo}
\end{figure}

Soft X-ray luminosities as a function of time, for all the models
excluding the resolution study models tbn1a \& tbn1b, 
are shown in Fig~\ref{fig:lx_lw}, in comparison to the starburst 
mechanical energy injection rate $L_{\rm W}$. Note the overall
similarity in form between $L_{\rm X}$ and $L_{\rm W}$, with periods of 
increased starburst energy injection are closely followed by periods of
increased soft X-ray emission (time lag $\Delta t \sim 0.5 \Myr$).

The thick disk models are clearly significantly more X-ray luminous
than the equivalent thin disk models. Lacking
significant amounts of dense gas for the wind to interact with, 
the thin disk models are very inefficient at radiating
the mechanical energy supplied by the starburst. Typically the thin
disk models radiate $\ltsimm 1$ per cent of $L_{\rm W}$ as soft X-rays in the
{\it ROSAT} band. In contrast the thick disk models typically radiate
$\sim 5$ per cent of $L_{\rm W}$, and up to $\sim 20$ per cent of $L_{\rm W}$ at
some epochs.

Note that the high X-ray luminosities at early times, $t \ltsimm 2 \Myr$,
are numerical artefacts due to poor numerical resolution
of the very young superbubbles when they only cover relatively
few computational cells.

The primary variable influencing the emerging X-ray luminosity of these
galactic winds is the ISM density distribution. The starburst 
energy injection rate $L_{\rm W}$ is of secondary importance
(in that $L_{\rm X} \propto L_{\rm W}$),
followed by the presence of mass-loading.

Both mass-loading (models tbn\_6, tbn\_9, mnd\_5 \& mnd\_7) 
and more distributed star formation histories (models tbn\_7 \& mnd\_6) 
increase the soft X-ray luminosities at later times with respect
to the luminosities found in the instantaneous starburst models.
Central mass-loading is most noticeable in altering the
X-ray emission in the thin disk models (model mnd\_5), 
where the soft X-ray luminosity
can be increased by typically an order of magnitude from the non-mass-loaded
models. Mass-loading, either distributed or central, is much less
significant in the thick disk models, where the soft X-ray luminosities
are typically 1 to 3 times those in the non-mass-loaded models.

The extremely high absolute values of some of these soft X-ray 
luminosities should be remarked upon. In the thick disk models
$L_{\rm X}$ is typically several times $10^{41} \ergps$, while in the
thin disk models it is approximately an order of magnitude lower.
The mass-loaded thick disk models
tbn\_6 and tbn\_9 have peak $0.1$ -- $2.4 \keV$ luminosities 
$L_{\rm X} \sim 2 \times 10^{42} \ergps$. This is considerably more
luminous than the majority of starburst galaxies, which typically 
have total X-ray luminosities in the 
{\it ROSAT} $0.1$ -- $2.4 \keV$ band in the range $10^{39}$ -- 
$10^{41} \ergps$ (\eg Read \etal 1997),
of which only a fraction is due to hot gas. Only the most luminous
starburst galaxies have soft X-ray luminosities are high we we find in some
of our models, \eg NGC 3690 has $L_{\rm X} \sim 5 \times 10^{41} \ergps$  
(Zezas, Georgantopoulos \& Ward 1998). 

Both TB and S94's simulations had high simulated X-ray luminosities
in the range $10^{41}$ -- $10^{42} \ergps$. With reasonably similar parameters
it is not surprising we find similar luminosities. That real starburst
galaxies do not typically have such high soft X-ray luminosities
 may be telling us that the
physical conditions assumed in the high $L_{\rm X}$ models are not a good
representation of the true conditions within starburst galaxies such as M82.
However, we must be careful in interpreting the absolute values of
$L_{\rm X}$ from numerical simulations, as low numerical resolution
leads to overestimated X-ray luminosities as discussed in 
Section~\ref{sec:results_numres}. The X-ray luminosities in higher 
resolution simulations will be lower, although similar in form,
than those in these simulations.

Another factor leading to high soft X-ray luminosities in these models
is the assumption of Solar metallicity, as discussed in 
Section~\ref{sec:cooling}. As we shall discuss below, when considering in
detail the origin of the X-ray emission from within the galactic wind
(Section~\ref{sec:results_physorigin}), 
the majority of the soft X-ray emission comes
from regions of interaction between swept-up disk material and the wind, 
and not directly from the hot metal-enriched shocked wind fluid. If 
the ambient ISM material has sub-solar abundances then the
resulting X-ray luminosities
will be reduced proportionally (see Fig.~\ref{fig:lambdax}).

A particularly interesting question not answered by the total X-ray 
luminosities given in Fig.~\ref{fig:lx_lw}, or in TB and S94, is
whether the X-ray emission is dominated by particular regions within the wind
(\eg gas within the disk or in the halo),
or is relatively uniformly spread throughout the wind volume.

A significant fraction of the X-ray emission from
these galactic winds comes from hot gas in the plane of the
galaxy (Fig.~\ref{fig:lx_diskhalo}). X-ray emission away from the
plane of the galaxy, in what would observationally be classed the ``wind,''
is typically much less luminous than that from the plane of the galaxy.
This is important as observational studies of the 
hot gas in the disk of galaxies with classic
starburst driven winds (\ie edge-on systems), such as M82, 
is hampered by source confusion
and high absorption columns. A result of this is that clear predictions
of the X-ray properties of galactic winds away from the plane of the
galaxy, where absorption and source confusion are less of a problem, 
are necessary.

\subsubsection{Hard X-ray emission}

Hard thermal X-ray emission from the galactic wind predominantly comes from
the starburst region itself, with a lesser contribution from the
free wind and shocked wind regions. 
Although the volume of the starburst region is small compared to
regions of hot or very hot gas in the free or shocked wind,
the density of the very hot gas in the starburst region is significantly
higher than that of the very hot gas elsewhere in the wind.

Hard X-ray luminosities for all the models
in the $2.4$ -- $15.0 \keV$ band are given in Table~\ref{tab:main_results}.
and lie in the range $L_{\rm X,hard} = 2 \times 10^{37}$ to 
$5 \times 10^{39} \ergps$. This is typically 2 orders of magnitude lower
than the soft X-ray luminosity of the wind. S94 had already
found similarly low ratios of thermal hard to soft X-ray luminosity, so
our results  are in good agreement with theirs.

Cappi \etal (1999) have argued that the hard X-ray emission from starburst
galaxies is from a very hot ($kT = 6$ to $9 \keV$) diffuse 
component of the ISM,
most likely associated with the starburst-driven wind. Based on
{\it BeppoSAX} observations, they find a ratio of hard X-ray to
soft X-ray luminosity (in the $2$ -- $10 \keV$ band
relative to the $0.1$ -- $2 \keV$ energy band) of $\sim 4$
for M82 and $\sim 2$ for NGC 253. This is totally inconsistent with
the ratio of $\sim 10^{-2}$ we and S94 have found for diffuse thermal
emission from starburst-driven winds.

{\em Non-thermal} processes associated with starburst-driven winds may
well increase the total hard diffuse X-ray emission from starbursts. Moran
\& Lehnert (1997) attribute the hard X-ray emission from the
nuclear region of M82 to inverse-Compton emission from IR photons 
scattered off relativistic electrons. This process is less likely
to be important for generating hard X-ray emission in the halo (Seaquist
\& Odegard 1991). An imminent solution to these uncertainties is
at hand, as {\it Chandra} observations of local starbursts
will have the spatial \& spectral resolution necessary to determine
the physical origin of the hard X-ray emission.

We therefore  
reiterate and emphasize S94's conclusion that {\em the hard X-ray emission
from starburst galaxies is not due to thermal emission from the
starburst-driven wind}.

\subsubsection{Phase distribution and filling factor of the X-ray emitting gas}
\label{sec:phase_distrib}
In almost all of the models we find that hot gas fills that majority of
the volume of the wind (see the hot gas filling factor $\eta_{\rm hot}$
in Table~\ref{tab:main_results}). 
The exception is the centrally mass-loaded
model tbn\_6, where warm and hot gas fill approximately equal fractions
of the wind volume. 

Although the distribution of gas volume
can always be approximated by a broadly peaked function of
temperature, centred at a temperature of $kT \sim 0.5$ to $1 \keV$,
it is be an over-simplification to say that gas at any one 
temperature fills the majority of the wind.

While the temperature of the gas that fills the majority of the
wind volume is similar to those derived from
fitting simple spectral models to {\it ROSAT} and {\it ASCA} spectra
of diffuse X-ray emission from starburst galaxies,
{\em the vast majority of the intrinsic soft X-ray emission comes from cooler
denser low filling factor gas}. The hot gas filling most of
the volume contributes only a small fraction of the soft X-ray
emission of the wind. 
This is true of all the models we have explored.

\begin{figure*}
 \centerline{
   \psfig{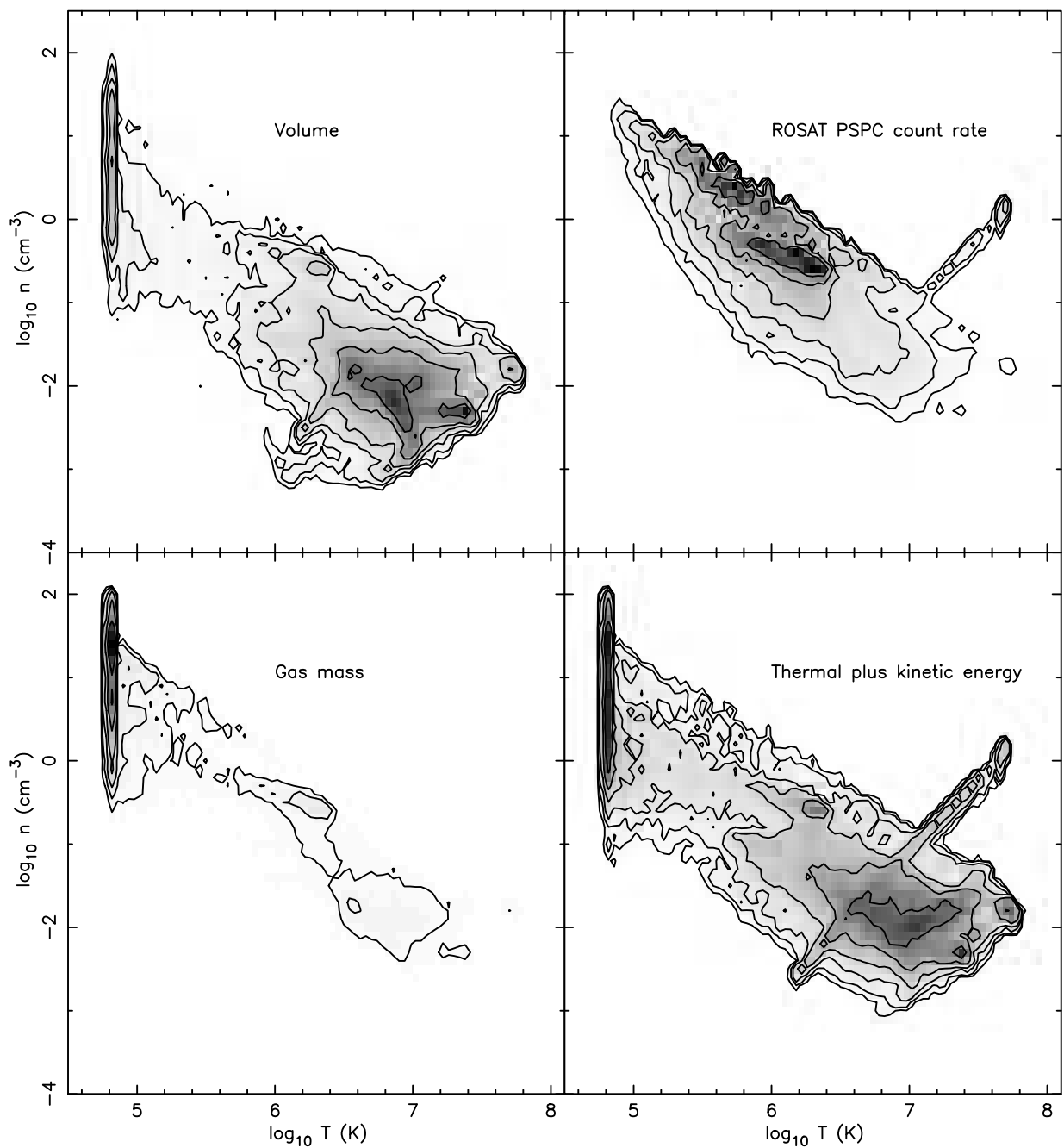}
	}
 \caption[]{The distribution of volume, soft X-ray emission 
({\it ROSAT} PSPC count rate), gas mass and total energy 
(thermal plus kinetic) as a function of gas temperature 
and density within the wind in model tbn1b at $t = 7.5 \Myr$,
binned logarithmically in temperature and density. 
The intensity
scale has been normalised to the peak value in each panel. Contours
begin at 0.1 per cent of the peak intensity, and increase in
0.5 dex intervals up to 31.6 per cent.
Note that the locus of maximum X-ray emission is not the
same as the locus of maximum volume, mass or energy --- 
the X-ray emission comes from low volume filling factor gas,
containing a relatively small fraction of the total mass and
energy content of the galactic wind. The starburst region and the
initially adiabatically expanding wind (for which $T \propto n^{2/3}$) 
produce the ridge running from
($\log T = 7.8$, $\log n = 0$) to ($\log T = 6.2$, $\log n = -2.5$)
in the X-ray emission and energy diagrams.
The use of a minimum allowed temperature leads to the spike of
material at $\log T = 4.8$, which contains the majority of the
total mass in the wind.}
 \label{fig:2dnt_images}
\end{figure*}

As an example of this we show the distribution of gas volume and X-ray emission
(detectable {\it ROSAT} PSPC count rate in the absence of absorption,
so as to include the energy-dependent sensitivity of the PSPC)
as a function of the gas temperature and number density
from model tbn1b at $t = 7.5 \Myr$ in Fig.~\ref{fig:2dnt_images}.
It is clear that the majority of the X-ray emission comes from
higher pressure ($P/k \sim 10^{6} \K \pcc$), 
denser and slightly cooler, material than the gas
that fills the majority of the volume (which has
$P/k \sim 10^{5} \K \pcc$). 

Just as the most X-ray luminous gas does not occupy much of the 
total wind volume, neither does it contain much of the mass
and energy (either thermal or kinetic) of the wind (see
Fig.~\ref{fig:2dnt_images}). 

\begin{table}
 \caption[]{Fraction of the total wind mass, volume and energy content
 in the gas contributing most to the intrinsic X-ray emission (in the {\it
 ROSAT} PSPC band) from model tbn1b at $t = 7.5 \Myr$. The most 
 X-ray-luminous gas occupies a small fraction of the wind volume, and contains
 little of the mass and energy of the outflow. For example, 95 per cent
 of the X-ray emission is produced by only 2.4 per cent of the wind's mass,
 occupying only 1.5 per cent of the total volume.}
 \label{tab:tbn1b_fraction}
 \begin{tabular}{lcccc}
  \hline
    & \multicolumn{4}{c}{Fraction (\%)} \\
X-ray emission  & 50 & 75 & 90 & 95 \\
  \hline
Mass                    & 0.17 & 0.38 & 0.83 & 2.40 \\
Volume                  & 0.08 & 0.34 & 0.78 & 1.45 \\
Thermal energy          & 0.69 & 2.21 & 5.04 & 7.93 \\
Kinetic energy          & 0.62 & 2.18 & 5.02 & 10.42 \\
  \hline
 \end{tabular}
\end{table}

The exact filling factors and mass and energy fractions depend
on what we define the majority of the X-ray emission to mean.
Clearly the hot gas filling most of the volume does 
emit soft X-rays at some level. Table~\ref{tab:tbn1b_fraction}
shows the percentage of the total gas mass, volume and energy
in model tbn1b as a function of the fraction of the total unabsorbed 
{\it ROSAT} PSPC count rate it is responsible for, from
the 50 per cent to the 95 per cent.

Even if we consider 95 per cent of all the X-ray emission
we are only sampling just over 1 per cent of the wind volume,
2 per cent of the total mass and 10 per cent of the energy of the
wind!

Note that although the fraction of the total mass in the X-ray dominant
gas is small, there is relatively little mass in the warm and
hot phases to begin with. For example, in model tbn1b only $\sim 8$
per cent of the total mass is in gas with $T \ge 10^{5.5} \K$ 
(Table~\ref{tab:main_results}). The
X-ray dominant gas in this model
contains about $\sim 0.8$ per cent of the total mass,
so X-ray observations would be probing of order 10 per cent of the
mass of the warm and hot gas.

\begin{table}
 \caption[]{Fractions of the
 total wind mass, volume, thermal and kinetic energy in the gas
 that produces the most luminous 90 per cent of the 
 intrinsic X-ray emission in the {\it ROSAT} PSPC band (see 
 Section~\ref{sec:phase_distrib}). In all models the X-ray dominant gas is
 of low volume filling factor, and allows us to probe only a small
 fraction of the mass and energy content of the wind.}
 \label{tab:ffactors}
 \begin{tabular}{lcccc}
  \hline
Model & $f_{\rm M}$ & $\eta$ ($f_{\rm V}$) & $f_{\rm TH}$ & $f_{\rm KE}$ \\
      & (\%) & (\%) & (\%) & (\%) \\
\hline
 tbn\_1 & 0.51 & 0.09 & 1.89 & 4.93 \\
 tbn1a  & 1.08 & 0.79 & 4.36 & 5.62 \\
 tbn1b  & 0.83 & 0.78 & 5.04 & 5.02 \\
 tbn\_2 & 0.05 & 0.01 & 0.19 & 8.09 \\
 tbn\_6 & 0.38 & 0.02 & 2.91 & 4.58 \\
 tbn\_7 & 0.12 & 0.01 & 0.53 & 2.02 \\
 tbn\_9 & 0.99 & 0.14 & 7.11 & 9.62 \\
 mnd\_3 & 2.21 & 1.61 & 4.24 & 12.76 \\
 mnd\_4 & 0.82 & 2.37 & 5.10 & 21.42 \\
 mnd\_5 & 0.35 & 0.02 & 1.81 & 3.49 \\
 mnd\_6 & 0.08 & 0.003 & 0.11 & 0.12 \\
 mnd\_7 & 0.63 & 0.40 & 2.85 & 2.53 \\
  \hline
 \end{tabular}
\end{table}

In all of the models we have run the gas contributing the majority of the
intrinsic soft X-ray emission comes from a very small fraction 
of the wind.
Table~\ref{tab:ffactors} shows the filling factors, mass
and energy fractions of the gas dominating the intrinsic
soft X-ray emission in all models, at a typical epoch once a
full galactic wind has developed ($t = 7.5 \Myr$).
The lowest filling factors in Table~\ref{tab:ffactors}, of
order $10^{-2}$ per cent, may be
numerical artefacts due to limited numerical
resolution. This is discussed further in Section~\ref{sec:results_numres}.

Nevertheless, it is clear that soft
X-ray observations only probe a minor fraction of the wind, 
whether measured
in terms of total volume, energy content, total mass or even mass of 
warm/hot gas.

\subsubsection{Temperature distribution of the X-ray-emitting gas}
 
\begin{figure}
\centerline{
 \psfig{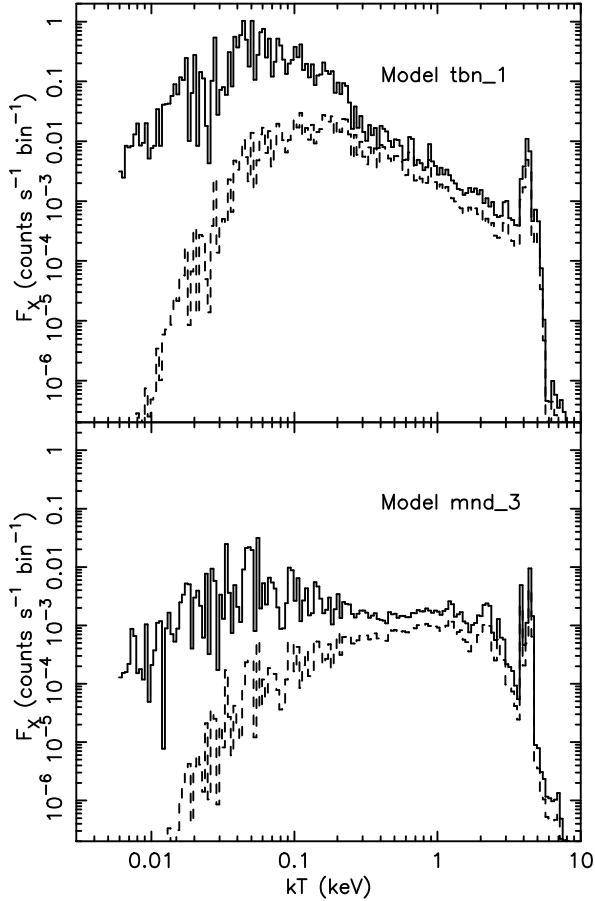}
	}
 \caption[]{The contribution from gas at each
logarithmic temperature bin to the total {\it ROSAT} PSPC count rate
in models tbn\_1 and mnd\_3 at $t = 7.5 \Myr$, assuming no
absorption (solid line) and $\nH = 4 \times 10^{20} \pcm2$ (dashed line).
A very wide range of gas temperatures is responsible for the emission
that would be detected by the {\it ROSAT} PSPC, even allowing for
a realistic amount of absorption.}
\label{fig:tkev_fx}
\end{figure}

From Fig.~\ref{fig:2dnt_images}, which is typical of the temperature-density
distributions found in all of these models, it is clear
that gas at a wide range of temperatures and densities contributes
to the X-ray emission in the {\it ROSAT} band.
In Fig.~\ref{fig:tkev_fx} we show the contribution from gas at each
logarithmic temperature bin to the total {\it ROSAT} PSPC count rate
in models tbn\_1 and mnd\_3 (the default thick and thin disk 
galactic wind models).
Even when the effect of realistic 
intervening absorption is included the PSPC would detect photons from
a very wide range of gas temperatures.

\begin{figure}
\centerline{
 \psfig{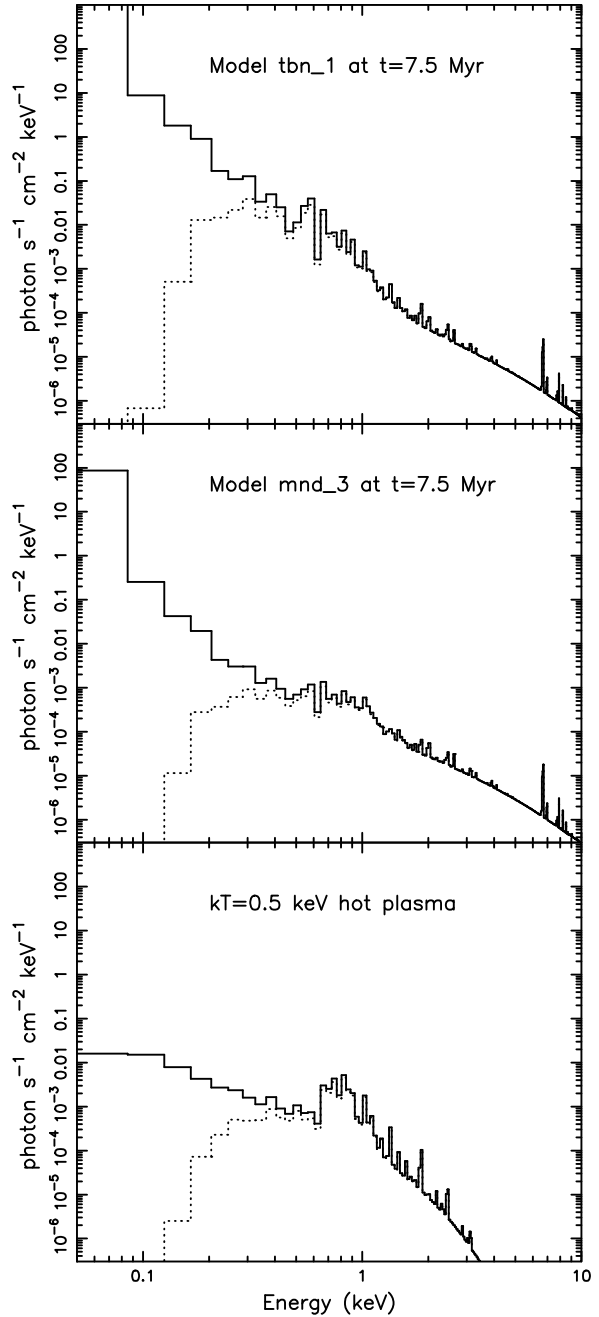}
	}
\caption[X-ray spectra for models tbn\_1 and mnd\_3 at $t = 7.5 \Myr$.]
 {Intrinsic X-ray spectra for models tbn\_1 (top) and mnd\_3 (bottom)
 at $t = 7.5 \Myr$, shown in comparison to the X-ray spectrum 
 for a hot plasma of temperature  $kT = 0.5 \keV$ (bottom panel,
 similar to the temperature of the hot gas in
 starburst galaxies inferred from X-ray observations). 
 The dotted lines show the same spectra after absorption
 by a hydrogen column of $\nH = 4 \times 10^{20} \pcm2$ (equivalent to the
 Galactic hydrogen column towards M82).}
\label{fig:true_spectra}
\end{figure}

Is this consistent with existing X-ray data on the hot gas in
galactic winds?
As a first-order approximation to a characteristic X-ray temperature
for these models we have calculated the average gas temperature 
$T_{\rm PSPC}$, which has been weighted by the {\it ROSAT} PSPC count rate 
(see Table~\ref{tab:xorigin}).
This assumes no absorption, or that given an observed X-ray spectrum we
could correct for any absorption to obtain the true temperature distribution
of the emitting gas.

These temperatures range from $T_{\rm PSPC} = 0.08 \keV$ to $0.55 \keV$,
similar if slightly lower than single temperature fits to {\it ROSAT}
PSPC spectra of galactic winds. For example, Strickland \etal (1997)
find best fitting temperatures ranging between $0.3$ to $0.7 \keV$
in M82's wind, and Dahlem \etal (1998) find $kT \approx 0.1$ to 
$0.7 \keV$ in NGC 253's
wind. Note that the characteristic temperatures in Table~\ref{tab:xorigin}
are not the result of fitting spectral models to observed spectra.
Nevertheless, they are in broad agreement with the observed characteristic
soft X-ray temperatures of galactic winds.

Although fitting simple spectral models (such as one or two temperature
thermal plasma models) to simulated {\it ROSAT} or {\it ASCA} X-ray spectra
from these models does give statistically good fits (as we shall show in Paper
II), the derived plasma properties are misleading. We urge caution in 
interpreting spectra of hot gas in starburst galaxies, observed with
either the preceding generation of X-ray telescopes ({\it ROSAT} 
or {\it ASCA}) or the latest generation ({\it Chandra} or {\it XMM}).

Very broad-band sensitive X-ray spectroscopy of the hot gas in these
outflows should show more evidence of the complex temperature structure
than current {\it ROSAT} or {\it ASCA} spectroscopy
(where the existing broad-band coverage is compromised by source 
confusion and limited sensitivity).
The simulated X-ray spectra for models tbn\_1 and mnd\_3 at $t = 7.5 \Myr$ 
are shown in Fig.~\ref{fig:true_spectra}. Note that both are 
softer and harder than the spectrum of a $kT = 0.5 \keV$ thermal
plasma, reflecting the wide range of temperatures in the gas in these
winds.

\subsection{Physical origin of the soft X-ray emission}
\label{sec:results_physorigin}

As discussed in Section~\ref{sec:introduction} the origin and
properties of the X-ray emitting gas in galactic winds is
unclear.
S94 had concluded that the majority of the soft X-ray emission in
their four models came from disk and halo gas shock-heated by the
wind. Recently D'Ercole \& Brighenti (1999) argued that
the soft X-ray emission in these models was not shock-heated
disk or halo gas, but was from warm gas in the numerically broadened
interfaces (\ie artificially smoothed-out 
contact discontinuities) between cool dense gas 
and the hot wind.

In all likelihood a mixture of these two effects are responsible for the
X-ray emission in these previous models, but it is very difficult to
separate the two effects. Regions of ambient ISM shock-heated by the wind
are also unavoidably interfaces between dense cool gas and hot tenuous
gas, which will be artificially smoothed out over a number of
computational cells (the exact number of which depend on the 
hydrodynamical scheme used. See Fryxell,
M\"{u}ller \& Arnett [1991] for a comparison of different codes,
which shows PPM-based codes that S94 and ourselves have used in
a favorable light!).

Without performing much higher resolution simulations
it is difficult to assess the relative significance of these
two effects. See Section~\ref{sec:results_numres}
for a more detailed discussion of the effect of numerical resolution
and numerical artefacts, as well as a comparison between the
models we have run at increasing resolution. 

To avoid a premature
choice between shock-heating of clouds and the numerical broadening
of contact discontinuities as the dominant source of the soft X-ray 
emission in these simulations,
we shall refer to ``the interaction of the wind with cool dense gas.''
X-ray emission from such interaction regions could then come
from one or more of any number of processes: In our numerical
simulations shock-heating and numerically broadened contact discontinuities,
and in reality shock-heating, conductive or photo-evaporative
interfaces, or turbulent mixing layers (Begelman \& Fabian 1990).

If we pick out the most X-ray luminous regions within the wind,
\ie those regions responsible for the most luminous 90 per cent of the
soft X-ray emission, we can find the physical origin of this emission.
As we have run a wider range of models than S94 or 
D'Ercole \& Brighenti (1999), and have included mass-loading 
(another process leading to soft X-ray emission),
this exercise is of some interest.

We have already shown above that the soft X-ray emission comes from
low filling factor gas, $\eta \sim 10^{-2}$ to $\sim 2$ per cent.
We find that in all the models interaction regions (as defined above)
are responsible for some, and in a few cases, all of the soft X-ray emission.

It is important to note that such interface regions do not always
dominate the X-ray emission. We provide a qualitative description of 
which regions within the wind are responsible for the most luminous
90 per cent of the soft X-ray emission in Table~\ref{tab:xorigin}.

\begin{table}
 \caption[]{Average gas temperature in each of the models at $t = 7.5 \Myr$
 (weighted by {\it ROSAT} PSPC count rate, and ignoring absorption),
 and a qualitative description of 
 what region the majority of the soft X-ray emission originates in.}
 \label{tab:xorigin}
 \begin{tabular}{lcccccc}
  \hline
Model & $T_{\rm PSPC}$ 
	& SB$^{a}$ & FW$^{b}$ & \multicolumn{2}{c}{IR$^{c}$} & SH$^{d}$ \\
       	& ($\keV$) & & & w/d & w/c & \\
\hline
 tbn\_1 & $0.09^{+0.20}_{-0.04}$ &     &     & yes & yes & \\
 tbn1a  & $0.11^{+0.26}_{-0.05}$ &     &     & yes & yes & \\
 tbn1b  & $0.13^{+0.27}_{-0.06}$ &     &     & yes & yes & \\
 tbn\_2 & $0.07^{+0.15}_{-0.03}$ &     &     & yes & yes & \\
 tbn\_6 & $0.21^{+0.27}_{-0.14}$ & yes &     & yes &     & \\
 tbn\_7 & $0.08^{+0.24}_{-0.03}$ &     &     & yes & yes & \\
 tbn\_9 & $0.16^{+0.21}_{-0.09}$ &     &     & yes & yes & \\
 mnd\_3 & $0.44^{+0.89}_{-0.33}$ & yes & yes & yes & yes & yes \\
 mnd\_4 & $0.35^{+0.42}_{-0.20}$ & yes &     & yes & yes & yes \\
 mnd\_5 & $0.55^{+0.23}_{-0.25}$ & yes & yes & yes &     & \\
 mnd\_6 & $0.14^{+0.65}_{-0.10}$ &     &     & yes &     & \\
 mnd\_7 & $0.19^{+0.40}_{-0.10}$ & yes &     & yes &     & \\
  \hline
 \end{tabular}
\\
$^{a}$ The starburst region. \\
$^{b}$ The freely expanding wind of SN and stellar wind ejecta. \\
$^{c}$ Interface regions, either between the wind and the ISM of the
       galactic disk (w/d) or between the wind and clumps of disk ISM entrained
       into the wind, such as superbubble shell fragments (w/c). \\
$^{d}$ Shocked halo gas. \\
\end{table}

In the mass loaded models tbn\_6, mnd\_5 \& mnd\_7
the starburst region itself is a
source of significant X-ray emission in the {\it ROSAT} band.
This is easily understandable in
the centrally mass-loaded models tbn\_6 \& mnd\_5, where the central
gas densities in the starburst region are an order of magnitude higher due
to the additional mass injection and lower outflow velocity.

Without additional mass loading we would not normally expect the
starburst region or free wind regions to be significant sources
of X-ray emission, due to the low density (\cf CC). 
Nevertheless,
the total X-ray luminosities of models mnd\_3 \& mnd\_4 are so
low (see Table~\ref{tab:main_results} or Fig.~\ref{fig:lx_lw})
that these regions are counted as being X-ray dominant. 
Shocked halo gas is also important in these two models. 
The thickness of the shell of swept-up
and shocked halo gas is large enough that there is no problem
of unresolved contact discontinuities, as this shell is very
well resolved in all the models.  
These models are so X-ray underluminous compared to M82 
(see Table~\ref{tab:main_results} or Fig.~\ref{fig:lx_lw}) 
that it is unlikely they are good models of M82's wind itself.

When interface regions and shock-heated halo gas dominate the soft
X-ray emission of galactic winds, abundance determinations based on
X-ray spectroscopy will reflect the metallicity of the ambient
disk and halo ISM to a significant degree. This is the same
conclusion as reached by D'Ercole \& Brighenti (1999).
As a result it seems unlikely that we can directly probe the metal-enriched
gas in these winds with X-ray observations, which will make it
difficult to measure the metal ejection efficiency of such outflows directly.

\begin{figure*}
 \centerline{
   \psfig{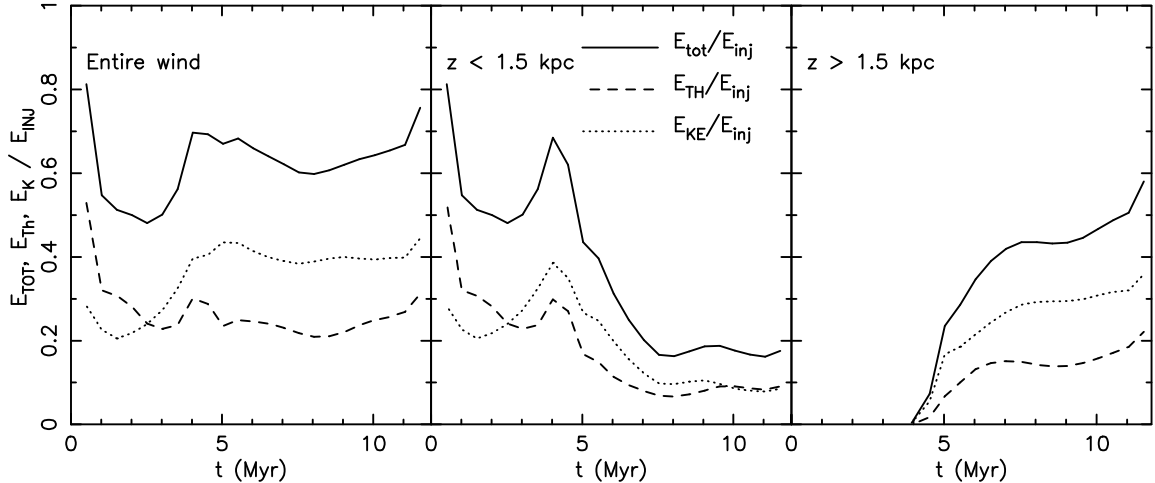}
	}
\caption[The energy content of the wind in model tbn\_1 as a function of time]
 {The energy content of the wind in model tbn\_1 as a function of time. Each of
 the three panels show the total (thermal plus kinetic energy, solid line), 
 thermal (dashed line) and kinetic energy (dotted line) content of the wind
 as fractions of the total energy injected into the wind by the starburst.
 The three panels show the  energy fractions within the entire wind, 
 energy fractions of the wind 
 within the plane of the galaxy ($|z| \le 1.5 \kpc$), 
 and  energy fractions of the wind 
 within the halo ($|z| > 1.5 \kpc$). Initially
 dominated by thermal energy, post-blowout kinetic energy dominates the
 energy budget of the wind. Note that the wind is an efficient mechanism of 
 transporting energy into the halo.}
\label{fig:energy_frac}
\end{figure*}

\subsection{Wind energy and mass transport}
\label{sec:results_energy}

The distribution of thermal and kinetic energy within
conventional wind-blown bubbles and superbubbles has been discussed
in the literature
(see Mac Low \& McCray [1988] and Weaver \etal [1997]), but the energy
distribution within galactic winds has not been
addressed until now.  As galactic winds are, by definition, bulk outflows,
their ``energy budget'' can differ substantially from
the energy distribution within superbubbles.

In what form (\ie thermal or kinetic) is the
energy within the wind stored, in which gas phases, 
and does this depend on time, position within
the wind and on the starburst or ISM? 
What fraction of the wind's energy is transferred to the swept-up ISM?
Which, if any, components of the wind have sufficient energy to
escape the host galaxy completely?
How efficient is the wind at transporting energy into the IGM?

The answer to the final question has important cosmological
consequences. 
Galactic winds may be important for the heating of the ICM in galaxy
groups and clusters (\eg Ponman \etal 1999) and the IGM,
but without knowing what fraction of the total wind energy budget can be
observed in a particular wave band (\eg soft X-ray emission, or 
observations of the optical emission line gas) it is difficult
to assess the impact of galactic wind heating.

\subsubsection{Wind energy budget and energy transport}
\label{sec:results_energy_budget}

In the standard model of a 
wind-blown bubble or superbubble (Weaver \etal 1977) expanding into a
uniform density medium, the majority
of the energy is in the thermal energy of the shocked wind region. Of the
total mechanical energy injected, $\sim 45$ per cent is thermal energy in the
hot shocked wind, $\sim 20$ per cent is in the form of the kinetic energy 
in the cool dense shell
and the majority of the remaining $\sim 35$ per cent has been radiated away,
primarily from the shell.

We find that 
blowout of the superbubble from the disk changes the energy balance of
a galactic wind,  from being dominated by thermal energy, to being dominated
by kinetic energy (Fig.~\ref{fig:energy_frac}). In the thick disk models
this can clearly be seen occurring at $t \sim 3 \Myr$. Blowout in our
thin disk models is almost instantaneous, and they become kinetic energy 
dominated within $\sim 0.5 \Myr$.

Decomposing the energy budget into disk ($|z| \le 1.5 \kpc$) and halo
($|z| > 1.5 \kpc$) components reveals that invariably the majority of
the energy injected by the starburst is transported out of the disk
and into the halo (a single example using model tbn\_1 is shown in
Fig.~\ref{fig:energy_frac}). The total energy within the disk remains
approximately constant after $t \sim 5 \Myr$, but this is a decreasing fraction
of the total energy injected by the starburst.

In superbubbles the thermal energy is in predominantly in the hot
($T \ge 10^{6} \K$) bubble interior, 
and the majority of the kinetic energy is in the cool ($T \sim 10^{4} \K$)
dense shell. In what temperature gas is the energy of a galactic wind
stored?

\begin{figure*}
 \centerline{
   \psfig{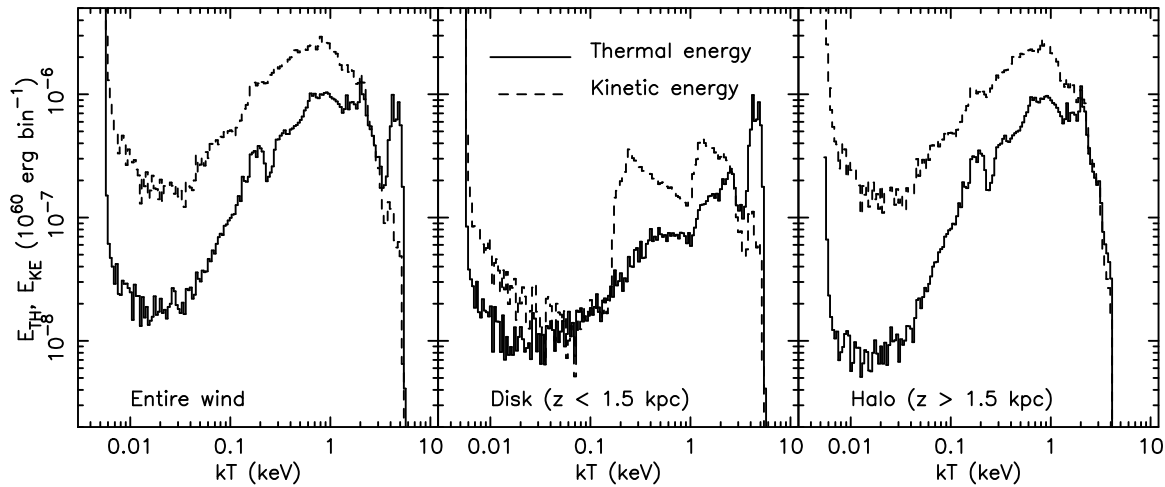}
	}
\caption[Temperature distributions of thermal and kinetic energy in model 
 tbn1b at $t = 7.5 \Myr$]
 {Temperature distributions of thermal (solid line) and kinetic (dashed line)
 energy in model tbn1b at $t = 7.5 \Myr$. The three panels show the  
 energy within the entire wind, 
 energy within wind material still in
 the plane of the galaxy ($|z| \le 1.5 \kpc$), 
 and the 
 energy of the wind in the halo ($|z| > 1.5 \kpc$) as a function of gas 
 temperature. The majority of the thermal and kinetic energy of the
 wind is in hot, $kT \sim 0.1$ to $2.0 \keV$, gas expanding out into the halo.}
\label{fig:tkev_energy}
\end{figure*}
 
In most of the models the majority of both the thermal and kinetic
energy outside the plane of the galaxy ($|z| > 1.5 \kpc$)
is in hot gas ($6.5 \le \log T (\K) \le 7.5$), with relatively little
thermal or kinetic energy in cool gas ($4.5 \le \log T (\K) \le 5.5$).
This can be seen in  
Fig.~\ref{fig:tkev_energy}, which shows the temperature
distribution of the thermal and kinetic energy, over the entire wind
and decomposed into disk and halo components, from model tbn1b.
Table~\ref{tab:main_results} gives the fractions of the total
thermal or kinetic energy in the warm and hot phases of the wind in
all the models.

It is interesting to note that a large fraction
of the total energy of the wind is in gas that can, {\em in principle}, 
be probed with
X-ray absorption line studies, where the relative dominance of
denser X-ray emitting gas is reduced. Such absorption line studies
would require bright X-ray sources in the background behind
a galactic wind, and a combination of high sensitivity and 
spectral and spatial resolution.

 In the centrally mass-loaded thick disk model tbn\_6,
there is very little hot gas ($6.5 \le \log T (\K) \le 7.5$). Instead,
the thermal energy is predominantly in warm gas 
($5.5 \le \log T (\K) \le 6.5$), while the kinetic energy of the
wind is evenly spread between warm and cool gas.

Distributed mass-loading (models tbn\_9 \& mnd\_7)
differs from central mass-loading in
 not having such an strong effect on the 
energy-temperature distributions, and these models
are quite similar to the non-mass-loaded
models in terms of energetics.

In the weak starburst models, such as tbn\_2 and mnd\_4, thermal energy begins
to dominate the total energy budget again after $t \sim 7.5 \Myr$. This
is  due to the
 thermal energy of the halo gas swept-up 
being a significant fraction of the total energy injected by the starburst
in these models. In more powerful starburst models the energy
content of the disk and halo ISM overrun by the wind is insignificant.

In all models the wind is efficient at transporting energy out
of the plane of the galaxy, and as the winds are generally inefficient at 
radiating this energy away, galactic winds can in principle be good
sources of heating for the IGM.

\begin{figure}
 \centerline{
   \psfig{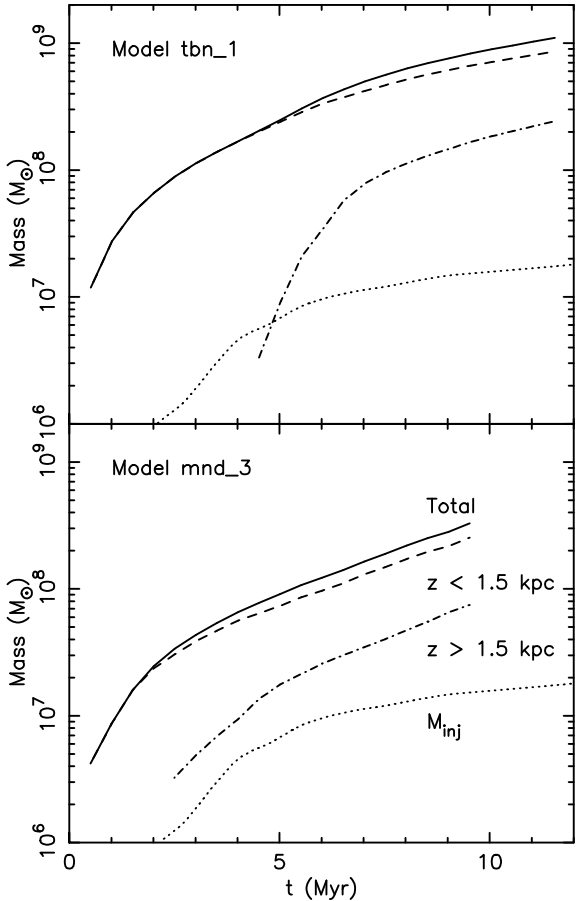}
	}
\caption[Wind mass as a function of time in models tbn\_1 and mnd\_3]{Wind 
 mass as a function of time in models tbn\_1 and mnd\_3 (solid line),
 compared to the total mass of gas injected by the starburst (dotted line).
 The wind mass within the plane of the galaxy ($|z| \le 1.5 \kpc$, dashed line)
 and above the plane ($|z| > 1.5 \kpc$, dot-dashed line) are also shown. The
 mass of enriched material injected by the starburst ($M_{\rm inj}$)
 is always small compared
 to the mass of disk and halo gas swept-up by the wind. Note that the
 majority of the gas in the ``wind'' remains in the plane of the galaxy ---
 galactic winds are relatively inefficient at removing 
 gas from the galaxy.}
\label{fig:mass_in_wind}
\end{figure}

\subsubsection{Mass transport and ejection}

Although the starburst can sweep up and shock large masses of
ambient gas ($\ltsimm 10^{2} \times$ the total mass of SN ejecta),
most of this mass is in, and remains in the  plane of the galaxy (see 
Fig.~\ref{fig:mass_in_wind}). 
Only a small fraction of the total mass can be found at heights above 
$z = 1.5 \kpc$ from the plane. Galactic winds
 of power similar to M82's are {\em inefficient at transporting gas
 out of the plane of the galaxy}. 

Note that the total mass of metal-enriched
gas injected by the starburst 
is significantly smaller than the total mass of wind material that
is transported out of the disk. If all of mass associated with
wind that is transported out of the disk were to make its way into the
IGM, then on average gas injected into the IGM by galactic winds
would not be highly metal enriched. 

\begin{figure}
 \centerline{
   \psfig{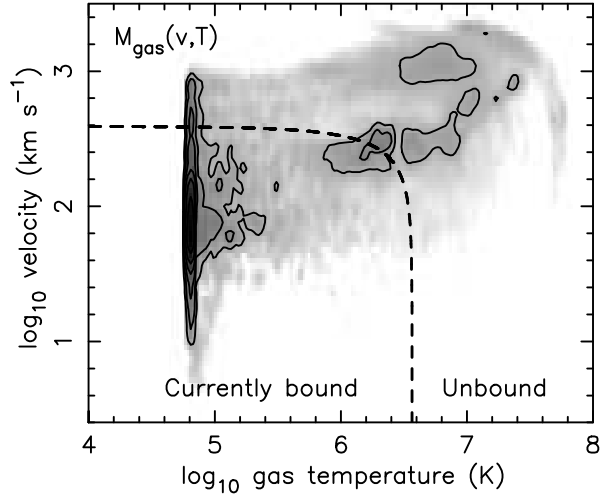}
	}
\caption[Mass in model tbn1b at $t = 7.5 \Myr$, as a function
  of gas temperature and velocity.]
  {Mass in model tbn1b at $t = 7.5 \Myr$, as a function
  of gas temperature and velocity. Gas above and to the right of the
  dashed line {\em currently} has sufficient total energy to escape
  the galaxy entirely, based on an escape velocity for M82 of
  $v_{\rm esc} = 390 \kmps$. Only a mass of $4 \times 10^{7} \Msol$
  out of a total gas mass of $5 \times 10^{8} \Msol$ is unbound.
  A logarithmic intensity scale has been
  used to display both the cool gas dominating the total mass and the
  hot but tenuous energetic gas.}
\label{fig:tbn1b_vt_mass}
\end{figure}

We cannot quantitatively assess the long term 
fate of the material in our simulations, given the necessarily
limited spatial and temporal domain of these high resolution
simulations. We can quantify what fraction of the mass of
the wind currently has energy sufficient to escape the gravitational
potential of the galaxy at any particular epoch. The distribution
of gas mass as a function of temperature and velocity in
model tbn1b at $t = 7.5 \Myr$ is shown in Fig.~\ref{fig:tbn1b_vt_mass}.
Several broad conclusions can be drawn from this figure:
\begin{enumerate}
\item Most of the wind's mass is in cool gas, expanding
  at $v \ltsimm 100 \kmps$. This represents the slow expansion
  of the wind within the plane of the galaxy, where the majority of the
  mass is swept-up.
\item Assuming conservatively that the escape velocity $v_{\rm esc}$
  from M82 is $v_{\rm esc} \sim 3 \times v_{\rm rot} \sim 390 \kmps$
  (see for example the arguments in Heckman \etal 1999, and M82's
  rotation curve in Fig.~\ref{fig:rotcurve}), then only
  gas above and the right of the dashed line in Fig.~\ref{fig:tbn1b_vt_mass}
  {\em currently} has sufficient total energy to escape the
  galaxy. This gas is only 8.0 per cent ($\sim 4 \times 10^{7} \Msol$)
  of the total mass associated with the wind. 
\item Interactions between the hot, energetic SN-ejecta phases of the wind
  and the ambient ISM has swept-up and accelerated 
  $\sim  6 \times 10^{6} \Msol$ of cool gas ($T \ltsimm 10^{5} \K$)
  up to velocities in the range $400 \le v \ltsimm 1000 \kmps$.
\item The concentration of mass at 
  $\log T \sim 4.8$ is due to minimum temperature allowed in these
  simulations. The distribution and dynamics of mass at temperatures lower than
  this limit is unknown, so the results above may overestimate the
  mass of gas with energy sufficient to escape.
  Note that even these high resolution
  simulations ($\Delta x = 4.9 \pc$)
  do not have the  numerical resolution necessary to 
  resolve the structure and dynamics of gas cooler than this
  limit (see the discussion in Section~\ref{sec:results_shellthick}), 
  so there is little
  point in simulations with lower temperature limits that do not
  have significantly enhanced numerical resolution.
\item Using the temperature of the X-ray-emitting gas 
   in starbursts to assess if this gas 
   can escape the host galaxy (the so-called ``escape temperature,''
   \eg Wang \etal 1995; Martin 1999)
   will underestimate the mass of gas that can escape, as it neglects
   the motion of the hot gas. As discussed in 
   Section~\ref{sec:results_energy_budget}, the majority of the energy
   in a galactic wind is in the kinetic energy of hot gas. 
\item Assessing what mass of gas can escape a galaxy in the long term, 
   based on some measure of gas energy or velocity at a particular 
   epoch, seems extremely difficult. As the wind evolves the hot energetic gas 
   within it will continue to entrain and accelerate cold gas.
   A major unknown is the effect of any halo medium (as assumed in these
   simulations), which acts to impeded and decelerate the wind's
   expansion. Given a dense enough halo environment none of the material
   in a galactic wind might escape into the IGM. 
   Treating this problem numerically is challenging, given it requires
   high numerical resolution to resolve the interaction between cloud
   and wind, as well as including the physics important to wind/cloud
   interactions (\eg hydrodynamic stripping of clouds, 
   thermal conduction, maybe even magnetic fields?).
\end{enumerate}

To summarise: (a) 
the mass of gas that can escape the galaxy in a starburst-driven
wind is low (perhaps a few times $10^{7} \Msol$ 
in a moderate starburst such as 
M82's), and (b) winds
are inefficient at transporting the ISM out of galaxies. 
These results  are qualitatively consistent with
those of Mac Low \& Ferrara (1999) or D'Ercole \& Brighenti (1999), 
although the outflows we have
considered are 2.0 -- 3.5 orders of magnitude more powerful than
the winds of those authors.

\section{Physical and numerical limitations}
\label{sec:results_numres}

A very important aspect of any numerical hydrodynamical simulation is
the degree to which the finite amount of physics included and
numerical limitations (in particular finite resolution) influence the results.
Understanding these effects is difficult for even the specialist, let alone
for the general astronomer.

We think it is important for readers to understand the basic problems
that can and do affect these and other hydrodynamical simulations.
Without an understanding the limitations of a simulation
it is difficult to judge what results are real and what are not.
We shall discuss qualitatively the effect and importance of
some of the physics that is not included in these
simulations. Numerical artefacts, in particular related to
resolution and numerical diffusion, are then
discussed in some detail. Finally we present 
a comparison between three models designed to study 
how the simulated properties of the hot gas in galactic winds
varies as we increase the numerical resolution.

\subsection{Missing physics}
\label{sec:limitations}

\subsubsection{Using effective instead of true gravitational potentials}

The use of an effective potential $\Phi_{\rm eff}$ and the resulting
effective gravity ${\bf g_{\rm eff}}$ to maintain the ISM in hydrostatic
equilibrium are artefacts of a 2-D approach. A 3-dimensional code would
incorporate the azimuthal velocity component without altering the
gravitational field. 3-D simulations would require significant increases
in terms of computation resources and code development, and are difficult
to justify given that the potential scientific return of 2-D simulations
has barely been tapped.

The use of a centrifugal potential means that
the gravitational force felt by material within the galactic wind will be
different from the true gravitational force it should feel. Put another
way, the effective gravitational force in these simulations has been tuned
to hold the ISM in its initial distribution given the assumption that at
any position ($r$,$z$) its rotational velocity $v_{\phi}$ 
around the $z$-axis and hence
angular momentum is known a priori. The problem is how to treat the
material within the galactic wind, given that it is a mixture of
material flowing outward from the central starburst region and ambient
ISM swept-up and entrained into the flow. Hot gas from the starburst
region itself can be presumed to have almost zero angular momentum,
given its origin at the centre of the galaxy and the violent dynamics
of the gas. As it flows radially and vertically outward in these 
2-D simulations it feels the effective, \ie modified by assumed rotation,
gravitational force, whereas it should feel only the true gravitational force.

What effect does this have on the dynamics of the material within the wind?
The effect on the very energetic free wind or shocked wind gas is negligible,
as the terminal velocity of this gas is $v_{\infty} \sim 3000 \kmps$,
so it hardly feels the gravitational field of the galaxy. The main
area where problems may arise is that the dynamics and
long term evolution (over 100s of Myr) of the dense disk material
dragged out of the disk and entrained into the flow (and which is 
assumed to be associated with the optical emission lines 
observed in local starbursts
such as M82) may well be distorted,
given that typical velocities for this gas are only several 
hundred kilometres per second.

As a very rough estimate of the magnitude of the force 
required to significantly
affect the dynamics of material within the galactic wind, a final
velocity of $500 \kmps$ (a velocity slightly greater than the escape
velocity and also significant in terms of the dynamics and 
morphology of the denser gas 
within the wind) after $10 \Myr$ of constant acceleration
requires an acceleration of 
$\approx 1.5 \times 10^{-7} \cm \thinspace {\rm s}^{-2}$.
As the difference between the true and effective gravitational fields
for the thick disk ISM model
is typically a tenth of this value, we can conclude that the use of an
effective gravitational field does not lead to a major distortion of the wind
dynamics over the time-scales of the simulations we are interested
in.

\subsubsection{Non-ionisation equilibrium}

The hot plasma emissivities used to calculate the radiative cooling
and X-ray emission from these models assume the gas is in collisional
ionisation equilibrium (CIE). 

For a suddenly heated plasma, the time required to
reach ionisation  equilibrium is approximately 
$t_{\rm ieq} \approx 0.03 n_{\rm e}^{-1} \Myr$ (Masai 1994) 
for a range of the important elements. 
The density and temperature
structure within the wind are complex, but typical electron
densities in the gas dominating the X-ray emission are of order
$n_{\rm e} \sim 0.3 \pcc$, implying an ionisation
time-scale of $t_{\rm ieq} \sim 0.1 \Myr$.
Although much shorter than the typical age of the wind, $t \sim 10 \Myr$,
the complex dynamical state within the wind probably means it
is rare that the conditions of any parcel of gas remain unchanged
for this amount of time. For example, gas returned to the ISM by
a SN in the starburst region is shocked within the starburst region to
$T \sim 10^{8} \K$, then flows out of the starburst
region in a $v \sim 3000 \kmps$ wind which rapidly adiabatically cools, 
before passing
through the reverse shock after $\ltsimm 1 \Myr$ 
and being reheated to $T \sim 10^{8} \K$. 

Much of the soft X-ray emission within a galactic wind will be due to
initially cool gas incorporated into the wind, \eg swept up and shock heated
ISM, cool gas mixed into the hot wind by conduction or mass-loading 
and instabilities (see Section~\ref{sec:results_physorigin}). 
The ionisation state of
this gas will lag behind its temperature rise, as discussed by Weaver
\etal (1977).  
Metal ions that in CIE
contribute significantly to the X-ray emission of plasmas  with $T
\sim 10^{6} \K$ will in practice be under-ionised. In general then the
true X-ray luminosity will be less than that obtained assuming CIE.

For a rapidly cooling plasma, for example the adiabatically expanding free wind
or some similar region of expansion, the gas will be over-ionised with
respect to its thermal temperature. The emissivity of this gas will be less
than that of a gas in CIE.

As a result of the complex ionisation histories of the gas within the
wind, it is difficult to quantify how non-ionisation equilibrium (NIE)
 effects would alter the 
properties of the wind from the CIE case we assume.
Calculating the ionisation state of the gas in these simulations would
require tracking the history of individual parcels of gas, and would be
prohibitively expensive computationally.

Given that the majority of observational analysis of real X-ray spectra
assumes CIE in using the Raymond-Smith or MEKAL plasma codes, assuming CIE
in creating artificial X-ray spectra does not seem too unrealistic.
Systematic biases in standard X-ray spectral fitting 
discovered in studying the artificial X-ray observations
of these wind simulations will only underestimate the true biases of
fitting complex, NIE X-ray spectra with simple CIE spectral models.

\subsubsection{Magnetic fields and cosmic rays}

These simulations, along with the previous work of TI, TB, S94, S96
\& Tenorio-Tagle \& Mu\~{n}oz-Tu\~{n}\'{o}n (1997), 
ignore both cosmic rays and magnetic fields, instead focusing
on a totally hydrodynamic model of galactic winds. One of the aims of
this work is to see if such pressure-driven winds can reproduce the
observed properties (primarily based on X-ray observations) of
the best studied galactic wind in M82.

Magnetic fields and cosmic rays are known be present within the wind,
but studying them is difficult and the results somewhat uncertain.
Magnetic fields may be important in galactic winds over large and/or small
scales. Large scale fields (over 100s to 1000s of parsecs) could potentially
alter the general dynamics and expansion of the wind, for example
confining its expansion in the plane of the galaxy or maybe acting
to collimate it (see Sections~\ref{sec:results_collimation} \&
\ref{sec:results_confinement}). Even if magnetic fields are not important
on such large scales, any smaller scale fields associated with the
clouds and clumps of ambient ISM entrained into these winds could
well be important. Such small scale fields might alter or inhibit
the hydrodynamical stripping or conductive evaporation of these
clouds, and hence affect the X-ray emission from galactic winds.

We can qualitatively assess the importance of large scale magnetic fields
and of cosmic rays,
relative to purely hydrodynamical processes, through comparison between the
gas properties in our simulations and the existing
observational data (summarised below):
\begin{enumerate}
\item Seaquist \& Odegard (1991) estimate the total energy injection rate into
  the halo in the form of relativistic particles is $\ltsimm 10^{40} \ergps$.
  Compared to the mechanical energy injection rate 
  from the starburst in these
  simulations of typically $\sim 10^{42} \ergps$, cosmic rays 
  are energetically unimportant in the halo. Cosmic rays also
  are unlikely to be an important source of non-thermal X-ray emission
  from the halo (but might be important within the starburst region
  itself, \eg Moran \& Lehnert 1997).
\item Klein \etal (1988) estimate a magnetic field strength of 
  $B \sim 50 \mu G$ within the central $650 \times 200 \pc$ starburst
  region, based on the assumption of equipartition. 
\item A rotation measure analysis by Reuter \etal (1994) gives a
  more direct estimate of the
  field strength $B \sim 10 \mu G$ (modulo uncertainties in the
  column density of ionised gas), 
  at one position at the periphery of the main 
  \halpha~emission, about $1 \kpc$ to the south west of the nucleus.
\end{enumerate}

Are fields of this range in strength dynamically important? Could such
high field strengths in the starburst region constrict and confine
the expansion of the wind in the plane of the galaxy?
Comparing magnetic pressure ($P_{\rm B} = B^{2}/8\pi$) to thermal
and ram pressures within the wind provides a simple basis for
comparison. A magnetic field strength of $B = 50 \mu G$ corresponds
to $P_{\rm B}/k \sim 8 \times 10^{5} \K \pcc$, a field of
$B = 10 \mu G$ to  $P_{\rm B}/k \sim 3 \times 10^{4} \K \pcc$.

In the plane of the galaxy, where confinement of the wind's
expansion is such a problem, typical thermal and ram pressures
are $P_{\rm TH}/k \sim 2 \times 10^{6} \K \pcc$ and
$P_{\rm RAM}/k \sim 7 \times 10^{6} \K \pcc$ for the powerful
starbursts we have simulated (\eg tbn1b, mnd\_3).
At a height of $z = 1 \kpc$ above the plane of the galaxy pressures
are typically lower ($P_{\rm TH}/k \sim 10^{6} \K \pcc$,
$P_{\rm RAM}/k \sim 2 \times 10^{6} \K \pcc$), 
but in some locations ram pressures
can reach exceedingly high values 
($P_{\rm RAM}/k \gtsimm 2 \times 10^{7} \K \pcc$).

Drawing some tentative conclusions from this, it appears that:
\begin{enumerate}
\item The estimated magnetic field strength of
  $B \sim 50 \mu G$ within M82's starburst region, although high
  by the standards of normal galactic magnetic fields, is still
  somewhat lower than that required 
  to make the magnetic field dynamically important in confining
  the expansion of the wind within the plane of the galaxy.
\item If large scale 
  field strengths are typically $B \ltsimm 10 \mu G$ within the wind,
  then they are dynamically unimportant.
\item However, if stronger large scale fields, of order $B \sim 50 \mu G$,
  exist throughout the entire 
  wind then the energy in such magnetic fields would be
  comparable to, or greater than,
  the thermal and kinetic energy of the hot gas.
\end{enumerate}

Hence it does not appear that large scale magnetic fields are dynamically
important for understanding this class of galactic-scale outflows.
The question of the
influence and importance of smaller scale fields in the interaction
between the wind and clouds embedded within it remains open.

\subsubsection{Thermal conduction}

In common with most previous simulations we ignore thermal conduction, as
it difficult to include in multidimensional hydrodynamic schemes
and computationally expensive. In the
standard wind-blown bubble model of Weaver \etal (1977) evaporation of the cold
bubble shell by thermal conduction
is the dominant source of mass in the hot bubble interior, increasing 
its density by approximately an order of magnitude. As a result, the
effects of thermal
conduction on the X-ray emission from wind-blown bubbles and superbubbles
are very important.

In galactic winds thermal conduction is expected to
 be important in regions where there
are reservoirs of dense cool gas in close proximity to hot gas, such
as the interface between the shocked wind and the disk within the plane of the
galaxy, or in the vicinity of superbubble shell fragments within the halo.    

One effect of thermal conduction is to increase the density of the
hot X-ray-emitting gas, in a similar manner 
to mass loading. Indeed, Mac Low \& Ferrara (1999)
add additional mass into their simulations of very weak winds in dwarf 
galaxies to approximate the effects of conduction in a manner similar
to our method of central mass loading. Our mass-loading
can be considered as a rough approximation to the effects 
of both hydrodynamical mass-loading and
thermal conduction, although a rigorous
treatment of thermal conduction is beyond the scope of this work.

Thermal conduction, unless inhibited in some manner, might
also totally evaporate the cool dense clumps and clouds
seen in our simulations. As described in Ferrara \& Shchekinov
(1993), clouds smaller than the Field length (Field 1965)
will suffer rapid conduction-driven evaporation, loosing a very
substantial fraction of their mass before achieving a 
semi-stable configuration. For the range of temperatures and
densities found in the hot gas filling the majority of the
wind's volume in our simulations, the Field length ranges
from several hundred parsecs to several kiloparsecs, so
all of the clumps seen in our simulations should suffer
conductive evaporation.

The net effect of conduction is difficult even to describe qualitatively.
The addition of mass to the hot phases of a galactic wind
will increase the X-ray emission from the wind itself, but if conduction
destroys clouds totally then the net effect might be a reduction
in total X-ray luminosity, given that cloud/wind interfaces appear to
be such strong sources of X-ray emission in these simulations.

Very recently D'Ercole \& Brighenti (1999) have performed a 2-D simulation
of a starburst-driven wind in a dwarf galaxy (based approximately on the dwarf
starburst NGC 1569) that includes thermal conduction.
Interestingly, thermal conduction leads to an increase
in the soft X-ray luminosity of this particular model 
by a factor of between 1 and 3 from the non-conductive case.

\subsubsection{Photoionisation by the wind}

These simulations assume all material within the wind is optically
thin to its own radiation, \ie  there is no absorption intrinsic
to the wind. This assumption is necessary to carry out these
simulations, as coupled multidimensional 
radiation-hydrodynamic problems are exceedingly challenging to solve.

For emission from, and propagating through, the hot tenuous material
of the wind (\eg free wind, shocked wind and shocked halo regions) 
this assumption is reasonable. Typical columns densities of the highly
ionised material that would be encountered by radiation traversing the
wind are in the range $10^{19}$ -- $10^{20} \pcm2$.

In the vicinity of cool dense clouds and the walls of the outflow
within the disk, column densities become more significant. Typical
column densities traversing a single clump range from $\sim 5 \times
10^{19}$ -- $10^{21} \pcm2$, depending on whether the line of sight crosses
the tail or head of a clump and what the angle of incidence is. 

The environment around these cool dense regions is the source of the 
majority of soft X-ray emission (and UV emission, which we have not
discussed) from galactic winds. Photoionisation of these clouds by the
energetic radiation produced through various cloud/wind interactions
may well have dynamical and observational consequences. This could
lead to further evaporation of the clouds, as well as to alter their
ionisation state further from the collision ionisation equilibrium
assumed in these calculations and in most X-ray spectral fitting.

Another important consequence of this will be a reduction in the
X-ray emission escaping the wind, due to the absorption of
a substantial fraction of the softest X-ray emission, produced at
cloud/wind interfaces, by the clouds themselves.
A column of $10^{20} \pcm2$ (assuming Solar abundances, at it is
the metals such as Carbon, Nitrogen and Oxygen that dominate the
photoelectric absorption cross-section) is optically thick
to radiation of energy below $E \sim 0.2 \keV$. The densest clouds,
with individual columns of $\sim 10^{21} \pcm2$ are optically thick
to X-rays with energy less than $\sim 0.4 \keV$. Such a reduction in
the ``escaping'' X-ray luminosity from the wind will bring
the soft X-ray luminosities of our simulations into closer agreement with the
observed soft X-ray luminosities of galactic winds 
(\cf Section~\ref{sec:results_softlx}).

\subsection{Numerical resolution and cell size}
\label{sec:results_cellsize}

For a structured Cartesian grid as employed in these simulations
the resolution is proportional to the cell size used. 
The properties of an unresolved
structure, \eg the cold dense superbubble shell, 
will be averaged out over the computational cell, as the finite
volume scheme employed ensures total mass and energy within
the cell are conserved correctly. For the example of a
unresolved superbubble shell, the density and thickness of the shell 
in the simulation will respectively be less than and greater than 
the true values.

In addition
to affecting the absolute values of the fluid variables, finite
numerical resolution can affect the dynamics of the fluid in other 
ways. Consider for example the fragmentation of the 
superbubble shell by Rayleigh-Taylor (RT) instabilities 
(\cf S94). 
The Rayleigh-Taylor
instability time-scale is $\tau_{\rm RT} \propto \lambda^{1/2}$,
where $\lambda$ is the wavelength. The perturbations that eventually
disrupt the shell have wavelengths similar to the shell thickness,
so unresolved shells are artificially stable against the RT 
instability, and eventually fragment into artificially large 
pieces.

\begin{figure*}
\centerline{
 \psfig{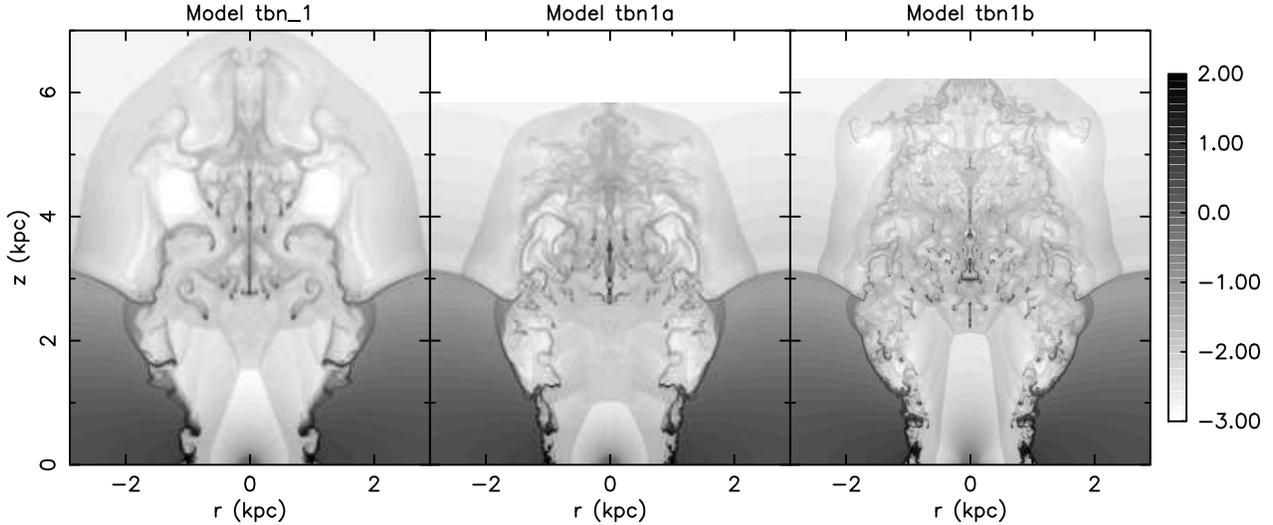}
	}
\caption[Wind number density distributions for models tbn\_1, tbn1a and tbn1b]
 {Wind number density distributions for models tbn\_1, tbn1a and tbn1b
 at $t = 7.5 \Myr$, shown as grey-scale images of $\log n (\pcc)$.
 Model tbn1a has twice the resolution of model tbn\_1, model tbn1b triple
 the resolution. Note the dramatic increase in the complexity of the
 structure of the densest gas as the resolution is increased.}
\label{fig:res_lnum}
\end{figure*}

This directly effects the dynamics and properties of the
coolest gas in these simulations. Fig.~\ref{fig:res_lnum} shows
images of the logarithm of the gas number density in
models tbn\_1, tbn1a (which has cells half the size of those
in model tbn\_1) \& tbn1b (cells one third the size of those in 
model tbn\_1). It is immediately clear that the coolest densest
gas in models tbn1a \& tbn1b is far more structured than that in model
tbn\_1, and that there are many more superbubble shell
fragments being dragged out with the wind.

The limitations of a finite cell size are worst for the densest,
coolest gas, as it is typically found in the smallest structures.
The hotter, tenuous gas, such as the thick shell of
shocked halo gas or regions of shocked wind material, 
is on the other hand most likely to be well resolved, as
it occurs in structures much larger than the cell size.

Nevertheless, poor numerical resolution of cool dense gas can affect
the properties of the hot X-ray emitting gas we are primarily
interested in. The main problems here are the artificial broadening
of contact discontinuities between cool dense gas and hot
tenuous gas, which creates regions of intermediate density warm gas
that will produce significant amounts of soft X-ray emission,
and numerical diffusion which further acts to increasingly broaden
contact discontinuities and spread material
over more cells with time.

\subsection{Cell size vs. shell thickness}
\label{sec:results_shellthick}
It is important to understand
that the true resolution of a hydrodynamic simulation is
not equal to the size of a single cell, but to the
 number of cells over which the code used can 
represent abrupt changes in the fluid variables 
such as shocks or contact discontinuities.

One of the great advantages of the Piecewise Parabolic Method 
(PPM) used by VH-1 it that it spreads out shocks over only
two cells, a major
advantage over many other numerical scheme which spread shocks
out over 4 -- 5 cells. Shocks are an integral part of galactic
winds, as can be seen from the many shocks visible in the
gas number density images shown in Figs.~\ref{fig:tbn_1_lnum}
 -- \ref{fig:mnd_4_lnum}. 

As a rough estimate of how well the densest structures
in our simulations are resolved 
we compare the cell size to analytical estimates
of the thickness of the superbubble shell before RT instabilities fragment it.
A simple method of estimating the shell thickness is to use the Weaver \etal
(1977) model of a constant mechanical luminosity wind blowing into a
constant density medium. Using the analytical solutions presented by
Weaver \etal, we can calculate the total mass in the shell and its pressure,
as a function of time. Assuming the shell cools to the minimum temperature
we allow on our computational grid, $T_{\rm disk} = 6.5 \times 10^{4} \K$,
the density of the shell follows directly from the pressure.
Given the total mass of the shell and its surface area, the shell thickness
follows trivially. Using the appropriate conditions that the
superbubble experiences before blowout at $t \sim 3 \Myr$ in the
thick disk models, the
shell thickness is typically $\Delta r_{\rm sh} \sim 10 \pc$. Comparing
this to the cell size in model tbn\_1 ($\Delta x = 14.6 \pc$), 
model tbn1a ($\Delta x = 7.3 \pc$) and model tbn1b ($\Delta x = 4.9 \pc$) 
shows that the superbubble
shell is under-resolved, although not drastically so.

Hence the shell fragmentation
process and the number and size of superbubble fragments is
controlled by the resolution of these simulations.

Note that this also means that without dramatic increases in
resolution there is no point in allowing gas temperatures below
the minimum gas temperature of
$T_{\rm disk} = 6.5 \times 10^{4} \K$ we impose (see
Section~\ref{sec:ism_distribution}).
A superbubble shell of temperature equivalent to the 
optical emission line gas observed in winds like M82's of 
$T \sim 10^{4} \K$ would be $\sim 2 \pc$ thick, and hence
totally unresolved.

\subsection{Numerical diffusion at contact discontinuities}

As we have mentioned, numerical broadening of contact discontinuities
will affect the properties of the X-ray emitting gas.
In reality physical processes such as thermal conduction and 
turbulent mixing layers 
will also create layers of intermediate temperature and
density gas between cool dense gas and hot tenuous wind material.
The problem with numerically broadened contact discontinuities
is that their width and structure (and hence the properties of 
any artificial X-ray emitting
region) is determined directly by the numerical scheme
employed, and not by real physical processes.

As discussed in Fryxell \etal (1991), contact
discontinuities continue to spread diffusively without limit in
many Eulerian hydrodynamical codes, as the contact region
moves over the computational grid. They show that of a series
of numerical methods PPM 
is the best at retaining sharp contact discontinuities. 
With the additional use of an contact discontinuity steepener 
algorithm, PPM spreads and maintains such discontinuities over a
width on only two cells. VH-1 does not use such a 
contact discontinuity steepener, although the Lagrangian remap scheme
used in VH-1 is believed to be better than 
the Eulerian PPM scheme studied by Fryxell \etal (1991) 
at maintaining sharp contact discontinuities without the use of a 
steepener (Blondin 1994).

We also employ a scheme similar to
that used by Stone \& Norman (1993) to reduce cooling in the region of
contact discontinuities to reduce the unphysical cooling
from the broadened discontinuity.

Although the efficient
shock and discontinuity capturing within PPM mitigate the
effects of finite numerical resolution, we can not
totally remove such effects.

\subsection{Increasing numerical resolution: models tbn\_1, tbn1a and tbn1b}

To investigate how sensitively our results depend on the numerical 
resolution of our models we ran three simulations using almost the
same starburst and ISM parameters but successively increasing the resolution.

Models tbn\_1 and tbn1a differ only in the cell size. Each computation
cell in model tbn\_1 representing a region $14.6 \pc \times 14.6 \pc$ 
in size, compared to cells of $7.3 \pc \times 7.3 \pc$ in model tbn1a.
Model tbn1b has triple the resolution of model tbn\_1, with
cells of $4.9 \pc \times 4.9 \pc$, but differs from the other
thick disk models in using a cylindrical rather than spherical 
starburst region.

The most dramatic differences between the three simulations is in the
amount of structure visible in the densest coolest gas. For example,
the number of superbubble shell fragments visible at $t = 7.5 \Myr$ 
increases significantly and their size decreases 
as the resolution increases (see Fig.~\ref{fig:res_lnum}),
for the reasons explained in Section.~\ref{sec:results_cellsize}.

Otherwise the overall morphology of the wind is very similar between
the different simulations, barring some minor differences in overall
volume occupied by the wind.

\begin{figure}
\centerline{
 \psfig{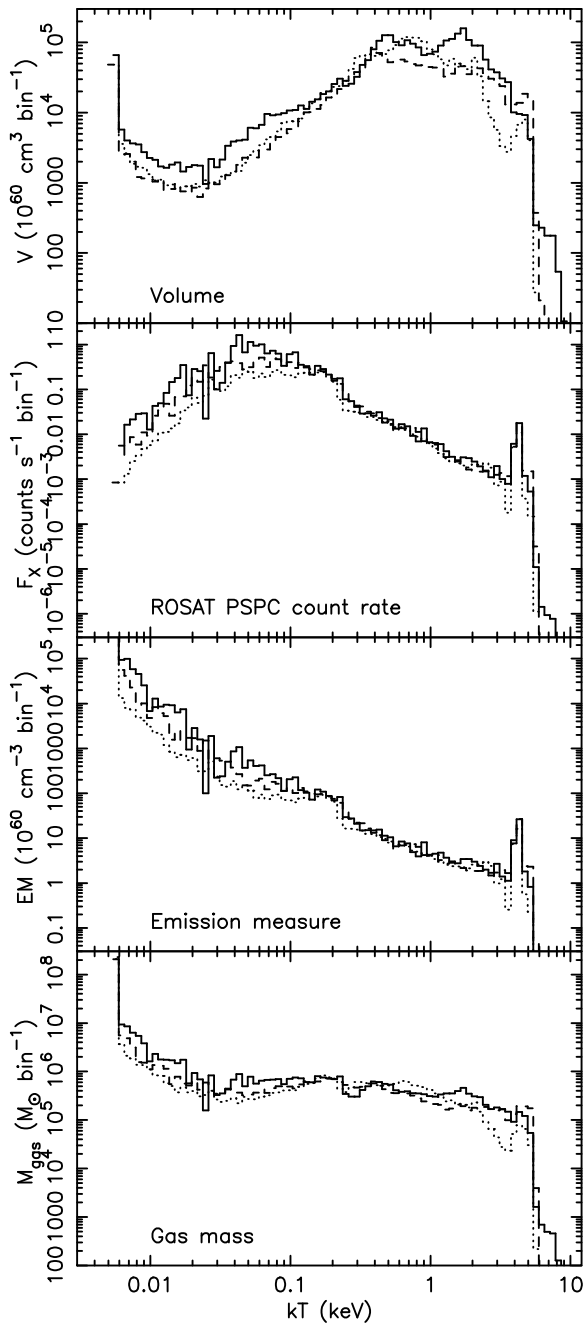}
	}
\caption[Temperature distributions at $t = 7.5 \Myr$ 
 in models tbn\_1, tbn1a and tbn1b]
 {The distribution of volume, X-ray flux, emission measure and mass with
 temperature at $t = 7.5 \Myr$ 
 in models tbn\_1 (solid line), tbn1a (dashed) and tbn1b (dotted).
 Note the extremely close agreement between the emission measure 
 and {\it ROSAT} count rate distributions
 for gas hotter than $kT \sim 0.1 \keV$.
 For gas cooler than this temperature the higher resolution models
 predict less emission.}
\label{fig:res_tkev}
\end{figure}

The properties of the X-ray emitting gas, whether measured in terms
of the distribution of volume, average density, emission measure 
or emitted X-ray flux as a function of temperature, are
all very similar between the three models when measured at the 
same epoch (see Fig.~\ref{fig:res_tkev} and Table~\ref{tab:main_results}).

Concentrating on the properties most important observationally,
the emission measure and {\it ROSAT} PSPC count rates, the agreement
between the different models is excellent at gas temperatures
above $kT \sim 0.1 \keV$. For gas with temperatures below this, the higher
resolution models predict systematically less emitting material
and less X-ray flux.

This is consistent with a picture of some, although not
necessarily all, of the
X-ray emission from gas at these lower temperatures being due to
numerically broadened contact discontinuities. 

The only other significant difference between these models we have
found is in the filling factors of the X-ray dominant gas 
(Table~\ref{tab:ffactors}), where model tbn\_1 has a much lower
filling factor than models tbn1a \& tbn1b. 
We do not believe the filling factors of the most of the other
models are as dependent on numerical resolution as model tbn\_1.
In particular, in the thin disk models the X-ray emission
is not purely from such interface regions (where numerically
broadened contact discontinuities are such a problem),
so increased numerical resolution should not lead to order of
magnitude increases in the filling factor of the X-ray
dominant gas. 

The models where the properties of the X-ray dominant gas
are most likely to be affected by resolution effects are
models tbn\_1, tbn\_2, tbn\_7 and mnd\_6. This is based on the 
number of computational cells the most X-ray luminous gas covers 
(\ie the gas contributing 
90 per cent of the soft X-ray counts in the {\it ROSAT}
band), which is much lower in these models than in any of the
other models.

\begin{figure}
\centerline{
 \psfig{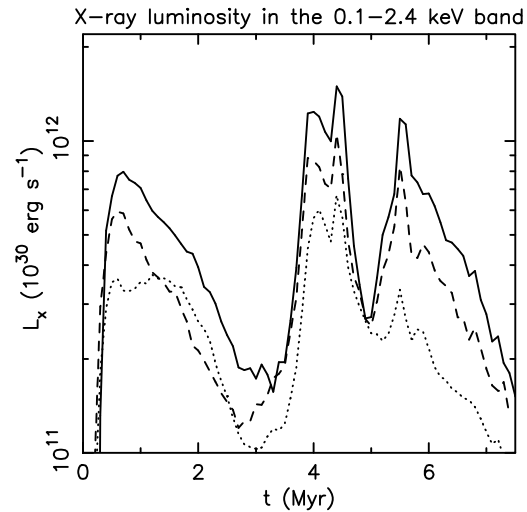}
	}
\caption[X-ray luminosity as function of time in models tbn\_1, tbn1a and tbn1b]
 {X-ray luminosity in the {\it ROSAT} $0.1$ -- $2.4 \keV$ energy
 band, as function of time in models tbn\_1 (solid line), tbn1a (dashed) 
 and tbn1b (dotted). Most of the variation in X-ray luminosity is
 due to changes in the properties of warm gas ($kT \ltsimm 0.1 \keV$)
 as the numerical resolution
 increases from model tbn\_1 to tbn1b.}
\label{fig:res_lx}
\end{figure}

Comparing the 
intrinsic X-ray luminosities in the
{\it ROSAT}  $0.1$ -- $2.4 \keV$ energy band as a function of time between the
three models also shows progressively lower X-ray luminosity 
accompanies higher numerical resolution (Fig.~\ref{fig:res_lx}). 
This figure probably exaggerates the importance of the
variation in X-ray luminosity
with resolution, as what can be measured in reality will be the
count rate of the attenuated emission, 
where the influence of the low temperature gas
so affected by numerical effects is significantly reduced.
There will be much less difference between the ``observable'' 
X-ray properties of these models than in their intrinsic
X-ray luminosities.

The coolest gas in these simulations is not resolved, and would
require quite substantial further increases in numerical resolution to
resolve if it were allowed to cool to $T \sim 10^{4} \K$.
The dramatic increase in the complexity of the structures associated
with the coolest gas with increasing resolution does suggest to
us that the complex filamentary structure of the optical
emission line gas observed in galactic winds (\cf McKeith \etal 1995; 
Lehnert \& Heckman 1996; Shopbell \& Bland-Hawthorn 1998)
will be produced as a natural
consequence of these models, once they
achieve the necessary resolution.

The level of variation found between these three simulations implies
that in these simulations, including the other thick and thin disk
models, the emission properties of the low filling factor warm
 gas responsible for
the majority of the
soft X-ray emission ($kT \ltsimm 0.1 \keV$) are partially controlled
by unresolved, numerically broadened, 
warm gas interfaces between cool dense gas and hotter wind material.
Nevertheless, the important finding that the X-ray dominant gas
has low filling factor ($\ltsimm 2$ per cent), is likely to be
robust.
	
Hotter gas within these winds ($kT \gtsimm 0.1 \keV$),
in particular volume filling components, 
seems well resolved, even in our lowest resolution simulations
and therefore its properties are reasonably quantitatively correct.

\section{Summary}
\label{sec:summary}

We have performed an extensive parameter study of starburst-driven
galactic winds using a high resolution 2-D hydrodynamical code, focusing
on the best studied galactic wind in M82. These simulations
allow us to study how the properties of galactic winds vary depending
on the influence of the host galaxy's ISM distribution,
starburst strength and star formation history, and the presence or
absence of additional mass-loading from clouds enveloped in the wind.

In this paper we have used this sample of twelve simulations to investigate
several aspects of galactic winds that are uncertain or have 
previously received little attention: the origin and filling factor of
the gas responsible for the observed soft X-ray emission; 
wind energetics and energy transport efficiencies and 
wind collimation and growth. In addition we have explored in detail
the influence of finite numerical resolution on the results of these
simulations.

The results of this study are summarised below:

\subsection{Wind collimation and confinement}

\begin{enumerate}
\item Thick ISM disks are required to reproduce the well collimated
wind observed in M82. Thin disk models 
(\eg Suchkov \etal 1994) have
wind opening angles that are too large, $\theta \gtsimm 90\deg$. Thick
disk models, such as that used by Tomisaka \& Bregman (1993), are much
better collimated, 
with typical opening angles $\theta \sim 40\deg$, much closer to the
observed value of $\theta \sim 30\deg$.

\item All the simulations, regardless of ISM distribution or star formation
history, fail to confine the size of the wind in the plane of the 
galaxy to the observed size of
roughly the size of the starburst region ($\sim 400 \pc$ in M82). 
The radius of the wind in the plane of the galaxy grows to between 1 -- 2 kpc
over the 10 Myr period covered by these simulations. The confinement of
galactic winds within the plane of the galaxy remains a problem, as initially
pointed out by Tenorio-Tagle \& Mu\~noz-Tu\~n\'on (1997).
There is clearly something interesting happening in the central regions
of starburst galaxies, perhaps related to the existing circum-nuclear molecular
rings, that must be able confine the base of the wind over
time-scales of $\sim 10 \Myr$.
\end{enumerate}

\subsection{X-ray emission in galactic winds}

\begin{enumerate}
\item The gas responsible for the majority of the soft X-ray emission
     detectable by {\it ROSAT} comes from very low filling factor
     ($\eta \sim 0.01$ to $2$ per cent) gas that contains very little
     of the total mass or energy of the wind. As a result 
     X-ray observations with {\it ROSAT}, {\it ASCA}, {\it Chandra} 
     or {\it XMM} will only give lower limits on the mass and energy 
     content of galactic winds.
\item The winds in these models contain gas at a wide range of temperatures
     and densities. In terms of a phase description
     there is a phase continuum of states rather
     than any particular well defined characteristic temperatures or 
     densities.
\item The majority of the wind volume in almost all of these models is filled
     with hot gas, covering the temperature range $10^{6.5}$ to 
     $10^{7.5} \K$. This gas is only a weak source of soft X-ray emission
     due to its low density.
\item The soft X-ray emission in these models does not come from
     any single well-defined temperature gas, but from a very
     wide range of temperatures from $T \sim 10^{5} \K$ to $10^{8} \K$.
     Fitting the resulting complex X-ray spectra with 
     standard simple spectral models
     is likely to give misleading results.
\item In all of our models, regions of interaction between the hot wind and
     cooler denser material (originally part of the ambient ISM) give
     rise to a large fraction of the soft X-ray emission. In these
     simulations this X-ray emission can be from shock-heated ambient
     gas,  and from the numerically broadened interfaces between cool
     dense and hot tenuous gas, although it is very difficult to
     discriminate between the two cases. In some of the
     models, in particular
     the mass-loaded models, the starburst region itself or the
     free wind region can be significant sources of soft X-ray
     emission. In the thin disk models shock-heated halo gas is also
     an important source of X-ray emission. 
\item The primary variable affecting the  soft X-ray luminosity
     of the galactic winds in these models is the density and distribution
     of the ambient, volume-filling, component of the ISM. The 
     energy injection rate from the starburst, and mass-loading from
     dense clouds, do have a significant but secondary effect on the
     X-ray luminosity.
\item Mass-loading of the wind, either within the starburst region
     itself, or from clouds distributed more generally throughout the ISM,
     is not necessary to produce galactic winds of M82's luminosity
     or characteristic soft X-ray temperature.
\item Hard X-ray emission from galactic winds is dominated by the
     starburst region itself. The ratio of hard X-ray to soft X-ray
     luminosity is typically $L_{\rm X,hard}/L_{\rm X,soft} \sim 10^{-2}$.
     We confirm S94's conclusion that the hard X-ray emission from
     starburst galaxies is {\em not} due to thermal emission
     associated with the starburst-driven wind. 
\end{enumerate}

\subsection{Wind energetics}

\begin{enumerate}

\item Galactic winds are efficient at transporting the energy supplied
by the starburst out of the plane of the galaxy. As radiative losses
are low, galactic winds seem likely to be good sources of heating
for the IGM and ICM. This is in contrast to their low efficiency
at moving mass out of the disk of starburst galaxies.

\item The energy within galactic winds is predominantly ($\sim 60$ per cent)
kinetic energy of
hot gas ($10^{6.5} \le T (K) \le 10^{7.5}$). 
This is significantly different from the standard superbubble, 
where the majority of the energy is in the 
thermal energy of the hot bubble interior. The change from 
thermally-dominated to kinetically-dominated energetics occurs 
when the superbubble breaks out of the disk.

\item Hot gas also contains the majority ($\sim 60$ per cent) of the total 
thermal energy content of these winds. Exceptions to this
rule are models with strong centralised mass-loading, where warm
gas ($10^{5.5} \le T (K) \le 10^{6.5}$) dominates the thermal energy content.

\item  The total energy content of galactic winds appears extremely difficult
to measure directly: Soft X-ray observations probe only denser low 
filling factor gas containing relatively little ($\ltsimm 10$ per cent)
of either the total thermal or kinetic energy. Measuring the
kinetic energy of the warm and hot phases is also hampered by the
lack of any direct measurements of this gas's velocity.

\end{enumerate}

\subsection{Concluding remarks}

Our findings have major implications for the 
ultimate aim behind the study of galactic winds: measuring
quantitatively the transport of mass, metal-enriched gas and energy
out of star-forming galaxies.

These simulations measure, for the first time, the observationally
important filling factor of the X-ray emitting gas.
In all models, even those including different forms of mass-loading,
we find that the soft X-ray emission from galactic winds comes 
from low filling factor ($\eta \ltsimm 2$ per cent) gas. This gas
contains only a small fraction of the mass and energy of the wind.
The majority of the thermal and kinetic
energy of these outflows is in a hot, volume filling, component,
which is extremely difficult to probe observationally due to its
low density and hence low emissivity.

We also find that galactic winds are efficient at transporting
energy out of the host galaxy, primarily in the form of the kinetic
energy of hot, $T \sim 10^{6.5}$ to $10^{7.5} \K$, gas. This is
an important finding, as it suggests that starburst-driven winds
are a good mechanism for reheating the IGM and ICM, as required by
recent observations.

In Paper II we shall present a 
direct comparison between the spectral properties
and spatial distribution of the soft X-ray emission in these
models and the existing X-ray data on M82, along with
a discussion of which of these models best reproduces M82's observed
properties.

It is a pleasure to thank the referee, A. Ferrara, 
both for the scientific value of his comments and the punctuality of
his response!
We would also like to thank the numerous people whose comments
on this work we have found enlightening, in particular Timothy Heckman, 
Martin Ward, Duncan Forbes, Trevor Ponman, Crystal Martin and
Martin Norbury. DKS acknowledges financial support from
a PPARC studentship, a Teaching Assistantship from the
School of Physics \& Astronomy at the University of Birmingham
and through NASA grant NAGW 3138. IRS acknowledges support from
a PPARC Advanced Fellowship.

\label{lastpage}

\begin{thebibliography}{}
\bibitem[]{}Allen M. L., Kronberg P. P., 1998, ApJ, 502, 218
\bibitem[]{}Arthur S. J., Henney, W. J, 1996, ApJ, 457, 752
\bibitem[]{}Begelman M. C., Fabian A. C., 1990, MNRAS, 244, 26P
\bibitem[]{}Bland-Hawthorn J., 1995, 
        Publ. Astron. Soc. Aust., 12, 190
\bibitem[]{}Blondin J. M., 1994, {\it The VH-1 Users Guide}, 
        University of Virginia
\bibitem[]{}Blondin J. M., Kallman T. R., Fryxell B. A., Taam R. E.,
        1990, ApJ, 356, 591
\bibitem[]{}Bradamante F., Matteucci F., D'Ercole A., 1998,
	A\&A, 337, 338
\bibitem[]{}Bregman J. N., 
        Schulman E., Tomisaka K., 1995, ApJ, 439, 155
\bibitem[]{}Calzetti D., Meurer G. R., Bohlin R. C., Garnett D. R.,
  Kinney A. L., Leitherer C., Storchi-Bergmann T., 1997,
  AJ, 114, 1834 
\bibitem[]{}Cappi M., \etal, 1999, A\&A in press (astro-ph/9908312)
\bibitem[]{}Chevalier R. A., Clegg A. W., 1985, 
        Nat, 317, 44 (CC)
\bibitem[]{}Cole S., Aragon-Salamanca A., Frenk C. S., Navarro J. F.,
	Zepf S. E., 1994, MNRAS, 271, 781
\bibitem[]{}Collela P., Woodward P. R., 
        1984, J. Comp. Phys., 54, 174
\bibitem[]{}Cottrell G. A., 1977, MNRAS, 178, 577 
\bibitem[]{} Cowie L. L., McKee C. F., Ostriker J. P., 
        1981, ApJ, 247, 908
\bibitem[]{}Dahlem M., 1997,  PASP, 109, 1298
\bibitem[]{}Dahlem M., Heckman T. M., Fabbiano G., Lehnert M. D., Gilmore D.,
        1996, ApJ, 461, 724
\bibitem[]{}Dahlem M., Weaver K. A., 
        Heckman T. M., 1998, ApJS, 118, 401
\bibitem[]{}de Gouveia Dal Pino E. M., Medina Tanco G. A., 1999,
	ApJ, 518, 129
\bibitem[]{}Dekel A., Silk J., 1986, ApJ, 303, 39
\bibitem[]{}Della Ceca R., Griffiths R. E., 
        Heckman T. M., 1997, ApJ, 485, 581
\bibitem[]{}De Young D. S., Heckman T. M., 1994, ApJ, 431, 598
\bibitem[]{}D'Ercole A., Brighenti F., 1999, MNRAS, in press
\bibitem[]{}Fabbiano G., 1988, ApJ, 330, 672
\bibitem[]{}Ferrara A., Shchekinov Yu., 1993, ApJ, 417, 595
\bibitem[]{}Field G.B., 1965, ApJ, 142, 531
\bibitem[]{}Freedman, W.L. \etal, 1994, ApJ, 427, 628
\bibitem[]{}Fryxell B., M\"{u}ller E., Arnett D., 1991, ApJ, 367, 619
\bibitem[]{}Godunov S. K., 1959, Mat. Sb. 47, 271
\bibitem[]{}Gonz\'alez Delgado R. M., Leitherer C., Heckman T., 
        Cervi\~no M., 1997, ApJ, 483, 705
\bibitem[]{}Gorjian V., 1996, AJ, 112, 1886 
\bibitem[]{}G\"{o}tz M., McKeith C. D., Downes D., Greve A., 1990,
        A\&A, 240, 52
\bibitem[]{}Hartquist T. W., Dyson J. E., Pettini M., Smith L. J., 
        1986, MNRAS, 221, 715
\bibitem[]{}Hartquist T. W., Dyson J. E., Williams R. J. R.,
	1997, ApJ, 482, 182
\bibitem[]{}Heckman T. M., 1998, in `Origins', ASP Conference Series 145,
      Eds. C. E. Woodward, J. M. Shull and H. A. Thronson Jr, 127
\bibitem[]{}Heckman T. M., 
        Armus L., Miley G. K., 1990, ApJS, 74, 833
\bibitem[]{}Heckman T. M., Lehnert M. D.,
        Armus L., 1993, in ``The Environment and Evolution of
	Galaxies,'' Eds. J. M. Shull and H. A. Thronson Jr,
	(Dordrect:Kluwer), 455
\bibitem[]{}Heckman T.M., Lehnert M.D., 
	Strickland D.K., Armus L., 1999, submitted to ApJ
\bibitem[]{}Kauffman G., Guiderdoni B., White S. D. M., MNRAS, 267, 981
\bibitem[]{}Klein R. I., 
        McKee C. F., Collela P., 1994, ApJ, 420, 213
\bibitem[]{}Klein U., Wielebinski R., Morsi H. W.,
	1988, A\&A, 190, 41
\bibitem[]{}Lehnert M. D., Heckman T. M., 1996, ApJ, 462, 651
\bibitem[]{}Leitherer C., Heckman T. M., 1995, ApJS, 96, 9 (LH95)
\bibitem[]{}Leitherer C., Robert C., 
        Drissen L., 1992, ApJ, 401, 596
\bibitem[]{}Loewenstein M., Mushotzky R. F., 1996, ApJ, 466, 695
\bibitem[]{}Lugten J. B., Watson D. M., Crawford M. K., Genzel R.,
  1986, ApJ, 311, L51
\bibitem[]{}McCarthy P. J., Heckman T., 
        van Breugel W., 1987, AJ, 92, 264
\bibitem[]{}Mac Low M.-M., 
  Ferrara A., 1999, ApJ, 513, 142
\bibitem[]{}Mac Low M.-M., McCray R., 1988, ApJ, 324, 776
\bibitem[]{}McKeith C. D., Greve A., Downes D., Prada F., 1995, A\&A, 293, 703
\bibitem[]{}McLeod K. K., Rieke G. H., Rieke M. J., Kelly D. M., 1993
        ApJ, 412, 111
\bibitem[]{}Martin C. L., 1999, ApJ, 513, 156
\bibitem[]{}Masai K., 1994, ApJ, 437, 770
\bibitem[]{}Meurer G. R., Heckman T. M., Leitherer C., Kinney A.,
         Robert C., Garnett D. R., 1995,  AJ, 110, 2665
\bibitem[]{}Mewe R., Kaastra J. S., 
        Liedahl D. A., 1995, Legacy, 6, 16
\bibitem[]{}Miyamoto M., Nagai R., 1975, PASJ, 27, 533
\bibitem[]{}Moran E. C., Lehnert M. D., 1997, ApJ, 478, 172
\bibitem[]{}Morrison R., McCammmon D., 1983, ApJ, 270, 119
\bibitem[]{}Muxlow T. W. B., Pedlar A., Wilkinson P. N., Axon D. J., 
        Sanders E. M., de Bruyn A. G., 1994, MNRAS, 266, 455
\bibitem[]{}Nakai N., Hayashi M., 
        Handa T., Sofue Y., Hasegawa T., 1987,
        PASJ, 39, 685
\bibitem[]{}Norman C. A., Ferrara A., 1996, ApJ, 467, 280
\bibitem[]{}O'Connell R. W., Mangano J. J., 1978, ApJ., 221, 62
\bibitem[]{}O'Connell R.W., Gallagher J.S., Hunter D.A.,
        Colley W.N. 1995, ApJ, 446, L1
\bibitem[]{}Pettini M., Steidel C. C., Adelburger K. L.,
	Dickinson M., Giavalisco M., 1999, ApJ, in press (astro-ph/9908007)
\bibitem[]{}Ponman T. J., Cannon D. B., Navarro J. F.,
	1999, Nat, 397, 135 
\bibitem[]{} Ptak A. F., 1997, Ph.D. Thesis, 
        University of Maryland
\bibitem[]{}Ptak A., Serlemitsos P., Yaqoob T., Mushotzky R., Tsuru T.,
        1997, AJ, 113, 1286
\bibitem[]{}Puxley P. J., Brand P. W. J. L., 
        Moore T. J. T., Mountain C. M.,
        Nakai N., Yamashita T., 1989, ApJ, 345, 163
\bibitem[]{}Raymond J. C., Smith B. W., 1977, ApJS, 35, 419

\bibitem[]{}Read A. M., Ponman T. J., 
        Strickland D. K., 1997, MNRAS, 286, 626
\bibitem[]{}Reuter H.-P., Klein U., Lesch H., Wielebinski R.,
        Kronberg P. P., 1994, A\&A, 282, 724
\bibitem[]{}Salpeter E. E., 1955, ApJ, 121, 161
\bibitem[]{}Satyapal S., Watson D. M., 
        Pipher J. L., Forrest W. J., Greenhouse M. A.,
        Smith H. A., Fischer J., Woodward C. E., 1997, ApJ, 483, 148
\bibitem[]{}Seaquist E. R., Odegard N., 1991, ApJ, 369, 320
\bibitem[]{}Seaquist E. R., Carlstrom J. E., Bryant P. M.,
  Bell M. B., 1996, ApJ, 465, 691
\bibitem[]{}Shopbell P. L., Bland-Hawthorn J., 1998, ApJ, 493, 129
\bibitem[]{}Silich S. A., Tenorio-Tagle G.,
	1998, MNRAS, 299, 249
\bibitem[]{}Somerville R. S., Primack J. R., 
	1999, MNRAS in press (astro-ph/9802268)
\bibitem[]{}Stark, A.A., Gammie, C.F., 
        Wilson, R.W., Bally, J., Linke, R.A.,
        Heiles, C., Hurwitz, M., 1992, ApJS, 79, 77
\bibitem[]{}Stevens I. R., Blondin J. M., Pollock A. M. T., 
         1992, ApJ, 386, 265
\bibitem[]{}Stone J. M., Norman M. L., 1993, ApJ, 413, 198
\bibitem[]{}Strickland D. K., Ponman T. J., 
        Stevens I. R., 1997, A\&A, 320, 378
\bibitem[]{}Strickland D. K., Stevens I. R., 1998, MNRAS, 297, 747
\bibitem[]{}Strickland D. K., Stevens I. R., 1999, MNRAS, 306, 43
\bibitem[]{}Strickland R., Blondin J. M., 1995, ApJ, 449, 727
\bibitem[]{}Suchkov A. A., Berman V. G., Heckman T. M., Balsara D. S., 
        1996, ApJ, 463, 528 (S96)
\bibitem[]{}Suchkov A. A., Balsara D. S., Heckman T. M., Leitherer C., 
        1994, ApJ, 430, 511 (S94)
\bibitem[]{}Swaters R. A., Sancisi R., van der Hulst J. M.,
	1997, ApJ, 491, 140
\bibitem[]{}Tenorio-Tagle G., Mu\~noz-Tu\~n\'on C., 
  1997, ApJ, 478, 134
\bibitem[]{}Tenorio-Tagle G., Mu\~noz-Tu\~n\'on C., 
  1998, MNRAS, 293, 299 (TT)
\bibitem[]{}Tenorio-Tagle G., Mu\~noz-Tu\~n\'on C., 
  P\'{e}rez E., Melnick J., 1997,
        ApJ, 490, L179
\bibitem[]{}Thornton K., Gaudlitz M., Janka H.-Th., Steinmetz M.,
	1998, ApJ, 500, 95
\bibitem[]{}Tomisaka K., Bregman J. N., 1993, PASJ, 45, 513 (TB)
\bibitem[]{}Tomisaka K., Ikeuchi S., 1988, ApJ, 220, 695 (TI)
\bibitem[]{}Tsuru T. G., Awaki H., Koyama K., Ptak A., 
  1997, PASJ, 49, 619
\bibitem[]{}Vader J. P., 1986, ApJ, 305, 669
\bibitem[]{}Wang Q. D., Walterbos R. A. M., Steakley M. F., Norman C. A., 
        Braun R. A., 1995, ApJ, 439, 176
\bibitem[]{}Watson M. G., Stanger V., Griffiths R. E., 1984, ApJ, 286, 144
\bibitem[]{}Weaver R., McCray R., Castor J., Shapiro P., Moore R.,
        1977, ApJ, 218, 377
\bibitem[]{}Zezas A. L., Georgantopoulos I., Ward M. J., 
	1998, MNRAS, 301, 915
\end{thebibliography}
\end{document}